\documentclass[twocolumn]{aastex631}

\usepackage{amsmath} 

\usepackage{color}

\newcommand{\Ax}{A^{\times}_{\mathrm{lens}}}
\newcommand{\Cov}{{\mathrm{Cov}}}

\usepackage{savesym}
\savesymbol{tablenum}
\usepackage{siunitx}
\restoresymbol{SIX}{tablenum}

\def\citejap#1{\citeauthor{#1}\ \citeyear{#1}}

\begin{document}

\title{The Atacama Cosmology Telescope DR6 and DESI: Structure growth measurements from the cross-correlation of DESI Legacy Imaging galaxies and CMB lensing from ACT DR6 and \textit{Planck} PR4}

\author[0000-0001-7805-1068]{Frank~J.~Qu}
\affiliation{Kavli Institute for Particle Astrophysics and Cosmology, 382 Via Pueblo Mall Stanford, CA 94305-4060, USA}
\affiliation{SLAC National Accelerator Laboratory 2575 Sand Hill Road Menlo Park, California 94025, USA}
\affiliation{DAMTP, Centre for Mathematical Sciences, University of Cambridge, Wilberforce Road, Cambridge CB3 OWA, UK}
\affiliation{Kavli Institute for Cosmology Cambridge, Madingley Road, Cambridge CB3 0HA, UK}

\author{Qianjun Hang}\affiliation{Department of Physics \& Astronomy, University College London, Gower Street, London, WC1E 6BT, UK}

\author{Gerrit Farren}\affiliation{Lawrence Berkeley National Laboratory, 1 Cyclotron Road, Berkeley, CA 94720, USA}

\author{Boris Bolliet}\affiliation{DAMTP, Centre for Mathematical Sciences, University of Cambridge, Wilberforce Road, Cambridge CB3 OWA, UK}
\affiliation{Kavli Institute for Cosmology Cambridge, Madingley Road, Cambridge CB3 0HA, UK}

\author{Jessica Nicole Aguilar}\affiliation{Lawrence Berkeley National Laboratory, 1 Cyclotron Road, Berkeley, CA 94720, USA}
\author[0000-0001-6098-7247]{Steven Ahlen}\affiliation{Physics Dept., Boston University, 590 Commonwealth Avenue, Boston, MA 02215, USA}

\author[0000-0002-3757-6359]{Shadab Alam}\affiliation{Tata Institute of Fundamental Research, Homi Bhabha Road, Mumbai 400005, India}

\author{David Brooks}\affiliation{Department of Physics \& Astronomy, University College London, Gower Street, London, WC1E 6BT, UK}

\author{Yan-Chuan Cai}\affiliation{Institute for Astronomy, University of Edinburgh, Royal Observatory, Blackford Hill, Edinburgh EH9 3HJ, UK}

\author{Erminia Calabrese}
\affiliation{School of Physics and Astronomy, Cardiff University, The Parade, Cardiff, Wales CF24 3AA, UK}

\author{Todd Claybaugh}\affiliation{Lawrence Berkeley National Laboratory, 1 Cyclotron Road, Berkeley, CA 94720, USA}

\author[0000-0002-1769-1640]{Axel de la Macorra}\affiliation{Instituto de F\'{\i}sica, Universidad Nacional Aut\'{o}noma de M\'{e}xico,  Cd. de M\'{e}xico  C.P. 04510,  M\'{e}xico}

\author[0000-0002-3169-9761]{Mark~J.~Devlin}\affiliation{Department of Physics and Astronomy, University of Pennsylvania, 209 South 33rd Street, Philadelphia, PA, USA 19104}

\author{Peter Doel}\affiliation{Department of Physics \& Astronomy, University College London, Gower Street, London, WC1E 6BT, UK}

\author[0009-0001-3987-7104]{Carmen~Embil-Villagra}
\affiliation{DAMTP, Centre for Mathematical Sciences, University of Cambridge, Wilberforce Road, Cambridge CB3 OWA, UK}

\author[0000-0003-4992-7854]{Simone Ferraro}\affiliation{Lawrence Berkeley National Laboratory, 1 Cyclotron Road, Berkeley, CA 94720, USA}

\affiliation{University of California, Berkeley, 110 Sproul Hall \#5800 Berkeley, CA 94720, USA}
\author[0000-0002-3033-7312]{Andreu Font-Ribera}\affiliation{Department of Physics \& Astronomy, University College London, Gower Street, London, WC1E 6BT, UK}
\affiliation{Institut de F\'{i}sica d’Altes Energies (IFAE), The Barcelona Institute of Science and Technology, Campus UAB, 08193 Bellaterra Barcelona, Spain}
\author[0000-0002-2890-3725]{Jaime E. Forero-Romero}\affiliation{Departamento de F\'isica, Universidad de los Andes, Cra. 1 No. 18A-10, Edificio Ip, CP 111711, Bogot\'a, Colombia}
\affiliation{Observatorio Astron\'omico, Universidad de los Andes, Cra. 1 No. 18A-10, Edificio H, CP 111711 Bogot\'a, Colombia}
\author{Enrique Gazta\~naga}\affiliation{Institut d'Estudis Espacials de Catalunya (IEEC), 08034 Barcelona, Spain}

\author[0000-0002-3589-8637]{Vera~Gluscevic}
\affiliation{Department of Physics and Astronomy,  University of Southern California, Los Angeles, CA, 90007, USA}

\author[0000-0003-3142-233X]{Satya Gontcho A Gontcho}\affiliation{Lawrence Berkeley National Laboratory, 1 Cyclotron Road, Berkeley, CA 94720, USA}
\author{Gaston Gutierrez}\affiliation{Fermi National Accelerator Laboratory, PO Box 500, Batavia, IL 60510, USA}
\author[0000-0002-1081-9410]{Cullan Howlett}\affiliation{School of Mathematics and Physics, University of Queensland, 4072, Australia}
\author{Robert Kehoe}\affiliation{Department of Physics, Southern Methodist University, 3215 Daniel Avenue, Dallas, TX 75275, USA}

\author[0000-0002-0935-3270]{Joshua~Kim}
\affiliation{Department of Physics and Astronomy, University of Pennsylvania, Philadelphia, PA 19104, USA}

\author[0000-0001-6356-7424]{Anthony Kremin}\affiliation{Lawrence Berkeley National Laboratory, 1 Cyclotron Road, Berkeley, CA 94720, USA}
\author{Andrew Lambert}\affiliation{Lawrence Berkeley National Laboratory, 1 Cyclotron Road, Berkeley, CA 94720, USA}
\author[0000-0003-1838-8528]{Martin Landriau}\affiliation{Lawrence Berkeley National Laboratory, 1 Cyclotron Road, Berkeley, CA 94720, USA}
\author[0000-0001-7178-8868]{Laurent Le Guillou}\affiliation{Sorbonne Universit\'{e}, CNRS/IN2P3, Laboratoire de Physique Nucl\'{e}aire et de Hautes Energies (LPNHE), FR-75005 Paris, France}
\author[0000-0003-1887-1018]{Michael Levi}\affiliation{Lawrence Berkeley National Laboratory, 1 Cyclotron Road, Berkeley, CA 94720, USA}

\author[0000-0002-6849-4217]{Thibaut Louis
}\affiliation{Université Paris-Saclay, CNRS/IN2P3, IJCLab, 91405 Orsay, France}

\author[0000-0002-1125-7384]{Aaron Meisner}\affiliation{NSF NOIRLab, 950 N. Cherry Ave., Tucson, AZ 85719, USA}
\author{Ramon Miquel}\affiliation{Instituci\'{o} Catalana de Recerca i Estudis Avan\c{c}ats, Passeig de Llu\'{\i}s Companys, 23, 08010 Barcelona, Spain}
\affiliation{Institut de F\'{i}sica d’Altes Energies (IFAE), The Barcelona Institute of Science and Technology, Campus UAB, 08193 Bellaterra Barcelona, Spain}
\author[0000-0002-2733-4559]{John Moustakas}\affiliation{Department of Physics and Astronomy, Siena College, 515 Loudon Road, Loudonville, NY 12211, USA}
\author[0000-0001-8684-2222]{Jeffrey A. Newman}\affiliation{Department of Physics \& Astronomy and Pittsburgh Particle Physics, Astrophysics, and Cosmology Center (PITT PACC), University of Pittsburgh, 3941 O'Hara Street, Pittsburgh, PA 15260, USA}
\author[0000-0002-1544-8946]{Gustavo Niz}\affiliation{Departamento de F\'{i}sica, Universidad de Guanajuato - DCI, C.P. 37150, Leon, Guanajuato, M\'{e}xico}
\affiliation{Instituto Avanzado de Cosmolog\'{\i}a A.~C., San Marcos 11 - Atenas 202. Magdalena Contreras, 10720. Ciudad de M\'{e}xico, M\'{e}xico}
\author{John Peacock}\affiliation{Institute for Astronomy, University of Edinburgh, Royal Observatory, Blackford Hill, Edinburgh EH9 3HJ, UK}
\author[0000-0002-0644-5727]{Will Percival}\affiliation{Department of Physics and Astronomy, University of Waterloo, 200 University Ave W, Waterloo, ON N2L 3G1, Canada}
\affiliation{Perimeter Institute for Theoretical Physics, 31 Caroline St. North, Waterloo, ON N2L 2Y5, Canada}
\affiliation{Waterloo Centre for Astrophysics, University of Waterloo, 200 University Ave W, Waterloo, ON N2L 3G1, Canada}
\author{Claire Poppett}\affiliation{Lawrence Berkeley National Laboratory, 1 Cyclotron Road, Berkeley, CA 94720, USA}
\affiliation{Space Sciences Laboratory, University of California, Berkeley, 7 Gauss Way, Berkeley, CA  94720, USA}
\affiliation{University of California, Berkeley, 110 Sproul Hall \#5800 Berkeley, CA 94720, USA}
\author[0000-0001-7145-8674]{Francisco Prada}\affiliation{Instituto de Astrof\'{i}sica de Andaluc\'{i}a (CSIC), Glorieta de la Astronom\'{i}a, s/n, E-18008 Granada, Spain}
\author[0000-0001-6979-0125]{Ignasi P\'erez-R\`afols}\affiliation{Departament de F\'isica, EEBE, Universitat Polit\`ecnica de Catalunya, c/Eduard Maristany 10, 08930 Barcelona, Spain}
\author{Graziano Rossi}\affiliation{Department of Physics and Astronomy, Sejong University, Seoul, 143-747, Korea}
\author[0000-0002-9646-8198]{Eusebio Sanchez}\affiliation{CIEMAT, Avenida Complutense 40, E-28040 Madrid, Spain}

\author{David Schlegel}\affiliation{Lawrence Berkeley National Laboratory, 1 Cyclotron Road, Berkeley, CA 94720, USA}

\author[0000-0001-6731-0351]{Neelima Sehgal}
\affiliation{Physics and Astronomy Department, Stony Brook University, Stony Brook, NY 11794, USA}

\author[0000-0001-6731-0351]{Shabbir Shaikh}
\affiliation{School of Earth and Space Exploration, Arizona State University, Tempe, AZ 85287, USA}

\author{Blake Sherwin}\affiliation{DAMTP, Centre for Mathematical Sciences, University of Cambridge, Wilberforce Road, Cambridge CB3 OWA, UK}
\affiliation{Kavli Institute for Cosmology Cambridge, Madingley Road, Cambridge CB3 0HA, UK}

\author[0000-0002-8149-1352]{Crist\'obal Sif\'on}
\affiliation{Instituto de F\'isica, Pontificia Universidad Cat\'olica de Valpara\'iso, Casilla 4059, Valpara\'iso, Chile}

\author{Michael Schubnell}\affiliation{Department of Physics, University of Michigan, Ann Arbor, MI 48109, USA}
\affiliation{University of Michigan, Ann Arbor, MI 48109, USA}
\author{David Sprayberry}\affiliation{NSF NOIRLab, 950 N. Cherry Ave., Tucson, AZ 85719, USA}
\author[0000-0003-1704-0781]{Gregory Tarl\'{e}}\affiliation{University of Michigan, Ann Arbor, MI 48109, USA}
\author{Benjamin Alan Weaver}\affiliation{NSF NOIRLab, 950 N. Cherry Ave., Tucson, AZ 85719, USA}
\author[0000-0002-7567-4451]{Edward J. Wollack}\affiliation{NASA Goddard Spaceflight Center, 8800 Greenbelt Rd, Greenbelt, MD 20771, USA}
\author[0000-0002-6684-3997]{Hu Zou}\affiliation{National Astronomical Observatories, Chinese Academy of Sciences, A20 Datun Rd., Chaoyang District, Beijing, 100012, P.R. China}

\correspondingauthor{Frank~J.~Qu}
\email{jq247@cantab.ac.uk}


\begin{abstract}
We measure the growth of cosmic density fluctuations on large scales and across the redshift range $0.3<z<0.8$ through the cross-correlation of the ACT DR6 CMB lensing map and galaxies from the DESI Legacy Survey, using three galaxy samples spanning the redshifts of $0.3 \lesssim z  \lesssim 0.45$, $0.45  \lesssim z  \lesssim0.6$, $0.6 \lesssim z  \lesssim 0.8$. We adopt a scale cut where non-linear effects are negligible, so that the cosmological constraints are derived from the linear regime. We determine the amplitude of matter fluctuations over all three redshift bins using ACT data alone to be $S_8\equiv\sigma_8(\Omega_m/0.3)^{0.5}=0.772\pm0.040$ in a joint analysis combining the three redshift bins and  ACT lensing alone. Using a combination of ACT and \textit{Planck} data we obtain $S_8=0.765\pm0.032$. The lowest redshift bin used is the least constraining and exhibits a $\sim2\sigma$ tension with the other redshift bins; thus
we also report constraints excluding the first redshift bin, giving $S_8=0.785\pm0.033$ for the combination of ACT and \textit{Planck}.
This result is in excellent agreement at the $0.3\sigma$ level with measurements from galaxy lensing, but is $1.8\sigma$ lower than predictions based on \textit{Planck} primary CMB data. 
Understanding whether this hint of discrepancy in the growth of structure at low redshifts arises from a fluctuation, from systematics in data, or from new physics, is a high priority for forthcoming CMB lensing and galaxy cross-correlation analyses.


\end{abstract}

\section{Introduction} \label{sec:intro}

The standard cosmological model, known as $\Lambda$CDM, has provided a remarkably successful framework for understanding the large-scale structure and evolution of the universe. One of its successes is its predictive power over a wide range of redshifts.
The way in which structure grows and clusters from initial primordial seeds depends sensitively on the parameters of the model, on the underlying theory of gravity \citep[e.g.,][]{PhysRevD.109.083540}, and on the details of the expansion history \citep[e.g.,][]{Sailer:2021yzm}. Thus, measuring the large-scale structure growth in the universe and comparing it with an extrapolation from CMB data is a powerful test for cosmological models.

    Multiple observables of cosmological large-scale structure provide a measure of density fluctuations at low redshifts. Some of these observables include gravitational lensing measurements, Sunyaev--Zeldovich (SZ) effects \citep{Carlstrom_2002,Horowitz_2017,bolliet2018cosmological}, redshift-space distortions \citep{Ivanov_2020,d_Amico_2020,Colas_2020}, galaxy cluster counts \citep{planckcluster2016,Haan_2016,Abbott_2020cluster, salcedo2023dark,bocquet2024spt} and peculiar velocities \citep{Howlett:2022len,Saulder:2023oqm,Stahl:2021mat}. These analyses cover a broad range of scales and redshifts, and are subject to different systematic effects. For ease of comparison, it is conventional to report the amplitude of matter fluctuations as $\sigma_8$ or $S_8\sim\sigma_8\sqrt{\Omega_m/0.3}$, where $\sigma_8$ is defined as the root-mean-square amplitude of linear fluctuations at present, smoothed on scales of $8h^{-1}\si{Mpc}$. 

    Several recent measurements have reported values of $S_8$ approximately $2-3\sigma$ lower than predicted from a $\Lambda$CDM fit to the primary CMB (e.g. $S_8=0.830\pm0.013$ from {\it Planck\/} PR4: \citejap{Rosenberg_2022}), raising questions about a potential breakdown of the $\Lambda$CDM model at low redshift. Examples of such results yielding low values of $S_8$ include galaxy weak lensing (in particular, the combination of cosmic shear and galaxy clustering): the Dark Energy Survey (DES), the Kilo-Degree Survey (KiDS), and the Hyper Suprime-Cam (HSC) obtain $S_8 =0.782\pm 0.019$,  $S_8 = 0.765^{+0.017}_{-0.016}$, $S_8=0.775^{+0.043}_{-0.038}$, respectively \citep{2022PhRvD.105b3520A,2021A+A...646A.140H,2023PhRvD.108l3521S}. A joint reanalysis of the DES and KiDS cosmic shear data \citep{2023DESKiDs} results in a slightly higher $S_8=0.790^{+0.018}_{-0.014}$.
    Cross-correlations between \textit{Planck} CMB lensing and 
    DESI Luminous Red Galaxy (LRG) targets \citep{White:2021yvw} or galaxies from the Baryonic Oscillation Spectroscopic Survey (BOSS) also indicate $S_8$ values that are $2-3\sigma$ low \citep{Singh:2018kmr, Chen_2022}. Results from the cross-correlation between various data sets from DES and CMB lensing from the South Pole Telescope (SPT) and \textit{Planck} \citep{2023PhRvD.107b3530C,2023PhRvD.107b3531A} have also found discrepancies at the $2.2$ and $3\sigma$ level. A similar result is found with some cross-correlations with ACT; albeit with large uncertainties,  CMB lensing from ACT DR4 and \textit{Planck} correlations with galaxy shear from KiDS-1000 \citep{Robertson_2021}, found $S_8=0.64\pm0.08$. A similar cross-correlation but using galaxy clustering from DES-Y3 \citep{2024JCAP...01..033M} instead results in $S_8=0.75^{+0.04}_{-0.05}$. \cite{ACT:2023skz} performed cross-correlation of ACT DR4 CMB lensing and DES galaxy shear, resulting in $S_8=0.782\pm0.059$. This collection of $S_8$ measurements that consistently return values lower than those measured by the primary CMB is known as the $S_8$ tension; while none of them are inconsistent with \textit{Planck} at high statistical significance, it is striking that they all tend to measure a low $S_8$.

It is however worth pointing out that not all low redshift measurements find low values of $S_8$. Notably,  the CMB lensing auto spectrum is in excellent agreement with the primary CMB predictions,  CMB lensing is arguably one of the cleanest probes of low redshift matter fluctuation, given that it probes directly the gravitational potential on mostly linear scales across a large range of redshifts ($z\approx0.5-5$), and is based on robust and well-understood statistical properties of the CMB. Recent measurements from the Atacama Cosmology Telescope \citep[ACT; ][]{qu2023atacama,ACT:2023kun} and Baryonic Acoustic Oscillation (BAO) data results in $S_8=0.840\pm0.028$ from ACT alone and $S_8=0.831\pm0.023$ in combination with the latest \textit{Planck} PR4 lensing \citep{Carron:2022}. Similar consistent measurements are obtained with SPT-3G \citep{Pan_2023}: $S_8=0.836\pm0.039$.

Furthermore, some improved analyses including more data of previous datasets suggesting low $S_8$ values have found less discrepant results compared to earlier measurements. Examples include the cross-correlation between CMB lensing from ACT DR6 and \textit{Planck} with unWISE galaxies \citep{farren2023atacama}, which resulted in $S_8=0.810\pm0.015$, the cross correlation between \textit{Planck} PR3 lensing and DESI LRGs with  $S_8=0.763\pm0.023$ \citep{Sailer2024,Kim2024}, and the cross-correlation between DESI Bright Galaxy Survey (BGS) and LRG, and the DES Year 3 galaxy shear in \cite{2024arXiv240704795C}, with $S_8=0.850^{+0.042}_{-0.050}$. We refer the interested reader to Fig. \ref{fig: compilation_S8} to a summary of the relevant $S_8$ measurements.

The present paper is especially concerned with
a further CMB lensing cross-correlation measurement that reported a low normalization. \cite{hang2021}  measured the cross-correlation between \textit{Planck } and the Dark Energy Spectroscopic Instrument
(DESI) legacy galaxy catalog, divided into four tomographic bins; from this, they inferred a constraint on $S^{\times}_8\equiv\sigma_8(\Omega_m/0.3)^{0.79}=0.758\pm0.023$, which is $2.8\sigma$ lower than the \textit{Planck} CMB prediction. Here, we re-examine the constraints obtained by \cite{hang2021},  motivated by the availability of lower noise CMB lensing maps from ACT DR6 and Planck PR4. In addition to these new datasets, we provide improved treatment of mode couplings across bandpowers due to the presence of a mask, a more accurate estimation of the covariance matrix based on simulations, a large suite of systematic tests, and use more conservative scale cuts choices used.

The paper is structured as follows: In Sec.~\ref{data} we present the datasets used in this analysis. In Sec.~\ref{ref.measurement} we discuss our measurement pipeline and scale cuts.  The verification of the pipeline and the covariance matrices are described in Sec.~\ref{sec. pipeline_verification}. We present a series of null and consistency tests in Sec.~\ref{sec.null}. In Sec.~\ref{sec.model} we discuss the modeling of the angular power spectra. Our results are given in Sec.~\ref{sec.results}, and we place these measurements in context in Sec.~\ref{sec.discussion}.

\section{Data used} \label{data}

We use the photometric DESI Legacy Imaging Survey galaxy catalog described in \cite{hang2021} and the corresponding galaxy density contrast maps in four tomographic bins. 
We conservatively restrict our analysis to linear scales (see Sec.~\ref{sec: scale cuts}) and therefore we exclude the lowest tomographic bin (bin 0) where all bandpowers receive significant non-linear contributions.
In Sec.~\ref{Sec. galaxy} we briefly describe this dataset, and refer the interested readers to \cite{hang2021} for more details. We then describe in Sec.~\ref{act_map} the lensing convergence map obtained using the ACT DR6 data and in Sec.~\ref{planck_map} the lensing convergence from \textit{Planck} PR4.

\begin{figure*}
 \centering
  \includegraphics[width=0.8\linewidth]{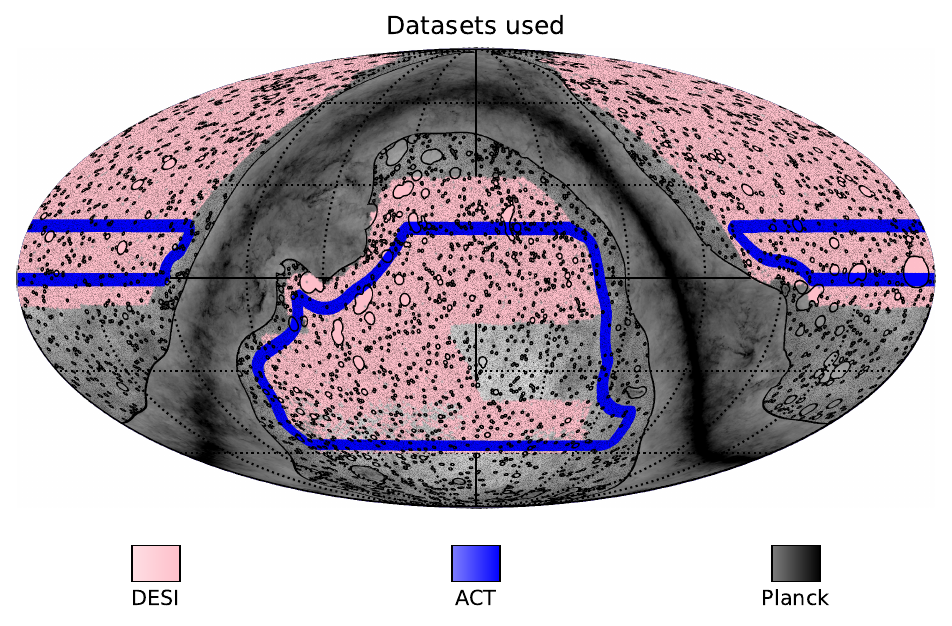}
 \caption{Overlap of the DESI imaging galaxies with the ACT DR6 lensing map (area within the blue contours) and the \textit{Planck} PR4 lensing footprint (black contours). The overlaid grayscale background is a Galactic dust map from \textit{Planck} \citep{1502.01588}.} \label{Fig.footprint}
  \end{figure*}

   \begin{figure}
 \centering
  \includegraphics[width=\linewidth]{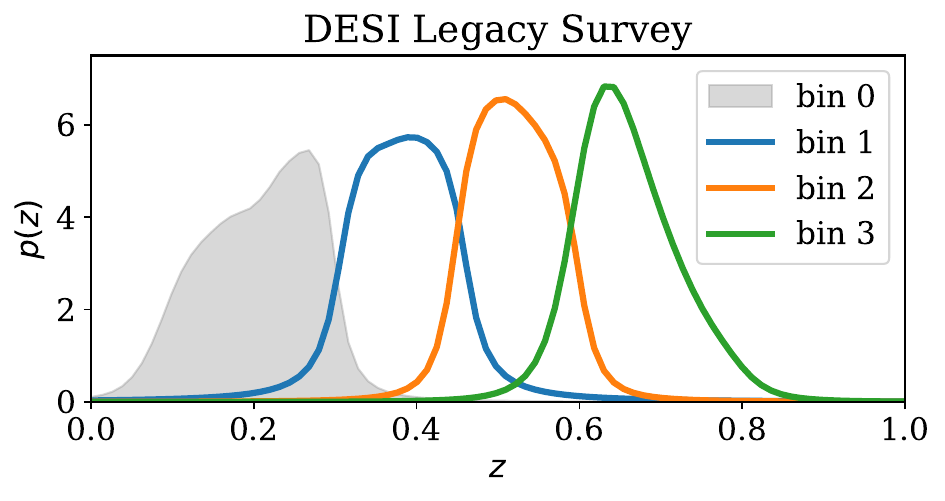}
 \caption{ The calibrated photometric redshift distribution of each tomographic bin of the DESI Legacy Imaging galaxy sample from \cite{hang2021}.} \label{Fig.dndz}
  \end{figure}

\subsection{DESI Legacy Imaging Surveys DR8}\label{Sec. galaxy}

The DESI Legacy Imaging Surveys \citep{DeyLegacy2019} are the union of three public imaging galaxy surveys: the Dark Energy Camera Legacy Survey \citep[DECaLS;][] {2015AJ....150..150F,2005astro.ph.10346T}, the Mayall $z$-band Legacy Survey \citep[MzLS;][]{2016SPIE.9908E..2CD}, and the Beijing-Arizona Sky Survey \citep[BASS;][]{2004SPIE.5492..787W}, altogether covering a total area of $\sim 14,000 \deg^2$. 
The photometry was measured in $grz$ bands, with the addition of the WISE fluxes ($W_1$, $W_2$, $W_3$) from the 4-year NEOWISE-Reactivation imaging, force-photometered in the unWISE maps at the location of the Legacy Survey sources \citep[][]{Wright2010, Schlafly_2019}.
This paper uses the public DR8\footnote{\url{http://legacysurvey.org/dr8/}} data, the first release to include both images and catalogs from all three of the Legacy Surveys. 
Note that this data set has been superseded by the subsequent data releases, DR9 and DR10\footnote{\url{https://www.legacysurvey.org/}}. The major updates include more recent data releases from the DECaLS, BASS, MzLS, and the NEOWISE-Reactivation, and additionally, DR9 improved in reduction techniques and procedures, and DR10 included additional DECam data from NOIRLab in $griz$ bands. Notably, DR9 was used to select DESI targets.
Because the purpose of this paper is to update the measurements from \cite{hang2021} with ACT DR6 CMB lensing, for consistency, we adopt the galaxy data from \cite{hang2021} instead of using a more recent data release.

The galaxy sample was selected with $g<24$, $r<22$, and $W_1<19.5$. 
Pixels contaminated by bright stars, globular clusters, and incompleteness in optical bands are masked by \texttt{bitmasks}\footnote{See \url{http://legacysurvey.org/dr8/bitmasks}. The different `bits' are flags at pixel level indicating different reasons for masking, e.g. bright stars or saturation at a certain band.} supplied by the Legacy Survey pipeline, with ${\rm bits}=(0, 1, 5, 6, 7, 11, 12, 13)$.
From this, the survey completeness map was computed which indicates the geometric completeness of the observation in the range $[0,1]$, with 0 indicating no observation and 1 indicating full coverage of the pixel. 
Regions with completeness $<0.86$ were masked. 
{This particular choice of completeness cut is based on the binned relation between the completeness and mean galaxy number density, such that the variation of the average $\delta_g$ is the galaxy overdensity in bins of completeness is $<0.1$.}

\begin{figure*}
 \centering
 \includegraphics[width=\linewidth]{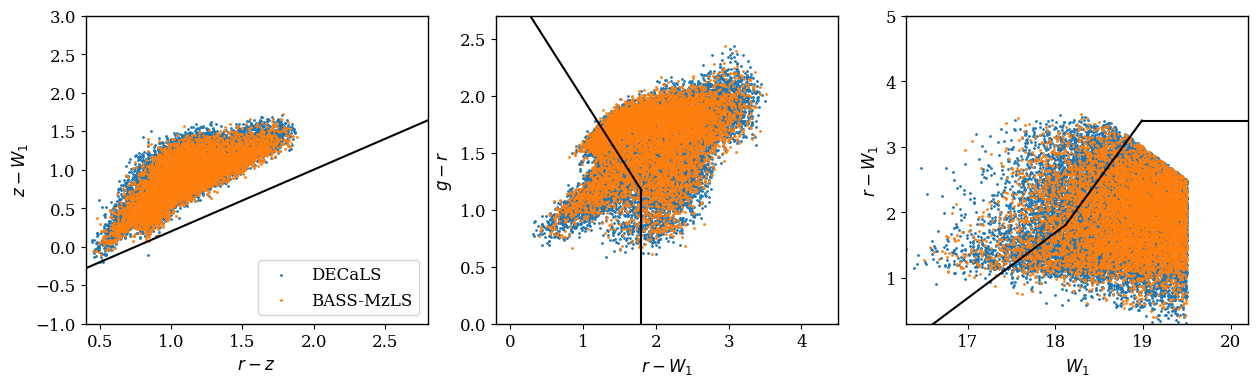}
 \caption{{The DESI Legacy Survey galaxy sample used in this paper in colour-colour and colour-magnitude space. The black lines indicate DESI LRG selection cuts, where the LRG sample is defined to the left of the lines in the first and third panel, and to the right of the line in the middle panel. DECaLS galaxies are shown in blue and BASS-MzLS galaxies are shown in orange.}} \label{Fig.gal_sel}
\end{figure*}

Photometric redshifts were calibrated using the following spectroscopic datasets: GAMA \citep[DR2; ][]{2015MNRAS.452.2087L}, BOSS LOWZ and CMASS samples \citep[DR12; ][]{2015ApJS..219...12A}, eBOSS LRG and ELG samples \citep[DR16; ][]{2020ApJS..249....3A}, VIPERS \citep[DR2; ][]{2018A&A...609A..84S}, DEEP2 \citep{2013ApJS..208....5N}, and COSMOS  \citep[][; with a magnitude cut of ${\tt r\_MAG\_APER2} \leq 23$]{2009ApJ...690.1236I}. 
{Notice that most of our calibration samples overlap with the DECaLS footprint, while the BASS + MzLS footprint is only partially covered by the BOSS sample.
We bin their spectroscopic redshifts in $g-r$, $r-z$, and $z-W_1$ color space with a bin width of about 0.03. We additionally use the DESY1A1 redMaGiC sample \citep{2018MNRAS.481.2427C} to fill in the color cells that do not have galaxies, thanks to their highly accurate photometric redshifts. 
The color cells populated with less than $5$ objects are excluded. For the remaining cells, we compute the average spectroscopic redshifts and their standard deviation. 
We do not apply further selections to the cells based on e.g. the standard deviation of the cells. This could potentially be applied to further increase the photometric redshift accuracy. Finally, the Legacy Imaging Survey galaxies are binned into the same color grid. They are either assigned with the mean redshift of the cell with spectroscopic samples or excluded from the analysis. This process acts effectively as an additional selection, and we retain $78.6\%$ of the above-selected galaxy sample.
Finally, we select galaxies with photometric redshifts consistent with those estimated in \cite{Zhou2020}, with a criterion of $|\Delta z|<0.05$, further removing 23.4\% of our galaxy sample. The total number of galaxies after these cuts is $20,120,352$ from DECaLS and $7,117,218$ from BASS + MzLS.

We split the galaxies into four tomographic bins with redshift bin edges: $[0, 0.3, 0.45, 0.6, 0.8]$. 
We construct maps of the galaxy density contrast, $\delta_g=n/\bar{n}-1$, in HEALPix \citep{2005ApJ...622..759G} format with pixel resolution \texttt{Nside}=1024. Here, $n$ is the number of galaxies in each pixel, and $\bar{n}$ is the mean number of galaxies per pixel within the mask.
The redshift distribution, $n(z)$, of each tomographic bin is then further calibrated using a `self-calibration' scheme, where the galaxy cross-power spectra between different tomographic bins were fitted simultaneously.

{The photometric redshift scatter is characterised by a modified Lorentzian function,
\begin{equation}
    L(z)=\frac{L_0}{[1+((z-z_0)/\sigma)^2/2a]^a},
    \label{eq: lorentz}
\end{equation}
where $z_0$, $\sigma$, and $a$ are free parameters for each tomographic bin, and $L_0$ is the normalisation such that $\int L(z)\,dz=1$. We also impose $\sum_i z_0^i=0$ for each bin $i$, such that the mean redshift of the full sample is not changed. This gives a total of seven nuisance parameters.
The calibrated $n(z)$ is computed by the convolution of the raw $n_{\rm raw}(z)$ with $L(z)$:
\begin{equation}
    n_{\rm cal}(z)=\int_{-\infty}^{\infty} n_{\rm raw}(z') L(z-z')\, dz'.
\end{equation}
In practice, the convolution can result in non-zero values of $n_{\rm cal}(z)$ at $z<0$, but this is in general negligible and thus set to zero ($n_{\rm cal}(z)$ is always normalised to $\int n_{\rm cal}(z) \, dz=1$). In \cite{hang2021}, we jointly constrained these nuisance parameters and galaxy biases using the galaxy auto- and cross-spectra. We found that marginalising over these free parameters gives the same posterior on the lensing amplitude, at fixed cosmology, as using the maximum a posteriori probability (MAP) estimate. Hence, in this work, we adopt the \textit{best-fit} model for the $n(z)$, as found in \cite{hang2021}.}
As shown in \cite{hang2021}, the cross-correlation coefficients, $r_{\ell}^{ij}=C_{\ell}^{ij}(C_{\ell}^{ii}C_{\ell}^{jj})^{-1/2}$, from neighboring tomographic bins $i$ and $j$ are quite flat up to $\ell_{\rm max}=500$. This means that the redshift calibration, which takes most information from these cross terms, has little dependence on the scales, and is independent of linear bias. Magnification was not included in the calibration, but there is little correlation between the highest and the lowest tomographic bins, suggesting that the effect is small. We also found that marginalising over the nuisance parameters made little change to the final result compared to simply adopting their best-fit values.
Here, we additionally check that the assumption of the fiducial \textit{Planck} cosmology does not affect our redshift calibration process. We verify the $n(z)$ calibration dependence on cosmology via variations in $r_{\ell}^{ij}$. We find that by adopting the Planck 2018 and DES Y3 best-fit cosmologies, the change in $r_{\ell}^{ij}<2\%$. 
We explore cases where we apply a shift in the mean redshift of all bins and the impact on cosmology in Appendix ~\ref{apdx: dndz shifts}.

Fig.~\ref{Fig.dndz} shows the calibrated redshift distributions of the galaxy samples used in this analysis. 
In this work, we omit the lowest redshift bin (bin 0) in \cite{hang2021} due to the choice of scale cuts (see Sec.~\ref{sec: scale cuts}), and refer to bins 1 - 3 as $z_1 - z_3$ hereafter.
We also compute and show in Table \ref{tab:sample_summary} the effective redshift for the angular auto-spectrum $C^{gg}_\ell$ and cross-spectrum $C^{\kappa{g}}_\ell$ 

\begin{equation} \label{eq.zeff}
    z^{xy}_\mathrm{eff}=\frac{\int{d\chi}z(\chi)W^x(z)W^y(z)/\chi^2}{\int{d\chi}W^x(z)W^y(z)/\chi^2},
\end{equation}
where $x,y\in\{g,\kappa\}$ and  $W^x(z)$ corresponds to the window function of the field that we will define in Sec. \ref{sec.model}. Note that these redshifts are quoted to aid the interpretation of the typical distance at which a given galaxy signal arises -- but they are not used in the modeling, which employs the full redshift distribution for each tomographic slice.

\begin{table*}
    \caption{
   Summary of the DESI legacy sample properties, {with the DESI x ACT mask}. The effective redshift $z_{\rm eff}$ is calculated with Eq.~\eqref{eq.zeff}. The shot noise $\mathcal{SN}$ is measured for each redshift bin from $4\pi{f}_\mathrm{sky}/N_\mathrm{gal}.$ 
   {The mean galaxy number density, $\bar{n}$, is shown in unit ${\rm deg}^{-2}$. The best-fit galaxy bias, $b_{\rm bfit}$, for each bin is shown with the $1\sigma$ uncertainty, marginalised over all model parameters. The magnification bias slope, $s$, is defined by $d\log N/dm$, where $N$ is the number of galaxies and $m$ is the magnitude limit adopted. $\ell_{\rm max}$ is the maximum multiple in the galaxy auto power spectra.}
   }
    \centering
    \begin{tabular}{c ccccccc }
     \hline
 \hline
         Sample & $z^{\mathrm{gg}}_{\rm eff}$ & $z^{\kappa{g}}_{\rm eff}$ & $10^7\,\mathcal{SN}$ & $\bar{n}$ [deg$^{-2}$] & $b_{\rm bfit}$ & $s$ & $\ell_{\rm max}$\\
         \hline
         $z_1$ & 0.37 &0.36& 6.58 & 460.8 & $1.39^{+0.94}_{-0.37}$ &0.29 &154\\
         $z_2$ & 0.51 &0.51& 5.27 & 579.6 & $1.37^{+0.92}_{-0.35}$ &0.41 &221\\
         $z_3$ & 0.65 & 0.65&8.41 & 363.6 & $1.72^{+1.12}_{-0.48}$ &0.57 &283\\
   \hline

    \end{tabular}
    \label{tab:sample_summary}
\end{table*}

The shot noise
and the effective redshifts of the galaxy samples used are summarised in Table \ref{tab:sample_summary}. The similarity between the $z^{gg}_\mathrm{eff}$ and $z^{\kappa{g}}_\mathrm{eff}$ justifies the use of a constant effective galaxy bias for each redshift bin in Sec.~\ref{sec.model}.

Two systematic corrections were applied at the level of galaxy density maps for each tomographic bin. Firstly, the correlation between the observed galaxy density and survey completeness was accounted for by including a completeness weight, defined as the inverse of the survey completeness for each pixel within the mask. 
Because the completeness is purely geometrical, it does not fully account for the number density variation with stellar density or extinction. Therefore, we apply a second correction to remove correlations with stellar density using the ALLWISE total density map. 
{The stellar density map was originally in $N_{\rm side}=512$, and then upgraded to the required $N_{\rm side}=1024$. Because of the relatively smooth, large-scale variation in number density, we do not think the difference in the initial $N_{\rm side}$ poses significant bias in our correction procedure. We fit the correlation between completeness weighted density $\delta_g$ and stellar number $N_{\rm star}$ with a 5th-order polynomial\footnote{{The high order polynomial is used here to provide a smooth interpolation between the $\delta_g$ vs $N_{\rm star}$ relation.}}. 
The particular choice of the functional form is simply to capture the smooth variation between $\delta_g$ and the systematics. 
The mean density of the galaxy field is computed using regions where completeness $>0.95$ and $N_{\rm star}<8.52\times10^3 {\rm deg}^{-2}$ (about 70\% of the total Legacy Survey footprint). Appendix~\ref{sec: stellar} shows that the level of stellar density correlation with the galaxy density maps within the ${\rm DESI}\times{\rm ACT}$ footprint is about $0.1\%$ at the scales we are interested in.} 
We further exclude pixels with stellar density $N_{\rm star}>1.29\times10^4$\,deg$^{-2}$ where the average $|\delta_g|>0.05$ in bins of $N_{\rm star}$.
Finally, we checked that the correlation between density fluctuations and the Milky Way extinction $E(B-V)$ map \citep{2018JOSS....3..695M} is consistent with zero.

We note that a possible consequence of applying the corrections directly to the density maps is introducing monopole power, which is coupled and propagated into higher $\ell$-modes via the mask. We checked that after applying the correction, the monopole is $< 0.3\%$, and we subtract the monopole before the analysis. 

{For comparison with \cite{Sailer2024}, Fig.~\ref{Fig.gal_sel} shows the selection of our galaxy sample compared to the DESI LRG target selection cuts. The major difference of the two samples are found in the $r-W_1$ and $W_1$ space. Our sample is deeper in $W_1$ compared to DESI LRGs. Compared to the full LRG sample, our sample is also at slightly lower redshifts - the sample drops off sharply beyond $z=0.8$, whereas the DESI LRG has a tail beyond $z=1.2$. Our sample is also $2 - 3$ times denser than the DESI LRG. 
The redshift ranges of our sample, {excluding bin 1,} are most closely matched to redshift bins 1 and 2 in \cite{Sailer2024,Kim2024}.  
This shows that we are not in the shot-noise dominated regime, and if the two samples are tracing the same large scale structure, we expect the cosmological results to be consistent in the two studies.}

\subsection{ACT CMB lensing}\label{act_map}
Our baseline analysis utilizes the CMB lensing map produced in the sixth data release of the Atacama Cosmology Telescope    
\citep[hereafter DR6 lensing map]{qu2023atacama,ACT:2023kun}. This lensing map is produced using night-time only CMB measurements made between 2017 and 2021 in the frequency bands of $90 ~\si{GHz}$ and $150 ~\si{GHz}$. 
The lensing map covers $9400\, \si{deg}^2$ of the sky and is signal-dominated on scales of $L<150$. Among many improvements, this map is produced using a cross-correlation-based estimator that makes use of several time-interleaved splits with independent instrument noise in each split to ensure that the lensing bias subtractions are insensitive to the modeling of the noise. 

The CMB scales used for the lensing analysis range from $600<\ell<3000$. The large scales of the input CMB maps are excluded due to the presence of an instrument-related transfer function \citep{Naess:2022jri}, Galactic foreground and large atmospheric noise. The cut on $\ell>3000$ was chosen to minimise contamination from extragalactic foregrounds like the thermal Sunyaev--Zeldovich effect (tSZ), the cosmic infrared background (CIB), and radio sources. Extragalactic foregrounds are further suppressed through the use of a profile-hardened lensing estimator \citep{PhysRevD.102.063517, PhysRevD.107.023504}. This involves assuming a typical cluster profile for the tSZ and constructing a quadratic estimator that is insensitive to the CMB mode couplings arising from objects with radial profiles similar to the tSZ.  In Sec.~\ref{sec. websky} we use simulations to show that the residual extragalactic contamination is negligible in our cross-correlation measurements.

The baseline ACT DR6 lensing mask is constructed from a Galactic mask that selects $60\%$ of the sky with the lowest dust contamination and is apodized using a cosine roll-off along the edges. For consistency tests, we also employ lensing maps produced using only $40\%$ of the sky, the region with the lowest dust contamination, or a slightly more restrictive $60\%$ mask compared to the baseline $60\%$ mask  in order to remove dust clouds along the edges of the baseline mask when using CIB deprojection for foreground mitigation. We will subsequently refer to these masks as the $60\%$, $40\%$ and CIB-depj masks, respectively.

\subsection{CMB Lensing from \textit{Planck}} \label{planck_map}
The PR4 \textit{Planck} lensing maps utilize CMB scales from $100\leq\ell\leq2048$ using the standard quadratic estimator \citep{Carron:2022}. This analysis improves over the PR3 analysis by using the reprocessed PR4 \texttt{NPIPE} maps that incorporate around $8\%$ more data than the 2018 \textit{Planck} PR3 release. Pipeline improvements, including optimal anisotropic filtering of both the CMB maps and the reconstructed lensing maps, resulted in an increase of the total signal-to-noise by around $20\%$ compared to the PR3 release.


\section{CMB lensing tomography measurements}\label{ref.measurement}
We measure the auto-spectra of the galaxy samples described in Sec.~\ref{Sec. galaxy} and their cross-correlation with the DR6 lensing map on $\sim20\%$ of the sky using a pseudo-C$_\ell$ estimator that appropriately accounts for the impact of the mask induced mode-coupling on the power spectrum between two fields. These pseudo-C$_\ell$ `$\tilde{C}_\ell$' differ from the true underlying power spectrum $C^{\mathrm{true}}_\ell$ due to the effects of the mode coupling and their expectation value $\langle\tilde{C}_\ell\rangle$ is related to the true spectra as \citep{2002ApJ...567....2H}
\begin{equation}
\langle\tilde{C}_\ell\rangle=\sum_{\ell^\prime}M_{\ell\ell^\prime}C^\mathrm{true}_{\ell^\prime},
\end{equation}
where $M$ is the mode coupling matrix that is purely a function of the mask. One can invert the above relationship approximately to extract the true power spectrum if the power spectrum is assumed to be piecewise constant across several discrete bins. We perform the mode decoupling operation on the binned power spectrum using the \texttt{NaMaster} code \citep{Alonso_2019}. For the galaxy field, we use an apodized mask that contains the joint overlap between the DESI Legacy Survey  and ACT lensing (see Fig.~\ref{Fig.footprint}).
The mode coupling matrix for the cross-correlation is thus computed using the galaxy mask and the square of the analysis mask used for lensing \footnote{This is a good approximation for the regime where the variations of the mask are on much larger scales than the CMB and lensing scales of interest. This approximation is done because the lensing signal is reconstructed using a quadratic estimator where each of the two CMB maps carries one power of the mask.}. This is to account for the fact that the lensing reconstruction with a quadratic estimator takes the product of two filtered, masked CMB fields.  We employ HEALPix maps with \texttt{Nside}$=1024$ \footnote{The original lensing maps comes with resolution \texttt{Nside}$=2048$ that we downgrade to match the resolution used in the analysis. } and run \texttt{NaMaster} with $\ell_\mathrm{max}=3000$  despite using only multipoles $\ell\leq300$ in our cosmology range.

We show the cross-correlation between the DESI Legacy galaxies and ACT DR6 lensing reconstructions in Fig.~\ref{fig:bandpower}. The best fits LCDM model obtained for the cross-correlation of DESI Legacy with ACT DR6 and \textit{Planck} PR4 CMB lensing are shown in red and grey respectively. We obtain a minimum $\chi^2$ of $17.5$ for ACT.  We estimate a probability to exceed (PTE) of 0.13 for 12 degrees of freedom and a $\chi^2$ of $6.34$ with a PTE of $0.9$ for the cross-correlation with \textit{Planck} PR4. For the combination of DR6+PR4$\times$DESI with 24 degrees of freedom, we obtain a minimum $\chi^2$ of $25.9$ corresponding to a PTE of 0.36. We note from the sixth panel of Fig.~\ref{fig:bandpower} that for the third redshift bin, the best fit $C^{gg}_\ell$ is significantly different for ACT and \textit{Planck} and this is mainly attributed to the difference in the footprint of the ACT and \textit{Planck} lensing maps, {as marked by the blue and black regions in Fig.~\ref{Fig.footprint}. }{The sky fraction for the ACT and {\it Planck} cases are $f_{\rm sky}=0.160$ and $f_{\rm sky}=0.334$, respectively. }{We speculate that this may be due to the difference in photometry in the DECaLS and BASS-MzLS regions, leading to a slight difference of galaxy selections in bin 3 that was absorbed into the galaxy nuisance parameters. \cite{Sailer2024} find a similar, larger than expected difference between the North and South  in the DESI LRG samples. However, these  differences in biases would not impact on giving consistent cosmology parameters if appropriately accounted for. } In Sec.~\ref{sec.null} we will show null tests of bandpower consistency between the different regions in the ACT footprint showing that the bandpower differences are consistent with null.

\begin{figure*}
    \centering
    \includegraphics[width=\linewidth]{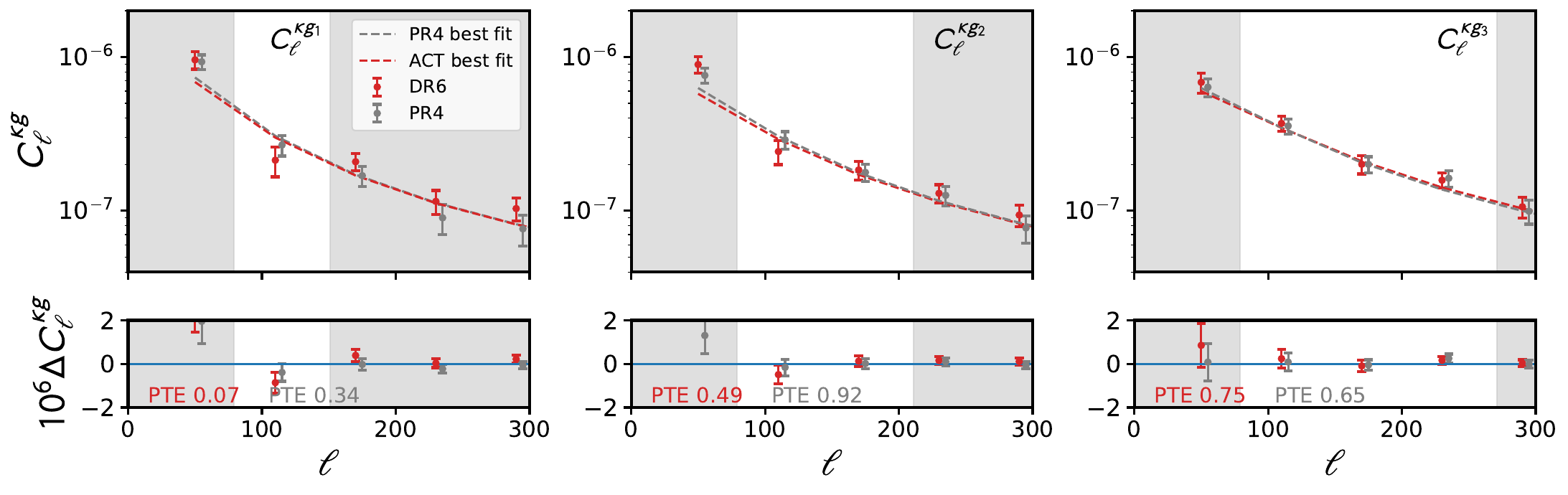} \\ 
    \vspace{-8pt} 
    
    \includegraphics[width=\linewidth]{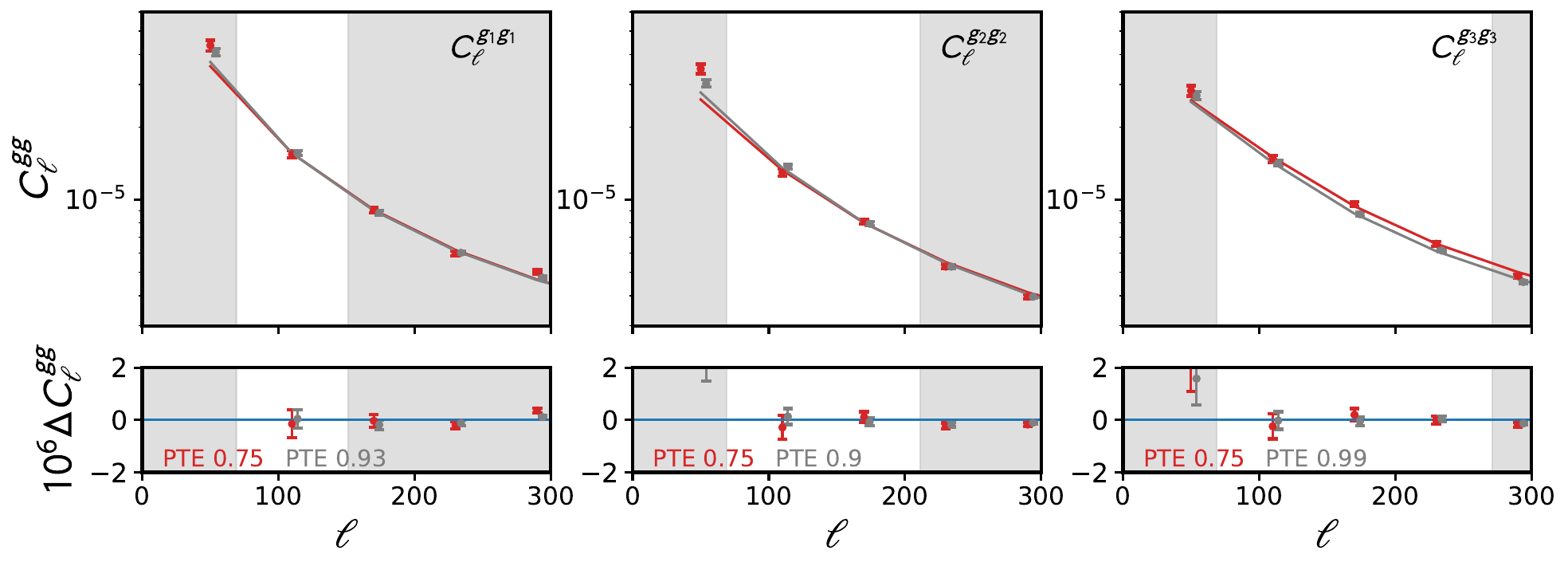} \\ 
    \vspace{-8pt} 
    
    \caption{ Measurements of $C^{\kappa{g}}_\ell$ (\textbf{top} row) and  $C^{gg}_\ell$ (\textbf{bottom} row) for the three redshift samples of DESI Legacy galaxies. Within our analysis range the cross-correlation between galaxies and CMB lensing is detected at an SNRs  of 5, 9 and 15, respectively. Combined with \textit{Planck} PR4 lensing, the cross-correlation of the joint redshift bins is measured at an SNR of 22. We show in dashed red (gray) the best fit from the joint fit to all redshift samples for the ACT DR6 (Planck PR4) analyses. The difference between the two datasets stems from \textit{Planck} having a larger overlap with the DESI galaxies compared to ACT, resulting in a different overall bias. The sub-panels show the model residuals. The total model $\chi^2$ for the joint fit is 25.9 and for 24 bandpowers we estimate to have a PTE of 0.36.}
    \label{fig:bandpower}
\end{figure*}

We show in Sec.~\ref{sec. pipeline_verification} that we can recover the theory spectra from simulations to within better than $1\%$ on the scales of interest and give details on the computation of the covariance matrix using simulations. 

\subsection{Scale cuts}
\label{sec: scale cuts}

We measure the cross-correlation between the ACT DR6 lensing reconstruction and the  DESI galaxies with a signal-to-noise ratio of 5, 9 and 15 for each of the three galaxy samples respectively within our cosmology analysis range of $50\leq\ell\leq\ell^{g_i}_\mathrm{max}$ with $\ell^{g_i}_\mathrm{max}=154,221,283$.

The large scale scale cut $\ell_\mathrm{min}=50$ is chosen due to potential mis-estimation of the lensing reconstruction mean-field signal, which can lead to underestimation of error, and to ensure insensitivity to potential large scale galactic systematics. We further check that on the scales used the contribution beyond the Limber effect from redshift space distortions (RSD) is negligible compared to the uncertainties at a level of at most $0.5\%$.

The maximum multipole scale cut, $\ell^{g_i}_\mathrm{max}$, was mainly chosen by requiring an unbiased recovery of the cosmological parameters; given the linear bias model adopted in this work (as described in Sec.~\ref{sec.model}), this becomes challenging on small scales. Specifically, the $\ell^{g_i}_\mathrm{max}$ are chosen by examining the fiducial $C^{\kappa{g}}_{\ell,\mathrm{fid}}$ and making the cut at the point where the non-linear part of the matter power spectrum\footnote{This is computed using \texttt{CAMB}, assuming a fiducial \textit{Planck} 2018 cosmology by subtracting the linear power spectrum from the total non-linear power spectrum obtained using HALOFIT \cite{Mead_2015}.} becomes $10\%$ of the linear power spectrum. These angular scale cuts correspond to spatial scales of $k_\mathrm{max}=0.15,0.16,0.17h/\si{Mpc}$  for the three redshift bins respectively. Given our {$\Delta\ell$ binning and the choices for $\ell_\mathrm{max}$, we tested that the actual non-linear contribution is less than $5\%$} for all data points. The non-linearity compared to the uncertainties on the data point, $\Delta{C}_\ell/\sigma_\ell\equiv(C^{\rm nl}_{\ell} - C^{\rm lin}_{\ell})/\sigma_\ell$, is most significant at highest $\ell$. For the smallest scale used we find $\Delta{C}_\ell/\sigma_\ell=0.21,0.30,0.41$ for tomographic bins 1 - 3 respectively. The same set of scale cuts are applied to the galaxy auto-correlation, for which we find $\Delta{C}_\ell/\sigma_\ell=0.71,1.46,1.84$. 
Although the nonlinear contribution is larger as expected given the smaller uncertainties in the galaxy auto-spectrum, the above values are expected to be smaller when we marginalize over the linear bias. The constraining power on $S_8$ comes mainly from the information of the cross-correlation, (i.e schematically by taking the ratio of $C^{gg}_\ell/(C^{\kappa{g}}_\ell)^2$ to break the degeneracy between bias $b$ and $\sigma_8$ that $C^{gg}_\ell$ is subjected to.This ratio is dominated by the uncertainties in the galaxy lensing cross-spectrum and not limited by the error in $C^{gg}_\ell$.)
{In practice, our baseline model when fitting the actual data uses the non-linear power spectrum, rather than purely linear theory.}

To ensure that this choice of $\ell_{\rm max}$ does not bias the recovery of cosmological parameters, in Appendix ~\ref{app.nonlinear}, we show that the cosmology constraints on ACT DR6 $\times$ DESI Legacy galaxies using more conservative $\ell_{\rm max}$ (discarding $z_1$ and limiting $k_\mathrm{max}$  of $z_2$ and $z_3$ to $k_\mathrm{max}=0.12,0.13j/Mpc$ respectively, results only in a shift of $S_8$ by $0.03\sigma$, consistent with our baseline measurements.

We also generated theoretical data vectors using the non-linear power spectrum, and obtained fits using the linear matter power spectrum\footnote{In both cases, we use the linear galaxy bias. In Appendix~\ref{sec. planck}, we also test the effect of a scale dependent bias using \textit{Planck} PR3 and adopting the empirical 2-bias model showing that the constraint is consistent with the one obtained with the baseline linear bias.}. We recovered unbiased $S_8$ within $0.1\sigma$ given the scale cuts.   Furthermore, we  perform the above consistency test on re-analysing the previous data set, using the {\it Planck} PR3 lensing map. As shown in Appendix~\ref{sec. planck}, at the $S_8$ level the difference between linear and non-linear modeling is small ($0.19\sigma$), providing further evidence that the scales chosen are insensitive to non linear modeling.


\section{Pipeline verification}\label{sec. pipeline_verification}

\begin{figure*}
    \centering
    \includegraphics[width=\linewidth]{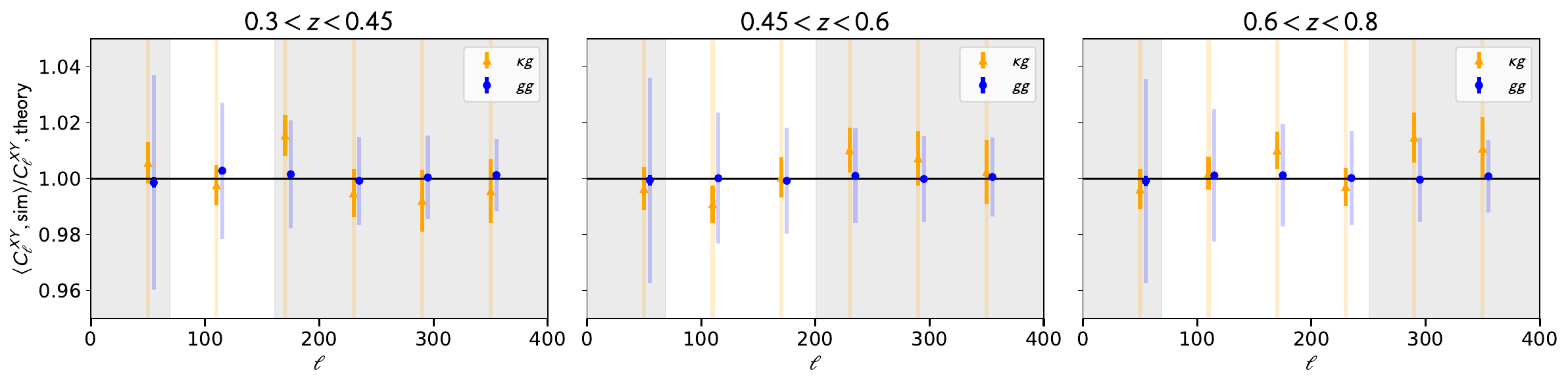}
    \caption{Recovery of $C^{gg}$ and  $C^{\kappa{g}}$ on simulations after correcting for the effect of mode coupling and inversion of mode coupling matrix is better than the per cent level. The lightly shaded error bars show the measurement errors, and the darker error bars show the error on the mean of 400 Gaussian simulations.}
    \label{fig:verification}
\end{figure*}

To test the accuracy of our pipeline, we measure the auto-correlation of the simulations described in Sec.~\ref{sec. sim} and their correlation with the lensing reconstruction simulations in the same manner as we treat the data. We then compare the average of the measured, binned and mode-decoupled spectra to the input power spectra. To make a fair comparison, we account for the fact that the mode decoupling is approximate and convolve the input  $C_{\ell,\mathrm{fid}}$ spectra with the appropriate mode-coupling matrix before binning them and applying the approximate decoupling matrix applied to the measured $\hat{C}_\ell$s. As can be seen in Fig.~\ref{fig:verification}, we can recover the inputs with our measured spectra $\langle{\hat{C}^{\kappa{g}}_\ell}\rangle$ and $\langle{\hat{C}^{g{g}}_\ell}\rangle$ to within $\lesssim
1\%$ well beyond our cosmology range.

\subsection{Simulations for the covariance matrix and pipeline verification}\label{sec. sim}

We generate a set of 400 ACT DR6 CMB lensing reconstruction simulations with appropriately correlated Gaussian simulations of galaxy number density fields. These simulations are used to test the recovery of the auto- and cross-power spectra, and to compute the covariance matrices for the baseline measurements and the null tests.

The lensing simulations are obtained by displacing a randomly drawn CMB realisation with a Gaussian lensing convergence field. We then add realistic survey noise to the lensed CMB \citep{Atkins:2023yzu} and mask the data, before passing the resulting CMB skies into the same lensing reconstruction pipeline that is applied to the real data \citep{qu2023atacama}.

We obtain Gaussian realizations of the galaxy field which are appropriately correlated  with the lensing simulations above by filtering the input lensing convergence as follows using fiducial $C^{gg}_{\ell,\mathrm{fid}}$,$C^{\kappa{g}}_{\ell,\mathrm{fid}}$.
\footnote{{These theory curves are obtained following the same prescription of \citep{hang2021}, based on fits to the data fixed to the \textit{Planck} cosmology using a two bias model.}}

\begin{equation}
    a^g_{\ell{m}}=\frac{C^{\kappa{g}}_{\ell,\mathrm{fid}}}{C^{\kappa\kappa}_{\ell,\mathrm{fid}}}a^{\kappa}_{\ell{m}}+a^{g,\mathrm{uncorrelated}}_{\ell{m}}+a^{g,\mathrm{noise}}_{\ell{m}} .
\end{equation}
{We added the shot-noise component as white noise on the galaxy spectra. Notice that aliasing and the pixel window function could lead to biases in cosmological parameters, as pointed out by \cite{2024JCAP...05..010B}. This effect is most significant for low resolution maps, and at $N_{\rm side}=1024$, the impact on the recovered power spectrum is $\sim 0.5\%$.}
The fiducial spectra are convolved with the appropriate pixel window function for our map resolution of \texttt{Nside}=$1024$.
The part of the galaxy field that is uncorrelated with lensing $a^{g,\mathrm{uncorr.}}_{\ell{m}}$  and the Gaussian noise $a^{g,\mathrm{noise}}_{\ell{m}}$ is obtained from

\begin{eqnarray}
    \langle{a_{\ell m}^{g, \rm{uncorr.}} (a_{\ell' m'}^{g, \rm{uncorr.}})^*}\rangle &=& \delta_{\ell \ell'} \delta_{m  m'} \left(C_{\ell, \rm{fid}}^{gg} - \frac{(C_{\ell, \rm{fid}}^{\kappa g})^2}{C_{\ell, \rm{fid}}^{\kappa \kappa}} \right)\nonumber\\
    &\\
    \langle{a_{\ell m}^{g, \rm{noise}} (a_{\ell' m'}^{g, \rm{noise}})^*}\rangle &=& \delta_{\ell \ell'} \delta_{m  m'} C_{\ell, \rm{noise}}^{gg}.
\end{eqnarray} 

We do not include correlations between the different galaxy samples, but simply  estimate the  Gaussian analytic approximation for the covariance between the different samples as described in Sec.~\ref{Sec. covariance}.

\subsection{Lensing Monte Carlo transfer function}
{As discussed in detail in \citep{qu2023atacama,farren2023atacama}, the lensing maps obtained from performing lensing reconstruction under the presence of a mask are misnormalised. This is corrected using Gaussian simulations by computing the ratio of the cross-correlation between the appropriately masked input lensing convergence with the  lensing reconstruction to the auto-correlation of the known input convergence. Specifically for the case of the cross-correlation this requires computing
}

\begin{equation}
    A^{MC}_\ell=\frac{C^{{\kappa_{\mathrm{in,}{\kappa-\mathrm{mask}}}}{\kappa_{\mathrm{in,}{g-\mathrm{mask}}}}}_\ell}{C^{\hat{\kappa}{\kappa_{\mathrm{in,}{g-\mathrm{mask}}}}}_\ell},
\end{equation}
{where $\hat{\kappa}$ is the masked CMB lensing reconstruction, $\kappa-\mathrm{mask}$ is the input lensing convergence masked with the lensing mask and $\kappa_{\mathrm{in,}{g-\mathrm{mask}}}$ is the input lensing convergence masked with the galaxy mask. Since this correction depends on the region of the overlap, we estimate this separately for the analyses that use the ACT DR6, \textit{Planck} PR4 and PR3 footprint. We estimate these using 480 DR6 Gaussian simulations and corresponding lensing reconstructions provided by \citep{Carron:2022} to obtain unbiased estimates of CMB lensing cross-correlation in the form of }

\begin{equation}
 C^{\hat{\kappa}g}_\ell\rightarrow{A}^{\mathrm{MC}}_\ell{C}^{\hat{\kappa}g}_\ell
\end{equation}

Fig.~\ref{fig:transfer} shows the size of these Montecarlo corrections for our analyses. On the scales of our analyses, these amounts to a change of at most $\sim3\%$ to the measured cross-correlation.

\begin{figure}
    \centering
    \includegraphics[width=\linewidth]{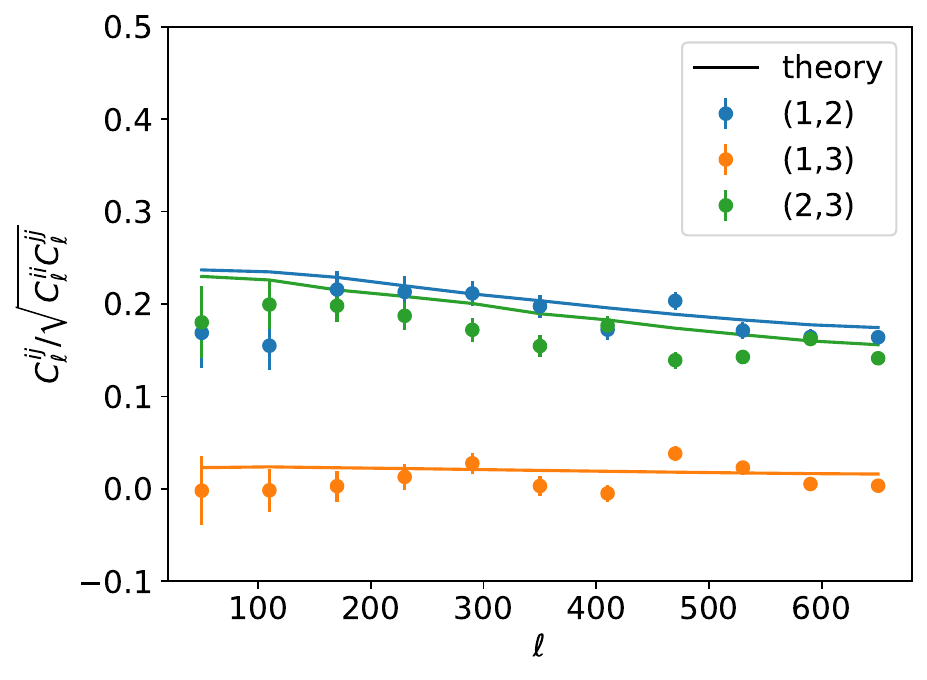}
    \caption{The theory (solid) and measured (dots) cross-correlation coefficients for galaxy clustering, $C_{\ell}^{ij}/\sqrt{C_{\ell}^{ii}C_{\ell}^{jj}}$, for redshift bins $i$ and $j$. The bin combination is shown in the legend as $(i,j)$. The smooth theory curves are generated using the best-fit \cite{hang2021} model to the $C_{\ell}^{gg}$ auto-correlations with the ${\rm ACT} \times {\rm DESI}$ joint footprint. Error bars are Gaussian computed using \texttt{NaMaster} assuming the theory curves.}
    \label{fig:clgg-off-diag}
\end{figure}

 \begin{figure}
  \includegraphics[width=\linewidth]{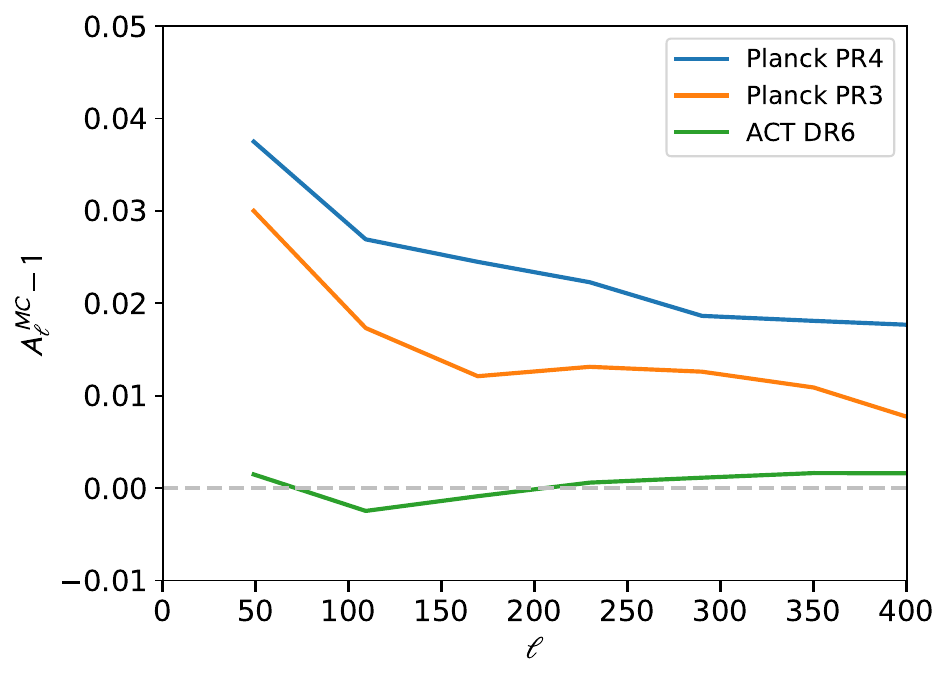}
  \caption{Size of the Monte-carlo transfer function correction for the different lensing maps estimated using simulations described in Sec. \ref{sec. sim}.}
  \label{fig:transfer}
\end{figure}

\subsection{Covariance Matrices} \label{Sec. covariance}

We compute the covariance $\Cov(C^{XY}_\ell,C^{AB}_{\ell^\prime})$ for $XY,AB \in \{gg,\kappa{g}\}$, using the suite of 400 Gaussian simulations of the galaxy and lensing fields discussed in Sec.~\ref{sec. sim}.  We take a hybrid approach where the covariance within each tomographic bin is estimated from the simulations, while the inter-tomographic bins correlation is computed analytically assuming Gaussian signal and noise.
The diagonal part of the covariance matrix, including the diagonals between $gg$ and $\kappa{g}$ between the same galaxy samples, are shown in Fig.~\ref{fig:covmat_diag}. The correlation matrix for one of the three galaxy samples is shown in Fig.~\ref{fig:corr_matrix}.

Our Gaussian simulations do not capture the correlations between the different galaxy samples. Therefore to measure the off-diagonal covariance blocks used in the joint analysis of the three galaxy samples we approximate these analytically using the Gaussian  covariance module implemented in \texttt{NaMaster} \citep{Alonso_2019, Garc_a_Garc_a_2019},
which requires the following as inputs: the fiducial input spectra used for our simulations; {theory curves for the galaxy cross-spectra between the different redshift bins obtained using the best fit biases used in the fiducial input spectra to obtain the input Gaussian simulations }(as described in Sec.~\ref{sec. sim}) and the assumed $n(z)$; and a curve of $C^{\kappa\kappa}_\ell$ including reconstruction noise appropriate to the level of ACT DR6. The different samples of $C^{\kappa{g}}_\ell$ are correlated at the level of ($26-20\%$) between neighboring bins and ($5-3\%$) between next to neighbor bins. The correlation between the different $C^{gg}_\ell$ are at the $5\%$ level for neighboring bins and almost negligible for next to neighbor bins.

\begin{figure}
    \centering
    \includegraphics[width=\linewidth]{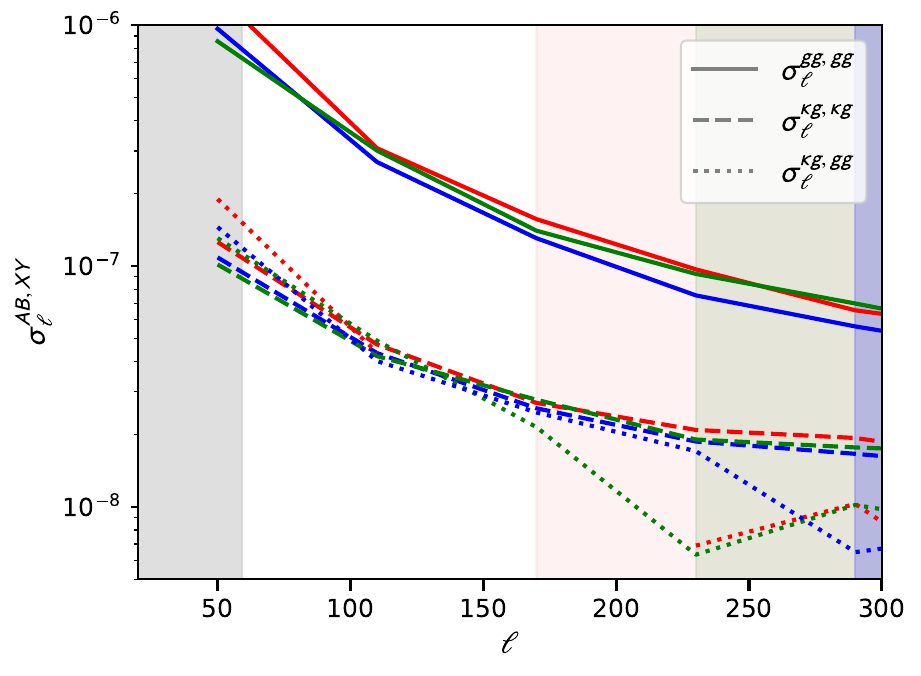}
    \caption{The diagonal elements of the covariance matrix, as well as the off-diagonal covariance between $C_\ell^{gg}$ and $C_\ell^{\kappa g}$ in dotted lines for the different galaxy samples. Shaded regions show the small-scale cut-off used for the different samples. Red, blue and green denotes the first $z_1$, second $z_2$ and third $z_3$ redshift bins respectively.}
    \label{fig:covmat_diag}
\end{figure}

 \begin{figure}
  \includegraphics[width=\linewidth]{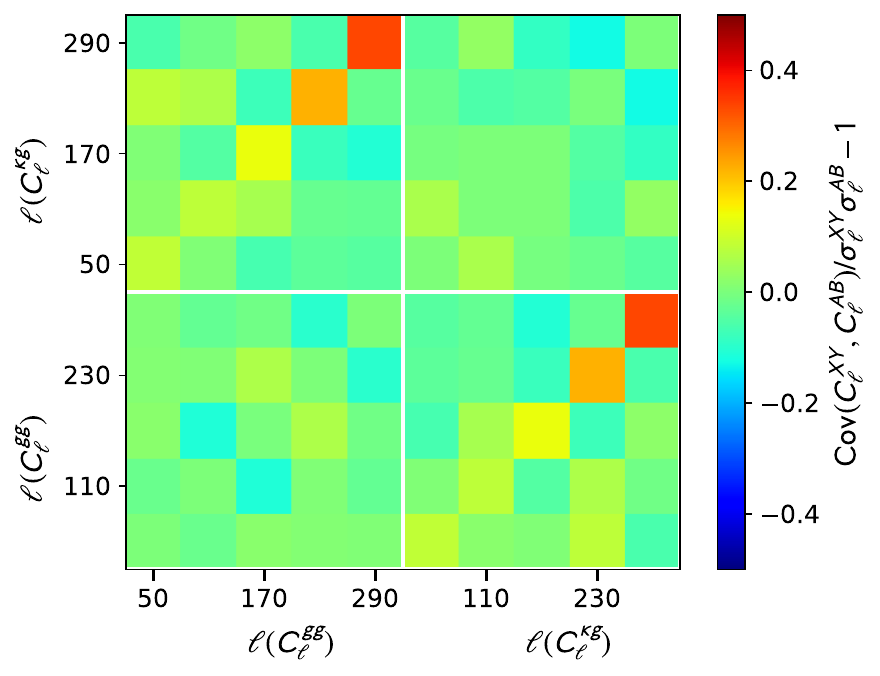}
  \caption{Correlation matrix for $z_3$ of the ACT$\times$DESI galaxies. The diagonal elements of the correlation matrix have been subtracted for improved legibility. While the correlations between different scales are small, there exist correlations between $C^{gg}_\ell$ and $C^{\kappa{g}}_\ell$ of up to $20\%$.} 
  \label{fig:corr_matrix}
\end{figure}

We verify that the diagonal elements of the covariance matrix computed using simulations agree well with the diagonal covariance estimated using the analytic estimation from the Gaussian covariance module implemented in \texttt{NaMaster}. 
{We additionally validate the error estimates from simulations using a Jackknife approach. We first downgrade the joint mask to $N_{\rm side}=8$, and exclude the downgraded mask pixels with value $>0.95$ one at a time, giving a total of 86 Jackknife samples.
We found that generally, this approach gives errors around $5\%$ larger than the analytical error bars and is consistent with our baseline simulation-based approach.}
We further verify that the binning introduced with $\Delta\ell=60$ helps to reduce the off-diagonal correlations of the covariance matrix compared to the $\Delta\ell=10$ adopted in \citep{hang2021} and yields a converged estimate of the covariance matrix given the finite number of simulations used.
We find correlations of up to $20\%$ between $C^{gg}_\ell$ and $C^{\kappa{g}}_\ell$ at the same $\ell$, while the off-diagonal correlations are smaller than $10\%$.
Fig.~\ref{fig:clgg-off-diag} shows the theory input for the off-diagonal covariance in the form of the correlation coefficients, $C_{\ell}^{ij}/\sqrt{C_{\ell}^{ii}C_{\ell}^{jj}}$ for bins $i$ and $j$, compared to the actual measured values. These theory inputs are based on the best-fit \cite{hang2021} models fixed at the \textit{Planck} 2018 cosmology to the auto-correlations $C_{\ell}^{gg}$, hence {not a fit to the measurements. However, the fact that they match closely means that our covariance estimate is relatively accurate. The only noticeable difference is the $(2,3)$ correlation, where the theory is slightly higher than the measurements. This implies that our uncertainties on the final results may be slightly over-estimated.}

Wherever we use the inverse of the covariance matrix, we account for the fact that the inverse of the above covariance matrix is not an unbiased estimate of the inverse covariance matrix, and we rescale the inverse covariance matrix by the Hartlap factor \citep{Hartlap_2006}:

\begin{equation}
    \alpha_\mathrm{cov}=\frac{N_s-N_\mathrm{bins}-2}{N_s-1},
\end{equation}
with $N_s=400$ simulations and the $12$ combined data points for $C^{gg}_\ell$ and $C^{\kappa{g}}_\ell$, this correction is $\alpha_\mathrm{cov}=0.97$ for the joint analysis of the three galaxy samples.

In Sec.~\ref{sec. Planck_ACT}  we also present a combined analysis of the ACT DR6 cross-correlation with the DESI galaxies and the equivalent cross-correlation analysis using \textit{Planck} PR4 lensing reconstructions. We outline the procedure used to estimate the covariance matrix for this joint analysis below.
We use the same set of $480$ FFP10 CMB simulations used in the \textit{Planck} PR4 lensing analysis \citep{Carron:2022}. These lensed Gaussian simulations  have corresponding lensing reconstructions generated using the PR4 lensing pipeline. In a similar manner as described in Sec.~\ref{sec. sim}, we obtain a set of correlated Gaussian galaxy realisations allowing us to estimate the covariance for the \textit{Planck} cross-correlation analysis.

As shown in \cite{qu2023atacama}, the correlation between the ACT and the \textit{Planck} lensing reconstructions is relatively small.  However, there exist large correlations of the order $40-50\%$ in the cross-correlation despite the partial overlap in survey areas and scales because of the identical galaxy sample used in both cross-correlations. The correlation between the galaxy auto-correlation measured on the ACT and \textit{Planck} footprints is up to $60\%$ (This is accounted for appropriately in the covariance matrix when combining the ACT DR6 and \textit{Planck} PR4 measurements). As in the ACT-only analysis we analytically estimate the covariance between the different galaxy samples; this is again small for the case of $C^{gg}_\ell$ $(<6\%)$ but non-negligible for $C^{\kappa{g}}_\ell$ (up to $30\%$).

\section{Tests for systematic errors}\label{sec.null}
We perform a series of comprehensive null and systematic tests that we discuss below. This test suite establishes that our data are free from significant systematic effects that may affect our measurements. For example, we test for contamination of the lensing reconstruction by extragalactic foregrounds (e.g., tSZ and CIB) that also correlate with the galaxy samples. Furthermore, we also examine contamination from galactic  dust by considering different galactic masks. Such contamination could correlate with the galaxy survey data due to correlations of the galaxy sample with galactic structure (e.g., stellar density and dust absorption).

The null tests aimed at systematic errors in the lensing reconstruction correlated with the galaxy samples are found in Sec.~\ref{null_len}; tests targeting the spatial inhomogeneity of the galaxy samples are found in Sec.~\ref{null_gal}. Finally, we estimate the biases in the lensing cross-spectra due to extragalactic foreground contamination of the lensing reconstruction using simulations in Sec.~\ref{sec. websky}.

\subsection{Null tests for contamination of the Lensing Reconstruction}\label{null_len}

We perform tests to show that our lensing reconstruction is free from systematic effects that are potentially correlated with the galaxy samples. Such contamination can in principle be caused by the thermal Sunyaev--Zeldovich (tSZ) effect which is produced by the inverse Compton scattering of CMB photons off hot electrons in thermal clusters, and by CIB contamination originating from unresolved dusty galaxies. Both of these astrophysical foregrounds produce non-trivial bispectra and trispectra which can bias the lensing reconstruction if not appropriately mitigated \citep{maccrann2023atacama}. They are also correlated with the large-scale matter distribution and thus with the galaxy densities, so cross-correlations are also susceptible to biases. 

The baseline DR6 lensing map used in this work uses profile-hardening to mitigate these biases (see \citejap{maccrann2023atacama} and \citejap{qu2023atacama} for details). We verify here that the contaminants are mitigated sufficiently in cross-correlations using this baseline lensing map. Furthermore, as well as extragalactic foregrounds, the correlation of galaxy density with Galactic structures can also bias the lensing reconstruction In particular, dust contamination has been tested extensively for the lensing reconstruction by using more conservative galactic masks and changing the minimum CMB multipole used for the lensing reconstruction \citep{qu2023atacama}. Their effect on cross-correlations can be tested with different galactic masks.

Null tests here are divided into two categories: first, cross-correlation of signal-nulled lensing reconstructions with the galaxy samples, where the signal-nulled lensing reconstructions are constructed  by taking the difference of the temperature  TT and polarization maps at $150$ and $90$ $\si{GHz}$ to obtain a map containing only noise and foreground residuals on which lensing reconstruction is performed.  The second category are bandpower difference tests where the cross-correlation of the same galaxy samples with lensing reconstructions obtained with different versions of CMB maps are taken and checked for consistency. 
The results for these null tests are summarised in Table \ref{null_len} and the histogram of Fig.~\ref{Fig.histnull}. We define the criterion for passing a null test to be that it returns a PTE greater than 0.05\footnote{We also flag tests passing with $PTE>0.95$ although these are not particularly worrying given that tests like curl$\times${galaxies} that specifically test for our covariance matrices give reasonable results}. 

The null tests specifically targeting the extragalactic foregrounds leverage the distinctive frequency dependence of these foregrounds. We  cross-correlate this signal-nulled lensing map with our galaxy samples. 
The results for these tests are passing except for a marginal failure with PTE$=0.03$ for the cross-correlation of redshift bin 2 with the temperature-only lensing null map (The bandpowers of this test can be seen in Fig. \ref{fig:nullfog} of Appendix ~\ref{appendix:null}.)  Given that the same test using the temperature and polarization null map passes and no failure is observed with the bandpower level test discussed below, we attribute this failure to random fluctuations. We also investigate the bandpower difference between the cross-correlations measured using the reconstruction performed only on the $150$ and $90$ $\si{GHz}$ data and find no failures for those tests.

We also explicitly test that profile hardening is effective in mitigating CIB foregrounds by performing a bandpower test difference between the cross-correlation of the DESI galaxies with the minimum variance temperature and polarization (MV) baseline lensing map and a lensing map that explicitly deprojects the CIB (cib-dpj) using the \textit{Planck} high-frequency channels, finding consistency in both approaches. In addition, we find from the consistent bandpower results using a more restrictive Galactic $40\%$ mask compared to our baseline $60\%$ mask that does not significantly affect our measurements.

We also cross-correlate the galaxy samples with the curl modes  of the baseline reconstruction. This test is primarily a test of our covariance estimation since we do not expect there to be any physical signal arising from the curl of the lensing reconstruction at the current levels of precision. This test passes for all redshift bins except for bin 3 which has a marginal failure of PTE$=0.04$ (see Fig. \ref{fig:nullcurl} in Appendix \ref{appendix:null}).

No significant failures are observed apart from two marginal failures of the curl$\times$galaxies test in redshift bin 3 at the level of 0.04 and the frequency nulled map in temperature$\times$galaxy test at redshift bin 2 which fails with 0.03. Further evidence that the above failures are consistent with fluctuations can be seen in Fig.~\ref{Fig.histnull}, the distribution of the PTE for the null tests is consistent with a uniform distribution and given the fact that we perform 36 null tests, observing 2 null tests at around the $5\%$ level is not unlikely.

\begin{table}
\centering
\caption{Summary of the lensing null tests described in Sec.~\ref{null_len} and galaxy homogeneity null tests of Sec.~\ref{null_gal}. For each test, we show the PTE values for the baseline range of each redshift bin. The first 9 tests are described in Sec.~\ref{null_len} and the last 3 tests are described in Sec.~\ref{null_gal}.}
\label{table:null} 
    \begin{tabular}{c c c c }
     \hline\hline 
     Null test & PTE $z_1$& PTE $z_2$ &  PTE $z_3$ \\ [0.5ex] 
     \hline
     Curl$\times{g}$&0.18&0.14&0.04
     \\
     $(f090-f150)\times{g}$ TT&0.12&0.03&0.49\\
     $(f090-f150)\times{g}$ MV&0.69&0.14&0.74\\
     $\kappa^\mathrm{MV}\times{g}-\kappa^\mathrm{cib-dpj}\times{g}$&0.13&0.18&0.18\\
     $\kappa^\mathrm{MV}\times{g}-\kappa^{40\% \mathrm{mask}}\times{g}$&0.22&0.84&0.44\\
     $\kappa^{f150\mathrm{MV}}\times{g}-\kappa^{f090\mathrm{MV}}\times{g}$&0.70&0.27&0.89\\
     $\kappa^{f150\mathrm{TT}}\times{g}-\kappa^{f090\mathrm{TT}}\times{g}$&0.90&0.89&0.96\\
     $\kappa^{\mathrm{MV}}\times{g}-\kappa^{\mathrm{TT}}\times{g}$&0.58&0.43&0.47\\
     $\kappa^{\mathrm{MV}}\times{g}-\kappa^{\mathrm{MVPOL}}\times{g}$&0.96&0.41&0.96\\
     \hline
     $C^{gg,\mathrm{north }}_\ell-C^{gg,\mathrm{south}}_\ell$ &0.22&0.88&0.12
     \\
    $C^{gg,60\%\mathrm{mask }}_\ell-C^{gg,40\%\mathrm{mask }}_\ell$ &0.68&0.79&0.23\\
    $C^{gg,60\%\mathrm{mask }}_\ell-C^{gg,cib-dpj\mathrm{mask }}_\ell$ &0.20&0.20&0.81\\
     \hline
     
    \end{tabular}
\end{table}

 \begin{figure}
  \includegraphics[width=\linewidth]{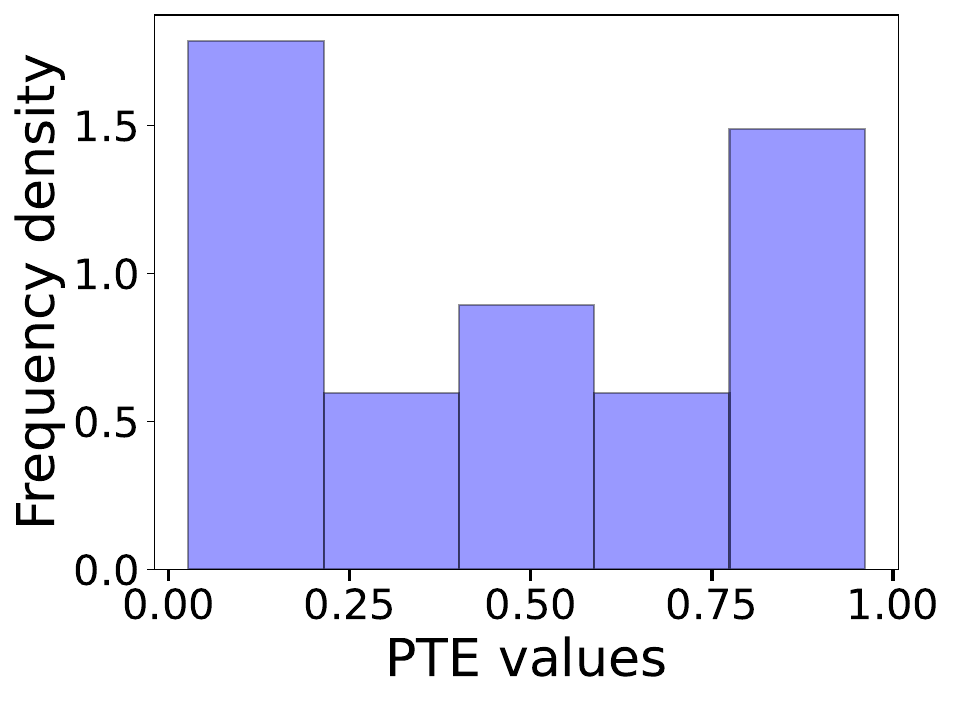}
  \caption{Distribution of the PTE of the 36 null tests. The distribution is consistent with uniform passing of the K-S statistic with a PTE of 0.15.} 
  \label{Fig.histnull}
\end{figure}

\subsection{Homogeneity and contamination tests of the galaxy samples} \label{null_gal}

We perform tests on the homogeneity of the galaxy sample by constructing null tests using different masks.
This is tested through the cross-correlation test using our $60\%$ and $40\%$ Galactic masks discussed in the previous section as well as bandpower level auto-spectra null tests where $C^{gg}_\ell$ is measured across different regions: in particular we compare the measurement using the $60\%$ and $40\%$ Galactic masks, the $60\%$ mask and the footprint covered by the CIB-deprojection mask and splitting the samples between the north and south galactic caps {(i.e., the disjoint regions marked by the blue lines in Fig.~\ref{Fig.footprint})}. All of these tests pass suggesting that the legacy sample used is uniform {within the ${\rm DESI}\times {\rm ACT}$ footprint} on the scales tested without large variations of galaxy bias, shot noise and redshift distribution across the sky. {However, as mentioned in Sec.~\ref{ref.measurement}, we do observe a fluctuation in the galaxy bias between the ${\rm DESI}\times {\rm ACT}$ footprint and the ${\rm DESI}\times {\rm PR4}$ footprint (roughly speaking, difference between the blue and pink regions Fig.~\ref{Fig.footprint}. This is mainly the BASS-MzLS footprint of the DESI Legacy Survey). }

\subsection{Simulation based tests for Extragalactic foregrounds}\label{sec. websky}
We further quantify the potential contamination from extragalactic foregrounds using a simulation-based approach with foreground simulations from \textsc{WebSky} \citep{Stein_2020} in an approach similar to \cite{maccrann2023atacama} by making the approximation that all the relevant extragalactic foregrounds are in the temperature channel such that the observed CMB temperature is given by $T=T_\mathrm{CMB}+T_\mathrm{fg}$. This assumption is valid as we do not expect the polarized tSZ and CIB to cause significant lensing biases at current observation levels \citep{qu2024impactmitigationpolarized}. Furthermore, any potential bright polarized sources in DR6 are masked and inpainted before lensing reconstruction \citep{qu2023atacama}.  The bias on the cross-correlation of the temperature-only lensing reconstruction with the galaxy samples due to unmitigated foreground contamination is then given by

\begin{equation} \label{Eq.bias}
    \Delta{C_\ell^{\kappa{g}}}=\langle\mathcal{Q}(T_{fg},T_{fg})g\rangle,
\end{equation}
where $\mathcal{Q}(T_A,T_B)$ denotes the quadratic estimator used to reconstruct the lensing convergence field from the two fields $T_A$ and $T_B$ , and the cross-correlation of a field $X$ with the galaxy field $g$  is denoted with the short-hand $C^{Xg}_\ell=\langle{X}g\rangle$. In Eq.~\ref{Eq.bias} we have assumed that the foregrounds are uncorrelated with the CMB, allowing us to neglect terms of the form $\langle\mathcal{Q}(T_{CMB},T_{fg})\rangle$.
We quantify the bias due to extragalactic contamination in terms of the bias in the cross-correlation lensing amplitude, $\Ax$. The bias $\Ax$ relative to the uncertainty  of this lensing amplitude is given by

\begin{equation}
    \frac{\Delta{\Ax}}{\sigma({\Ax})}=\frac{\sum_{\ell,\ell^\prime}\Delta{C^{\kappa{g}}_\ell}\mathbb{C}^{-1}_{\ell\ell^\prime}C^{\kappa{g}}_{\ell^\prime}}{{\sqrt{\sum_{\ell,\ell^\prime}{C^{\kappa{g}}_\ell}\mathbb{C}^{-1}_{\ell\ell^\prime}C^{\kappa{g}}_{\ell^\prime}}}},
\end{equation}
where  $C^{\kappa{g}}_{\ell^\prime}$ is the true, baseline cross-correlation signal and $\mathbb{C}_{\ell\ell^\prime}$ is the associated covariance matrix.

We test the foreground mitigation strategies employed in the lensing maps by cross-correlating lensing reconstructions performed on foreground-only maps with \textsc{WebSky} and galaxy samples prepared by populating the \textsc{WebSky} halo catalog with galaxies using the HOD described in \cite{hang2021}. The resulting galaxy catalog is sampled to match the redshift distribution of our samples. 

We find that the baseline foreground mitigation strategy adopted for the ACT DR6 lensing maps is effective in suppressing biases in the cross-spectra, resulting in negligible bias levels of $\Delta{\Ax}/{\sigma({\Ax})}=-0.01$ for all the galaxy samples. Additionally, explicitly deprojecting the CIB does not help in reducing foreground biases further although it is effective in suppressing the foreground biases compared to the no mitigation case. This is consistent with other findings  \citep{maccrann2023atacama,farren2023atacama,Kim2024}, indicating that profile hardening is effective in suppressing not only tSZ clusters but also diffuse foregrounds from the CIB (see \cite{PhysRevD.107.023504} for a detailed explanation). For comparison, the resultant biases when not performing profile hardening are at the $\Delta{\Ax}/{\sigma({\Ax})}=-0.19,-0.18,-0.07$ level for the three galaxy samples respectively (See Fig.~\ref{fig:websky_biases} for a summary). From this test, we have evidence that extragalactic biases are negligible compared to the size of our statistical error, after profile hardening.

\begin{figure*}
    \centering
    \includegraphics[width=\linewidth]{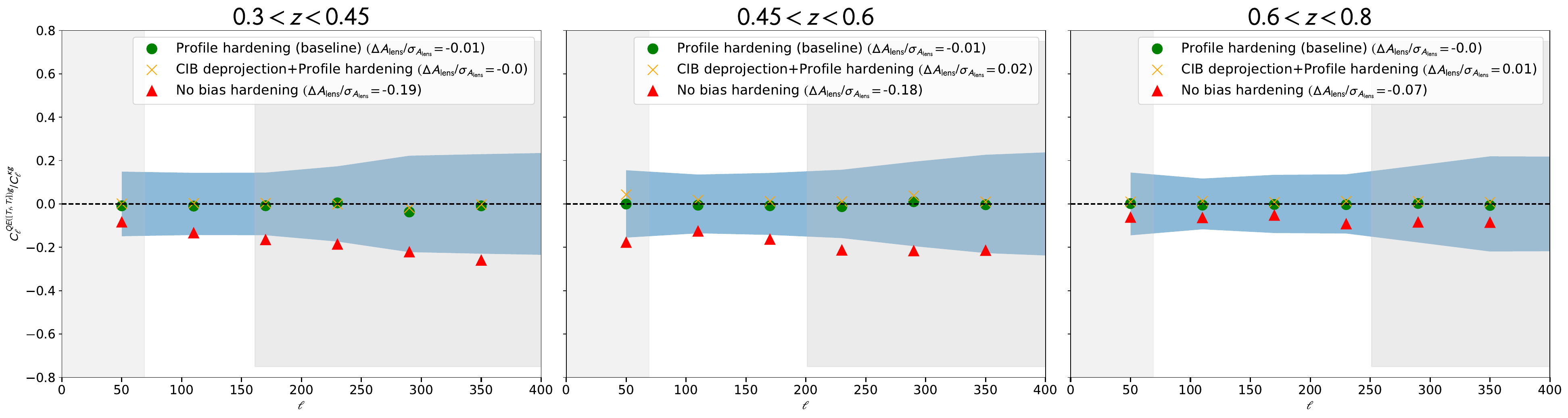}
    \caption{We estimate biases due to extragalactic foregrounds using realistic foreground simulations from \textsc{WebSky} \citep{ Stein_2020}. We perform lensing reconstruction on foreground-only maps using different foreground mitigation strategies and cross-correlate them with galaxy number density maps that we obtain by populating the \textsc{WebSky} halo catalog using a HOD. We find that our baseline analysis reduces all biases to $<0.2\sigma$ while the analysis without any mitigation yields significant biases (up to $\sim 3\sigma$).}
    \label{fig:websky_biases}
\end{figure*}

\section{Modeling and analysis choices}\label{sec.model}

In this section we introduce the model used to compare to the measured spectra presented in Sec.~\ref{ref.measurement} when measuring cosmological parameters. We further present our blinding strategy, and discuss the likelihood and priors adopted for the analysis.

\subsection{Models for galaxy-clustering and CMB lensing power spectra}

We use \texttt{HALOFIT} to model the three-dimensional power spectra of the clustering of matter $P_{mm}(k,z)$. We model the linear galaxy bias $b(z)$ in each redshift bin as a single effective number that is  scale independent. Such that in each tomographic redshift bin we have $P_{gg}=b(z)^2P_{mm}$ and $P_{{g}m}=b(z)P_{mm}$. 

We use the Limber approximation \citep{1953ApJ...117..134L,2008PhRvD..78l3506L} to project the three-dimensional matter power spectra along the line of sight to obtain the angular galaxy auto and galaxy-lensing cross spectra.

\begin{align}
    C^{gg}_\ell&=\int\frac{d\chi}{\chi^2}[W^g(z)]^2P_{gg}\left(k=\left(\ell+\frac{1}{2}\right)/\chi,z\right)\nonumber,\\C^{\kappa{g}}_\ell&=\int\frac{d\chi}{\chi^2}W^\kappa(z)W^g(z)P_{gm}\left(k=\left(\ell+\frac{1}{2}\right)/\chi,z\right),\label{eq:ggkg}
\end{align}
where $z(\chi)$ is the redshift and is implicitly a function of $\chi$,  the comoving distance.

As shown in \cite{hang2021}, bias evolution is important when considering wide tomographic bins, for example, when all four tomographic bins are combined in a single redshift bin. For the individual redshift bins, however, the bias evolution is not strong, and a constant mean bias gives consistent lensing amplitude $A_{\kappa}$ to the case of accounting for the bias evolution. 

 The projection kernels used for the galaxies and CMB lensing are given by
\begin{align}
    W_g(z)&=\frac{H(z)}{c} n(z)\nonumber,
    \\
    {W_{\kappa}}(z)&=\frac{3}{2}\Omega_m\Big(\frac{H_0}{c}\Big)^2(1+z)\frac{\chi(\chi_*-\chi)}{\chi_*},
\end{align}
where $n(z)$ is the normalized redshift distribution for the galaxy sample, $\Omega_m$ is the total matter density that includes the density of neutrinos, $H(z)$ is the Hubble rate with $H_0=H(z=0)$ and $\chi_*$ is the conformal distance to the surface of last scattering.

The observed angular power spectrum also contains contributions from the lensing magnification bias. This effect arises from the gravitational lensing of galaxies by foreground structures, inducing a magnification (or demagnification)
affecting the sample selection by artificially increasing (decreasing) the magnitude of a galaxy in a correlated way with the large-scale structure.
Denoting quantities related to the magnification bias by $\mu$, we model the contributions from the magnification bias as

\begin{align}
    C^{g\mu}_\ell&=\int{d\chi}\frac{W_g(z)W_\mu(z)}{\chi^2}P_{gm}\left(k=\left(\ell+\frac{1}{2}\right)/\chi,z\right),\nonumber\\    C^{\kappa\mu}_\ell&=\int{d\chi}\frac{W_\kappa(z)W_\mu(z)}{\chi^2}P_{mm}\left(k=\left(\ell+\frac{1}{2}\right)/\chi,z\right),\nonumber\\  \nonumber\\  C^{\mu{\mu}}_\ell&=\int{d\chi}\frac{W^2_\mu(z)}{\chi^2}P_{mm}\left(k=\left(\ell+\frac{1}{2}\right)/\chi,z\right),
\end{align}\\
with the lensing magnification kernel given by
\begin{align}
    W_\mu(z)=(5s_\mu-2)\frac{3}{2}\Omega_mH_0^2(1+z)\times \nonumber\\
    \int^{\chi_*}_{\chi}d\chi^\prime\frac{\chi(\chi^\prime-\chi)}{\chi^\prime}H(z^\prime)n(z\prime).\label{eq:wmu}
\end{align}

The parameter $s_\mu\equiv{d\log_{10}N/dm}$ is the response of the galaxy number density to a change in magnitude and is measured from the data by perturbing the photometry of the DESI galaxies and reapplying the selection criteria \citep{hang2021}. The total, observed, galaxy and galaxy-CMB lensing spectra are then given by $C^{gg}_\ell+2C^{g\mu}_\ell+C^{\mu\mu}_\ell$ and 
$C^{\kappa{g}}_\ell+C^{\kappa\mu}_\ell$. 

To evaluate Eq.~\eqref{eq:ggkg}-\eqref{eq:wmu}, we use \texttt{class\_sz} \citep{Bolliet:2017lha,Bolliet:2023eob}, a machine-learning accelerated CMB and Large Scale Structure code written in Python and C, that builds on top of the \texttt{class} \citep{2011arXiv1104.2932L, 2011JCAP...07..034B} infrastructure. \texttt{class\_sz}\footnote{\href{https://github.com/CLASS-SZ}{https://github.com/CLASS-SZ}}  has a parallelized implementation of the Limber integrals and uses neural network emulators for the matter power spectrum. Together, this makes the model evaluation optimally fast (roughly, 0.3s for evaluation of all $C^{gg}_\ell$ and $C^{\kappa g}_\ell$ in three redshift bins). The emulators are presented in detail in \cite{Bolliet:2023sst} and are based on \texttt{cosmopower} \citep{SpurioMancini:2021ppk}.

We find that the impact of the magnification term, $C_{\ell}^{\kappa m}$, compared to the signal $C_{\ell}^{\kappa{g} }$, is about $0.5\% - 1\%$ for the third redshift bin (where the effect is largest), and including this term only shifts the cosmological parameters by $\sim0.1\sigma$, given a uniform prior of $(0.9s, 1.2s)$ on the magnification coefficient, where $s$ is given in Table~\ref{tab:sample_summary} for bin 3. We thus decide not to include this effect in the subsequent modeling for all the redshift bins. 
Since we do not correct the effect of pixelization at the measurement level, we forward model the effect of pixelization by convolving the theory curves by two powers of the pixel window function for \texttt{Nside=1024} $({p_w}_\ell)^2$ for the galaxy-auto spectra and ${p_w}_\ell$ for the galaxy cross spectra. To compare the theory predictions with the observed spectra, we further convolve the theory spectra with the bandpower window that captures the effect of the approximate mode decoupling applied to the data \citep{farren2023atacama}. The final theory auto and cross spectra that we compare with the measurements are given by

\begin{widetext}
\begin{align}
    C^{\mathrm{th},gg}_b&=\sum_{b^\prime}\widehat{M^{-1}}_{b b'}\sum_{\ell} w_\ell^{b'}\left[\sum_{\ell'} \left(M_{\ell \ell'} ({p_w}_{\ell'})^2C^{gg}_{\ell'} + M_{0, \ell'} N_{\rm{shot}} \right)\right], \nonumber \\
        C^{th,\kappa{g}}_b&=\sum_{b^\prime}\widehat{M^{-1}}_{b b'}\sum_{\ell} w_\ell^{b'}\left[\sum_{\ell'} M_{\ell \ell'} {p_w}_{\ell'} C^{\kappa{g}}_{\ell'}  \right], 
\end{align}
\end{widetext}
where $\widehat{M^{-1}}_{b b'}$ is the inverse of the binned mode coupling matrix \footnote{Obtained under the assumption that the power spectrum is piecewise constant and is the default approximation implemented in \texttt{NaMaster}.} and $w_\ell^{b}$ are the uniform weights associated with each multipole in bin $b$. The shotnoise level $N_{\rm{shot}}$ for the data is sampled with a prior centred at the inverse of the galaxy number density in the respective galaxy footprints.

We test the impact of photometric redshift uncertainties on our results in Appendix~\ref{appendix:null} showing that they do not affect the constraints on $S_8$.

\subsection{Blinding Policy}

We adopt a blinding policy that is intended to be a reasonable compromise between reducing the effects of confirmation bias and improving our ability to diagnose issues with the data and the pipeline. During the preparation of the analysis presented in this work, constraints on cosmological parameters were blinded until we demonstrated that a sequence of tests described below were passed. We were not blind to the measured spectra which, in the case of $C^{gg}_\ell$, had already been present in  \citep{hang2021}. We followed the procedure below before unblinding the ACT DR6 lensing and DESI galaxies cross-correlations analysis:
\begin{itemize}
    \item To verify our data are not contaminated by systematic effects, in particular galactic and extragalactic foregrounds, we run a series of null tests described in Sec.~\ref{null_len}. We also perform some tests for the galaxy-galaxy auto spectra in Sec.~\ref{null_gal} although the bulk of this work was already presented in \cite{hang2021}. We classify a test to be passing if it yields a PTE greater than 0.05. We qualify this stage to pass and that our mitigation strategies are sufficient if the number of failures is consistent with what is expected from random statistical fluctuations given the number of tests performed. 
    \item After passing the null tests and verifying the parameter recovery using simulations with our model, we perform a reanalysis of the \textit{Planck} PR3 lensing map cross-correlated with the DESI galaxies discussed in Appendix ~\ref{sec. planck}. We do not consider this to affect blinding as this measurement has already been looked at in \citep{hang2021}. Using PR3$\times$DESI we investigate the consistency of the parameters recovered with different analysis choices, including different sky masks and consistency of parameters when using subsets of the data.
    \item Before unblinding the results using the ACT DR6 and \textit{Planck} PR4 lensing maps, we freeze all the baseline analysis choices. These include the scale cuts used in $C^{gg}_\ell$ and $C^{\kappa{g}}_\ell$, the priors on the cosmological parameters and the nuisance parameters for the galaxies' linear biases and shotnoise.
\end{itemize}

\subsection{Likelihood and priors} \label{sec: likelihood_prior}
We provide constraints on cosmological parameters by constructing a Gaussian likelihood

\begin{equation}
    -2\ln\mathcal{L}\propto\sum_{bb^\prime}\begin{bmatrix}
\Delta{\hat{C}^{gg}_b(\boldsymbol{\theta})} \\
\Delta{\hat{C}^{\kappa{g}}_b(\boldsymbol{\theta})}
\end{bmatrix}\mathbb{C}^{-1}\begin{bmatrix}
\Delta{\hat{C}^{gg}_{b^\prime}(\boldsymbol{\theta})} \\
\Delta{\hat{C}^{\kappa{g}}_{b^\prime}(\boldsymbol{\theta})}
\end{bmatrix}
\end{equation}
where $\Delta\hat{C}^{gg}_b$ and $\Delta{\hat{C}^{\kappa{g}}_b}$ are the residuals between the binned observed galaxy-galaxy auto and galaxy-lensing cross spectra $\hat{C}^{gg}_b$ and $\hat{C}^{\kappa{g}}_b$, and the respective window convolved theory spectra, ${C}^{gg}_b$ and ${C}^{\kappa{g}}_b$. The covariance matrix $\mathbb{C}$
has the form

\begin{equation}
    \mathbb{C}=\begin{bmatrix}
        \mathbb{C}^{gg,gg}_{bb^\prime}&\mathbb{C}^{gg,\kappa{g}}_{bb^\prime}\\(\mathbb{C}^{gg,\kappa{g}}_{bb^\prime})^T&\mathbb{C}^{\kappa{g},\kappa{g}}_{bb^\prime}
    \end{bmatrix}
\end{equation}
where $\mathbb{C}^{gg,gg}_{bb^\prime}$, ${C}^{\kappa{g},\kappa{g}}_{bb^\prime}$, and $\mathbb{C}^{gg,\kappa{g}}_{bb^\prime}$ are the galaxy-auto spectrum covariance, the galaxy-lensing cross-spectrum covariance and the cross-covariance between them. These covariance blocks are estimated from simulations as described in Sec.~\ref{Sec. covariance}.

We use the Markov Chain Monte Carlo code \texttt{cobaya} \citep{Torrado_2021} to perform the sampling and infer parameters from our galaxy-galaxy and galaxy-lensing data using the model described in Sec.~\ref{sec.model} evaluated with \texttt{class\_sz} for a fast and accurate inference. We consider the chains to be converged when the Gelman-Rubin statistic \citep{Gelman1992,Gelman1998} satisfies $R-1\leq0.01$ (which is reached within roughly one hour). 

The dataset used here is insensitive to the optical depth to reionization and similar to \citep{qu2023atacama}, we fix this at the best-fit value of \textit{Planck} \citep{planck2015param}. Table \ref{table:priors} shows the whole set of cosmological priors used, which largely follows from the priors assumed in the most recent \textit{Planck}
and ACT lensing analyses, except for slightly more restrictive $A_s$ and $H_0$ priors. 
Similar also to the analysis of \cite{farren2023atacama}, we fix $\Omega_bh^2$ to the central value from \textit{Planck} of $\Omega_bh^2=0.2242$ \citep{Planck2018Param}.  We also fix the tilt of the primordial power spectrum to $n_s=0.9665$ \citep{Planck2018Param} and assume the minimum neutrino mass allowed in the normal hierarchy ($\sum m_\nu = 0.06$ eV) \footnote{We verified that when we enable those parameters fixed to the \textit{Planck} value to vary with an uncertainty set by the \textit{Planck} measurement only degrades $S_8$ by $\sim10\%$.}. 

For the shot-noise term, we choose a Gaussian prior on SN, the shot-noise amplitude, centered on the Poisson value with a width of $30\%$. We have checked that changing the width to $60\%$ has no significant effect (at the $0.08\sigma$ level) the mean of the posteriors. Notice that due to our conservative binning scheme and hence the small number of data points, large values of shot noise could become degenerate with the signal amplitude, hence affecting the $\sigma_8$ values. This happens when the fitted shot noise is about 20 times that from the $1/N_{\rm gal}$ expectation. Since these shot noise values are nonphysical, we limit our shot noise prior to the above range, rather than using a wide uniform prior.
The linear bias is sampled with a uniform distribution from $0$ to $3$. 

\begin{table}
\centering
\caption{Priors used in the cosmological analysis of this work. Uniform priors are shown in square brackets and Gaussian priors with mean $\mu$ and standard deviation $\sigma$ are denoted $\mathcal{N}(\mu,\sigma)$. Nuisance parameters are marginalised. The expected shot noise for each redshift bin is centered on the values listed in Table \ref{tab:sample_summary}.}
\begin{tabular}{cc}
\hline\hline
Parameter       & Prior      \\ \hline
\multicolumn{2}{c}{Cosmological}\\
$\ln (10^{10}A_s)$ & $[2.5,3.5]$           \\ 
$H_0$           & $[50,80]$        \\ 
$n_s$           & $0.9665$     \\ 
$\Omega_bh^2$   & $0.02242$ \\ 
$\Omega_ch^2$   & $[0.08,0.20]$    \\ 
$\tau$          & $0.055$   \\
$\sum{m_\nu}$   & $0.06$ eV \\ 

\\ \hline
\multicolumn{2}{c}{Nuisance}\\
$b_1$ & $[0,3]$\\
$b_2$ & $[0,3]$\\
$b_3$ & $[0,3]$\\
$s_1$ & $\mathcal{N}(\mathrm{Table} \ref{tab:sample_summary},0.3)$\\
$s_2$ & $\mathcal{N}(\mathrm{Table} \ref{tab:sample_summary},0.3)$ \\
$s_3$ & $\mathcal{N}(\mathrm{Table} \ref{tab:sample_summary},0.3)$ \\
\hline
\end{tabular}
\label{table:priors}
\end{table}

\section{Results}\label{sec.results}

\begin{deluxetable}{lCCCCC}
    \caption{Summary of the $1\sigma$ constraints on cosmological parameters obtained from the cross-correlation of DESI Legacy galaxies with ACT DR6 lensing reconstruction. We also present constraints from the joint analysis of the \textit{Planck} and ACT cross-correlations. We incorporate BAO data to break the degeneracy between the matter density, $\Omega_m$ and the amplitude of fluctuations $\sigma_8$, shown in the third and fourth blocks of the Table.}\label{table.results}
\centering
    \tablehead{
    \nocolhead{}    & \multicolumn{1}{c}{$\Omega_m$} & \multicolumn{1}{c}{$\sigma_8$} & \multicolumn{1}{c}{$S_8$}}
    \startdata
    &\multicolumn{4}{c}{ACT DR6 $\times$ DESI Legacy only}\\
    $z_1$ & 0.207\pm0.063 & 0.712\pm0.125  & 0.579\pm0.095 \\
    $z_2$ & 0.246\pm0.060 & 0.863\pm0.107 &  0.768\pm0.062\\
    $z_3$ & 0.324\pm0.108 & 0.814\pm0.142 & 0.812\pm0.048 \\
    \textbf{Joint} & 0.240^{+0.014}_{-0.046} & 0.872^{+0.088}_{-0.061}  & 0.772\pm0.040 \\
    \\ \hline
      &\multicolumn{4}{c}{(ACT DR6  + \textit{Planck} PR4) $\times$ DESI Legacy}\\ 
    $z_1$ &  0.237\pm0.067& 0.714\pm0.124 & 0.619\pm0.077 \\
    $z_2$ & 0.247\pm0.060 & 0.850\pm0.111 &  0.756\pm0.051\\
    $z_3$ &  0.304\pm0.080& 0.824\pm0.120  & 0.807\pm0.039 \\
    \textbf{Joint} &   0.271^{+0.029}_{-0.075} & 0.821^{+0.11}_{-0.095} & 0.765\pm0.032 \\
        \\ \hline
      &\multicolumn{4}{c}{ACT DR6$\times$ DESI Legacy+ BAO}\\ 
    $z_1$ &  0.290\pm0.015& 0.644^{+0.041}_{-0.065} & 0.633^{+0.053}_{-0.067} \\
    $z_2$ & 0.309\pm0.015 & 0.717^{+0.046}_{-0.064} &  0.727^{+0.054}_{-0.062}\\
    $z_3$ &  0.315\pm0.015& 0.770^{+0.049}_{-0.056}  & 0.788\pm0.051 \\
    \textbf{Joint} &   0.319\pm{0.013} & 0.708^{+0.031}_{-0.041} & 0.731\pm0.038 \\
    \\ \hline
      &\multicolumn{4}{c}{(ACT DR6  + \textit{Planck} PR4) $\times$ DESI Legacy + BAO}\\ 
    $z_1$ &  0.291\pm0.014& 0.645^{+0.037}_{-0.057}& 0.635^{+0.047}_{-0.059} \\
    $z_2$ & 0.310\pm0.014 & 0.714^{+0.038}_{-0.053}&  0.726^{+0.043}_{-0.050}\\
    $z_3$ &  0.318\pm0.014& 0.764^{+0.040}_{-0.045}  & 0.785\pm0.040 \\
    \textbf{Joint} &   0.322\pm{0.011} & 0.714^{+0.026}_{-0.034} & 0.739\pm0.029 \\
    \\ \hline
    \enddata
\end{deluxetable}

\subsection{DESI galaxies $\times$ ACT DR6} \label{result_ACT}

We jointly analyze the  auto-correlation of the three DESI Legacy galaxy samples and the cross-correlation of each with the ACT DR6 lensing reconstruction to obtain a $5.1\%$ ($68\%$) constraint on $S_8\equiv\sigma_8(\Omega_m/0.3)^{0.5}$ of

\begin{equation}
    S_8=0.772\pm0.040 \quad \text{(DESI$\times$ACT DR6)}.
\end{equation}

 The posterior in the $\sigma_8-\Omega_m$ plane is shown in the left panel of Fig.~\ref{fig: banana plot - act pr4 split zbin} (purple contours). The constraints from the individual redshift bins are summarized in Table \ref{table.results}.

\begin{figure*}
    \centering
    \includegraphics[width=0.5\linewidth]{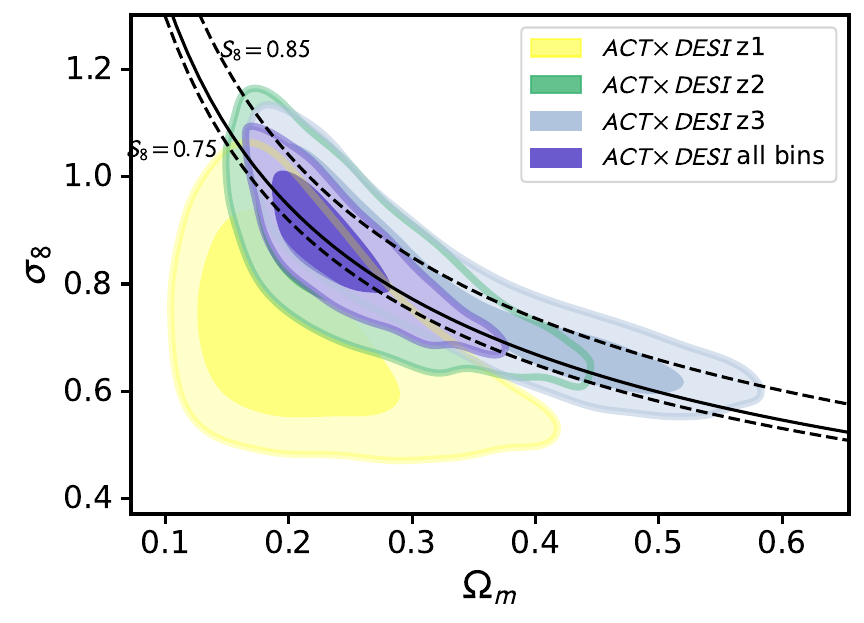}%
    \includegraphics[width=0.5\linewidth]{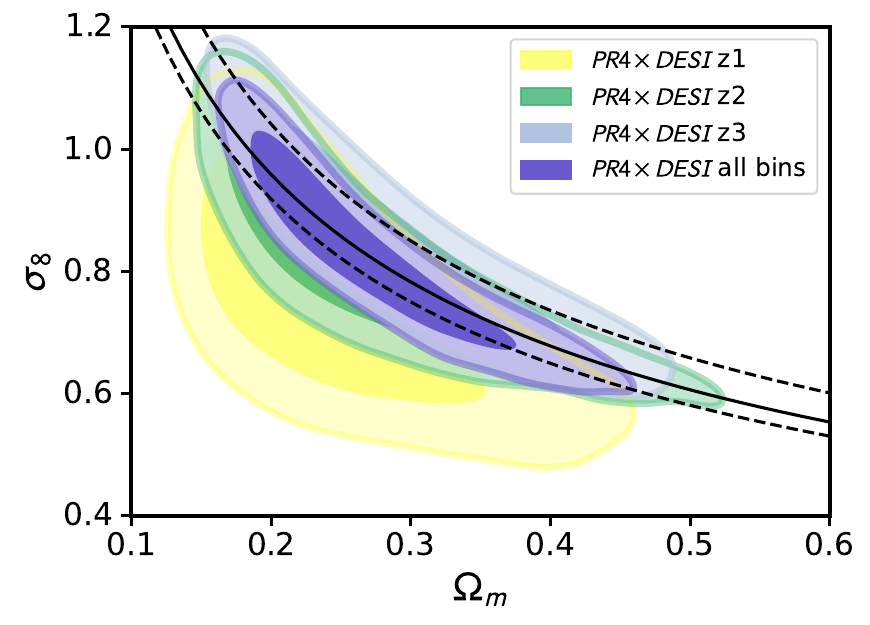}
    
    \caption{Cosmological constraints of the DESI Imaging galaxies. \textbf{Left} panel shows the 1 and 2-$\sigma$ contours using $C^{gg}_\ell$ and $C^{\kappa{g}}_\ell$ with ACT DR6 only. \textbf{Right} panel shows the equivalent constraints for \textit{Planck} PR4. The black solid line shows the best fit $S_8$.
    As a reference, we also show as dashed lines of constant $S_8=0.85$ and $S_8=0.75$. Our measurement comfortably sits in between these two lines as can be seen as well with the best fit of $S_8=0.772$ and $S_8=0.783$ for ACT and \textit{Planck} respectively. }
    \label{fig: banana plot - act pr4 split zbin}
\end{figure*}

\subsubsection{Combination of ACT DR6 and \textit{Planck}} \label{sec. Planck_ACT}

In Appendix ~\ref{sec. planck}, we present a reanalysis of the cross-correlation between \textit{Planck} PR3 CMB lensing and the four DESI Legacy galaxy samples using our more conservative linear model, new pipeline and previously neglected Monte Carlo lensing norm corrections. Fig. ~\ref{fig.s8evol} shows in detail how the above changes affected the parameter $S^{\times}_8\equiv\sigma_8(\Omega_m/0.3)^{0.78}$ constrained by \cite{hang2021}. Below, we update the analysis results using our scale cuts and the new \textit{Planck} PR4  lensing map. The $\sigma_8-\Omega_m$ contours are shown on the right panel of Fig.~\ref{fig: banana plot - act pr4 split zbin}. We find a $4.8\%$ constraint on $S_8$:
\begin{equation}
    S_8=0.776\pm0.037 \quad \text{(DESI$\times$\textit{Planck} PR4)}.
\end{equation}

Although the reconstruction noise is significantly lower for ACT, using the \textit{Planck} lensing map constrains $S_8$ more tightly than the results presented in ~\ref{result_ACT}, owing to the area gain of the \textit{Planck} lensing map. ($33\%$ vs $18\%$ in overlap).

The consistency between the ACT and \textit{Planck} results motivates us to present the joint analysis of the cross-correlation of the DESI legacy galaxies with the ACT DR6 and \textit{Planck} PR4 lensing. We describe in Sec.~\ref{Sec. covariance}  the estimation of the joint covariance, accounting for the correlation between the ACT and \textit{Planck} cross-correlations. Given that we find the bias preferred by the region of the  \textit{Planck} footprint is different to that of ACT in Fig ~\ref{fig:bandpower}, we do not assume them to be identical in the ACT and \textit{Planck} footprints. Hence we include $C^{gg}_\ell$
measured using both the ACT and \textit{Planck} masks in our analysis. We also include the significant cross-covariance between $C^{gg,\mathrm{ACT}}_\ell$ and $C^{gg,\mathrm{Planck}}_\ell$, which we also estimate from our Gaussian simulations.

The combination of ACT DR6 and \textit{Planck} PR4 CMB lensing with DESI legacy results in a $4.2\%$ joint constraint of
\begin{equation}
    S_8=0.765\pm0.032 \quad \text{(DESI$\times$\textit{Planck} PR4 + ACT DR6)}
\end{equation}
This constitutes a $24\%$ and a $13\%$ improvement compared to ACT DR6 and \textit{Planck} PR4 alone respectively. The joint analysis in black is shown alongside the constraints from ACT and \textit{Planck} in the top panel of  Fig.~\ref{bao_robustness}. 
The best-constrained parameter in our analysis differs slightly from $S_8$ due to the redshifts and scales used and we determine this empirically to be closer to $S^{\times}_8\equiv\sigma_8(\Omega_m/0.3)^{0.58}$, which we constrain to  $4.1\%$ $S^{\times}_8=0.757\pm0.031$.

A feature that can be seen in Fig.~\ref{bao_robustness} is that the lowest redshift bin is $\sim1.8\sigma$  lower than the baseline constraint mean for both the ACT, \textit{Planck} and combined ACT+\textit{Planck} results. This corresponds to a shift in $S_8$ at the $2.21$ and $2.08\sigma$  using the subset test proposed by \cite{Gratton_2020} given that the constraints from $z_1$ are a subset of that of the joint analysis. We thus also proceed to report joint constraints of ACT+\textit{Planck} $\times$ DESI excluding the first redshift bin

\begin{equation}
    S_8=0.785\pm0.033.
\end{equation}
This only degrades the baseline $S_8$ constraint by $3\%$ but results in a shift of $S_8$ upwards by $0.45\sigma$. 

Overall, our cross-correlation analysis is robust to the data and analysis choices used. A summary of the $S_8$ constraints using different analysis variations is shown in the top panel of Fig.~\ref{bao_robustness} and we see that, excluding the analysis using only the first redshift bin, all the other choices are within the $1\sigma$ level of the baseline measurements. In green we show that we obtain consistent results when adding a prior on the angular size of the sound horizon $\theta_{\rm{MC}}$, that is predominantly sensitive to the combination of $\Omega_m h^3$ \citep{Percival_2002}. Taking the mean value measured by \textit{Planck} \citep{Planck2018Param} of  $\Omega_m h^3=0.09635$ we obtain a $S_8$ value of

\begin{equation}
    S_8=0.747\pm0.028.
\end{equation}

\begin{figure*}
    \centering
    \includegraphics[width=\linewidth]{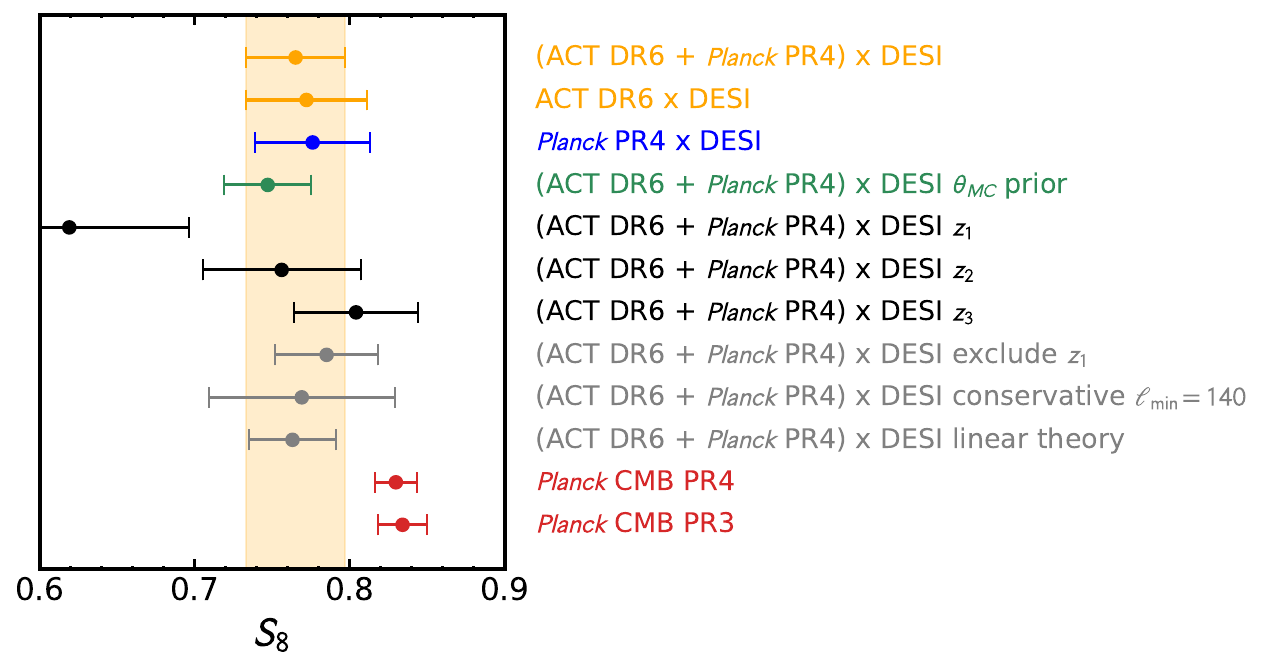} \\ 
    \vspace{-8pt} 
    
    \includegraphics[width=\linewidth]{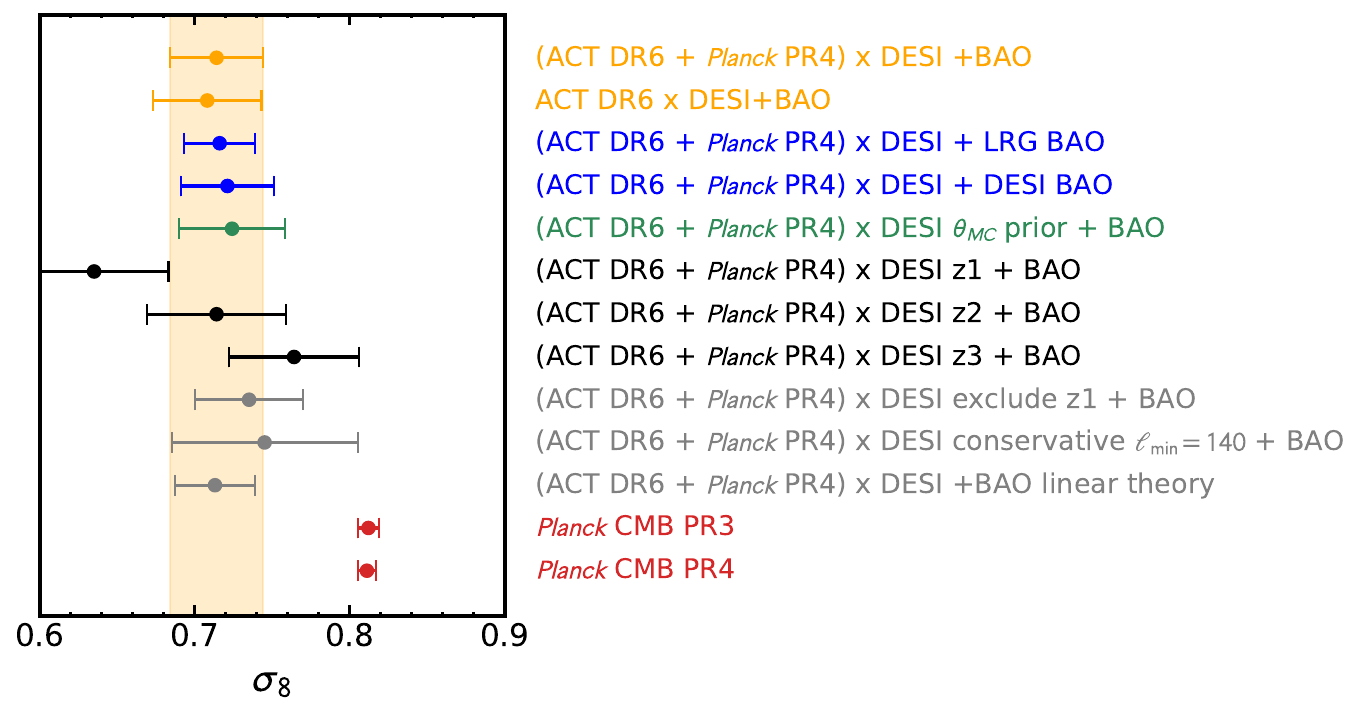} \\ 
    \vspace{-8pt} 
    
    \caption{We show the $S_8$ (\textbf{top panel}) and $\sigma_8$ $1\sigma$ constraints when combined with BAO  (\textbf{bottom panel}) using different data and analysis variations. Gold points correspond to constraints using the baseline analysis for the combination of ACT and \textit{Planck} lensing (top) row and ACT DR6 only (second row). Black data points show the joint ACT DR6 + \textit{Planck} PR4 constraints cross-correlated with each redshift bin. We see that the mean of the first bin lies outside the $1\sigma$ interval of the baseline analysis although given that we have only bandpower for this redshift bin, it has a small weighting in the overall constraint. This can be seen from the gray point where removing the first redshift bin causes a $\sim0.45\sigma$ shift with respect to the baseline in gold. Shown as reference in red are the $S_8$ and $\sigma_8$ values obtained from the \textit{Planck} primary CMB.
}
    \label{bao_robustness}
\end{figure*}

\subsubsection{Combination with BAO}

\begin{figure*}
    \centering
    \includegraphics[width=0.5\linewidth]{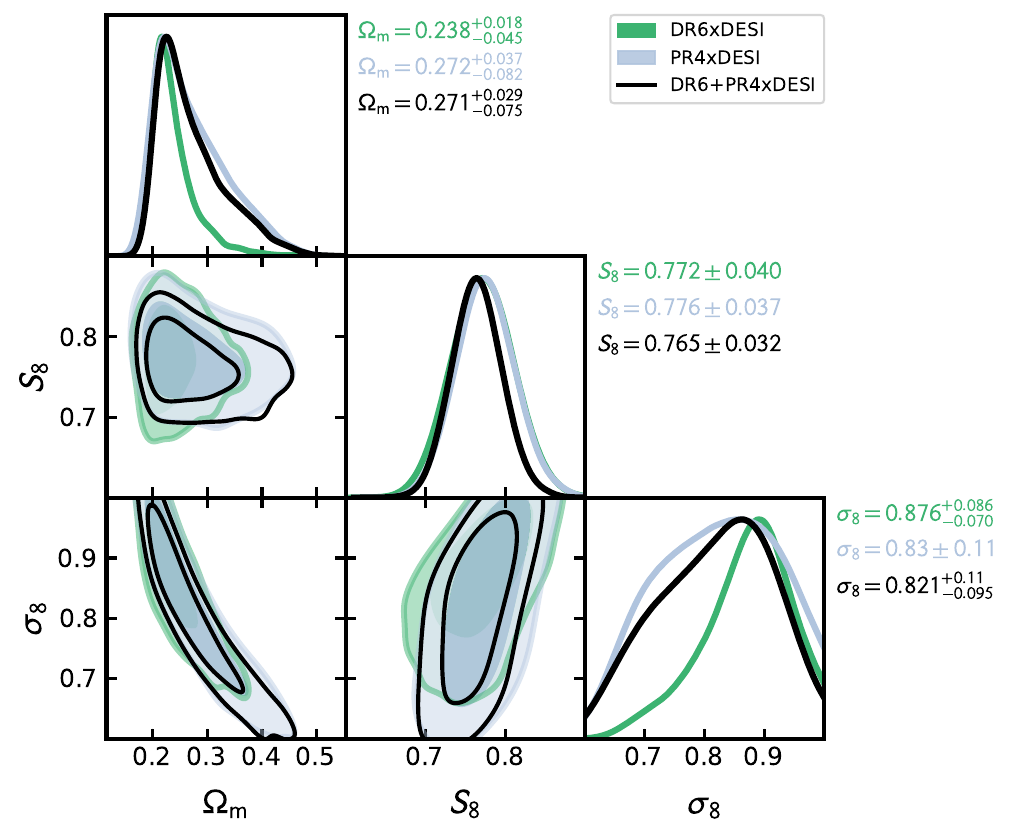}%
    \includegraphics[width=0.5\linewidth]{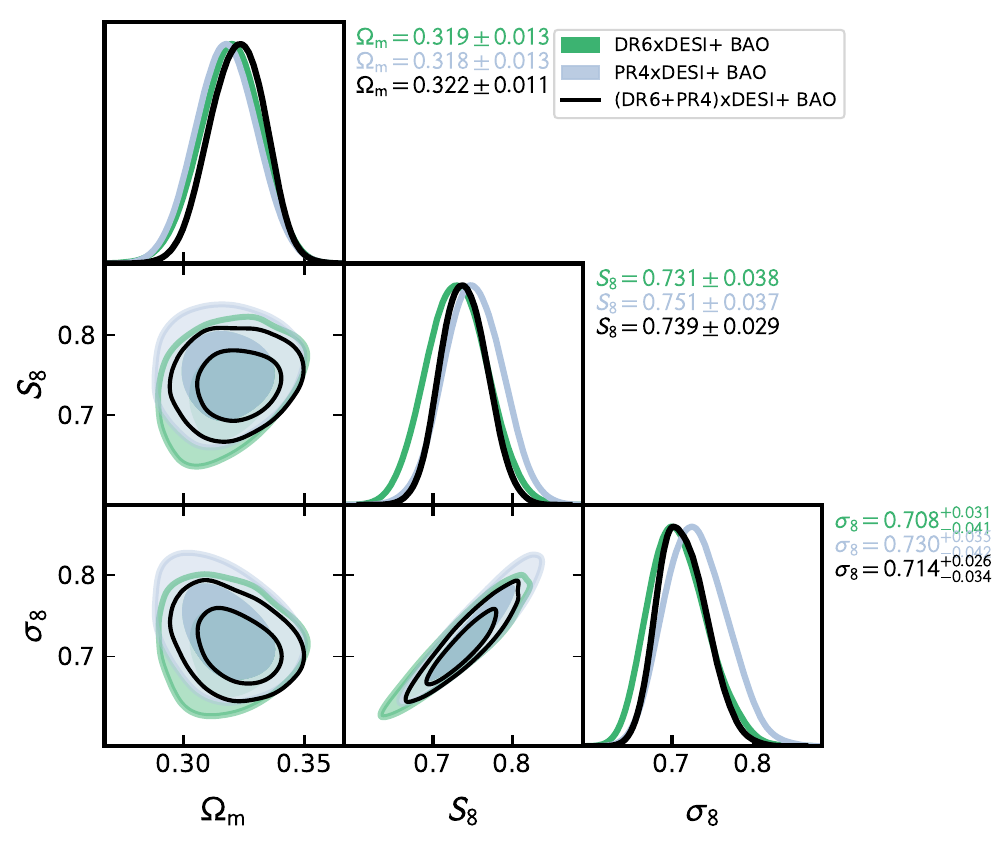}
    \caption{Parameter constraints from the cross-correlation of ACT DR6 lensing, PR4 lensing, their combination and DESI galaxies (\textbf{left}). We break the degeneracy between $\Omega_m$ and $\sigma_8$ with additional information on $\Omega_m$ from BAO (\textbf{right}). }
    \label{fig:planckXcorner with bao}
\end{figure*}

The tomographic lensing measurements provide information on a 3-dimensional volume comprising the amplitude of matter fluctuations $\sigma_8$, the matter density $\Omega_m$, and the Hubble constant $H_0$. To reduce the degeneracies of $\sigma_8$ with the other parameters and enable comparisons with other probes of weak lensing, CMB lensing auto-spectra and CMB primary information, we include expansion information from the 6dF and SDSS surveys. The included data measures the BAO in the clustering of galaxies up to $z\approx1$; the data are from 6dFGS \citep{1106.3366}; SDSS DR7 Main Galaxy Sample \citep{1409.3242}; BOSS DR12 luminous red galaxies \citep{Alam17}; and eBOSS DR16 LRGs \citep{Alam2021}. Additionally, we include the higher-redshift Emission Line Galaxies ELGs  \citep{2016A+A...592A.121C}, Lyman-$\alpha$ forest \citep{2020ApJ...901..153D}, and quasar samples \citep{Alam2021} from eBOSS DR16. We do not include additional information that constrains structure growth from RSD. This choice allows us to isolate information on structure formation purely from the galaxy samples and its cross-correlation with CMB lensing. Fig. ~\ref{Fig. 3dplot} shows a breakdown of the constraints of the lensing-galaxy cross correlation, BAO and their intersection in the three dimensional $\sigma_8-H_0-\Omega_m$ space.

\begin{figure}
    \centering
    \includegraphics[width=\linewidth]{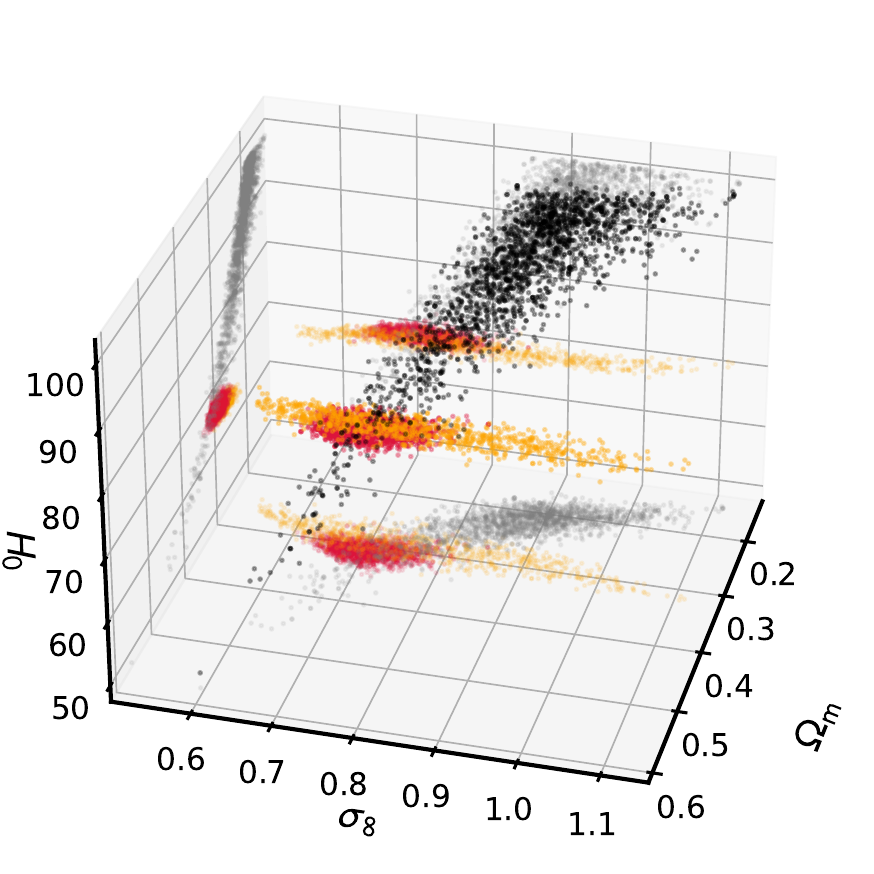} \\ 
    \caption{Distribution of MCMC samples for our constraints with ACT DR6 in the $\sigma_8$-$\Omega_m$-$H_0$ space in black. BAO samples are shown in orange, the intersection between the BAO and the lensing plane (red) provides constraint on $\sigma_8$.}
    \label{Fig. 3dplot}
\end{figure}

The lower panel of Fig.~\ref{bao_robustness} demonstrates again the robustness of the results against different analysis choices.
Fig.~\ref{fig:planckXcorner with bao} shows the marginalized contours of $\Omega_m$, $\sigma_8$ and $S_8$ with BAO. With ACT DR6 we find a $5\%$ constraint of 

\begin{equation}
    \sigma_8=0.708^{+0.031}_{-0.041}.
\end{equation} 
Combining with \textit{Planck} PR4 lensing again results in a improvement from the ACT DR6 only measurement by $16\%$ giving
\begin{equation}
    \sigma_8=0.714^{+0.026}_{-0.034}.
\end{equation}
The combined constraints alongside the individual ACT and \textit{Planck} constraints are shown on the right of Fig.~\ref{fig:planckXcorner with bao}. We find that the $\sigma_8$
constraint after combining with BAO is $3.2\sigma$ lower than the CMB 2pt value obtained from \textit{Planck} PR4 NPIPE. 

Similar to the lensing only measurements (made without the inclusion of BAO), we find that the first redshift bin is discrepant at the $2.02\sigma$ with respect to the third redshift bin. Excluding this first bin to ensure a more internally consistent dataset results in a slightly larger $\sigma_8$ value of 

\begin{equation}
    \sigma_8=0.735\pm0.035 ,
\end{equation} improving the consistency with the 2pt \textit{Planck} PR4 NPIPE result to $2.14\sigma$.

It is worth noting that by building a normalization flow distribution for the 
parameter shift between the BAO and the cross-correlation dataset using \textsc{tensiometer}\citep{PhysRevD.104.043504}, we find a discrepancy of $1.94\sigma$ between these two datasets. This tests quantifies the discrepancy between the orange bands from BAO and the black bands from the ACT DR6 $\times$ galaxies cross-correlation in the 3D space of $\sigma_8-\Omega_m-H_0$  in Fig. \ref{Fig. 3dplot}. This discrepancy hence cautions the naive combination of the two datasets and the constraints obtained when including BAO. We additionally perform a range of systematic tests in Fig.~\ref{bao_robustness} and find that a test with a larger $\ell_\mathrm{min}=140$ causes the posterior mean of $\sigma_8$ to move up by $0.5\sigma$ in the direction that reduces the tension to $1.1\sigma$. Although the error bars nearly double given that the most signal-dominated scales are discarded, this shift suggests that if large-scale systematics are the cause of the low $\sigma_8$ values obtained, they will primarily impact the lowest redshift bin. The scale cut essentially removes the constraint provided by the $z_1$ bin. As a result, the outcome aligns with the analysis where $z_1$ is simply discarded, but no scale cut is applied to the other bins. The above instabilities and internal tensions suggest further that investigation into large-scale systematics is needed before external datasets are added in. In what follows we will proceed with the more robust lensing only results that provide a $\sim4\%$ constraint on structure growth.

\subsection{Consistency with predictions based on \textit{Planck} CMB anisotropies}\label{sec.lcdm}

Our combined $S_8$ constraint is $1.93\sigma$ lower than the one obtained from \textit{Planck} 2018 CMB anisotropies \citep{Planck:2018vyg} and $1.88\sigma$ lower than preferred by the latest PR4 \citep{Rosenberg_2022} + \texttt{SRoll2} EE \citep{Pagano_2020} analysis that is based on an extrapolation assuming the $\Lambda$CDM model .

We also assess the consistency of our dataset with the preferred parameters obtained from \textit{Planck} 2018 by fixing the cosmological parameters to those reported by \textit{Planck} 2018, namely with $ H_0= 67.4$, $\omega_\mathrm{c}=0.120$ and $\ln{10^{10}A_s}=3.044$ and comparing this with our baseline model. We find a $\Delta\chi^2=11.6$  ($\Delta\chi^2=12.4$ with the NPIPE parameters) going from the fixed to the free cosmology case. With a PTE of 0.04, the LCDM model provides a fairly reasonable fit to the data. In Fig.~\ref{Fig. redevol} we show the redshift evolution predicted by \textit{Planck} 2018 and the values measured by our baseline analysis. We rescale the $S_8$ at each redshift bin by $\sigma_8(z^{i}_\mathrm{eff})/\sigma_8(0)$. Apart from the constraint from the first fit bin, deviating from the  \textit{Planck} 2018 prediction by $\sim 2\sigma$, the joint constraint (orange line) and the constraints from redshift bins 2 and 3 are consistent with the cosmology preferred by \textit{Planck}.

We show in Fig.~\ref{fig:bandpower LCDM fit} the best-fit obtained when fixing the cosmology parameters to the values preferred by \textit{Planck} 2018 and varying only the nuisance shot-noise and bias parameters. Apart from the first redshift bin, our data show a good fit to the \textit{Planck} 2018 cosmology.

\begin{figure*}
    \centering
    \includegraphics[width=\linewidth]{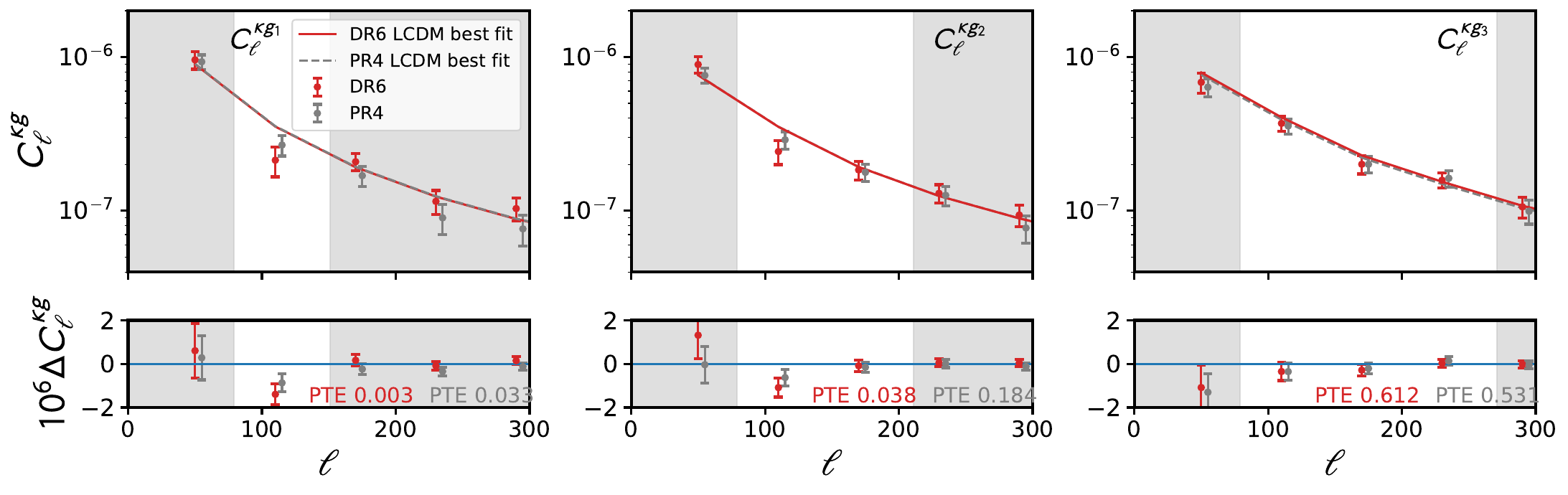} \\ 
    \vspace{-8pt} 
    
    \includegraphics[width=\linewidth]{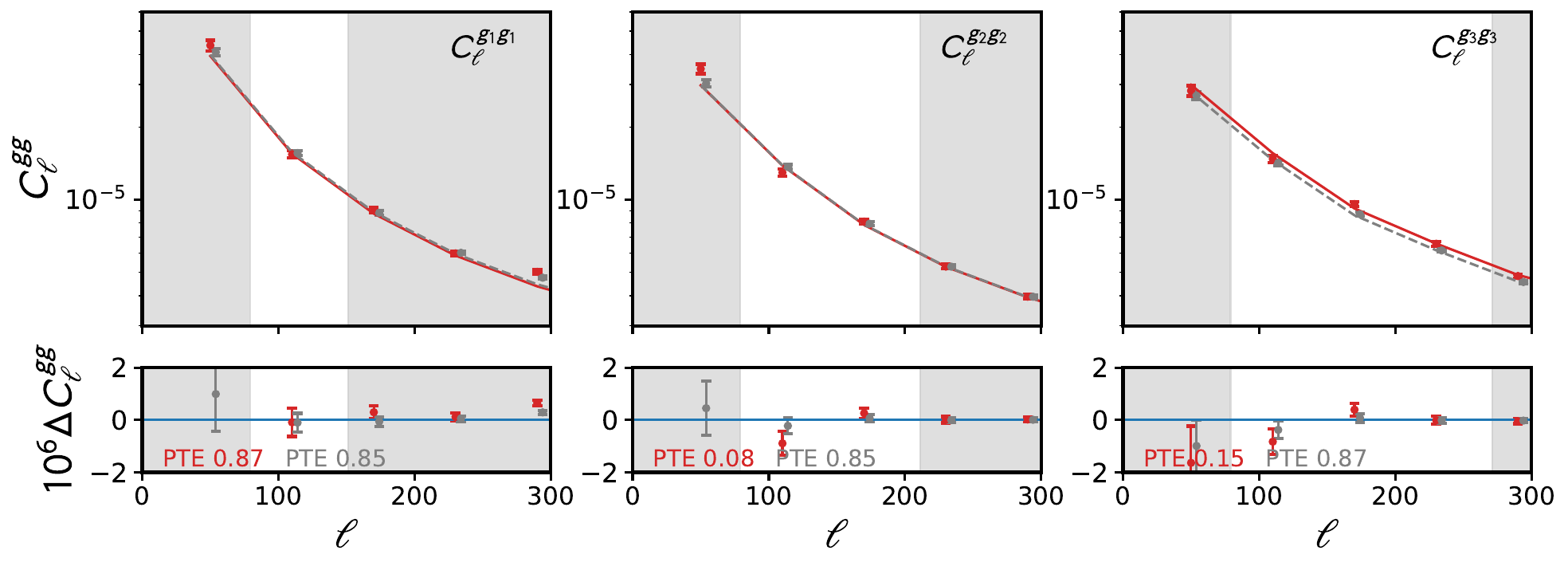} \\ 
    \vspace{-8pt} 
    
    \caption{Measurements of $C^{\kappa{g}}_\ell$ (\textbf{top} row) and  $C^{gg}_\ell$ (\textbf{bottom} row) for the three redshift samples of DESI Legacy galaxies. Shown are also the best fit lines where the cosmological parameters are fixed to that of \textit{Planck} 2018 CMB anisotropies and only the bias and shotnoise parameters are allowed to vary.
}
    \label{fig:bandpower LCDM fit}
\end{figure*}

\begin{figure*}
    \centering
    \includegraphics[width=\linewidth]{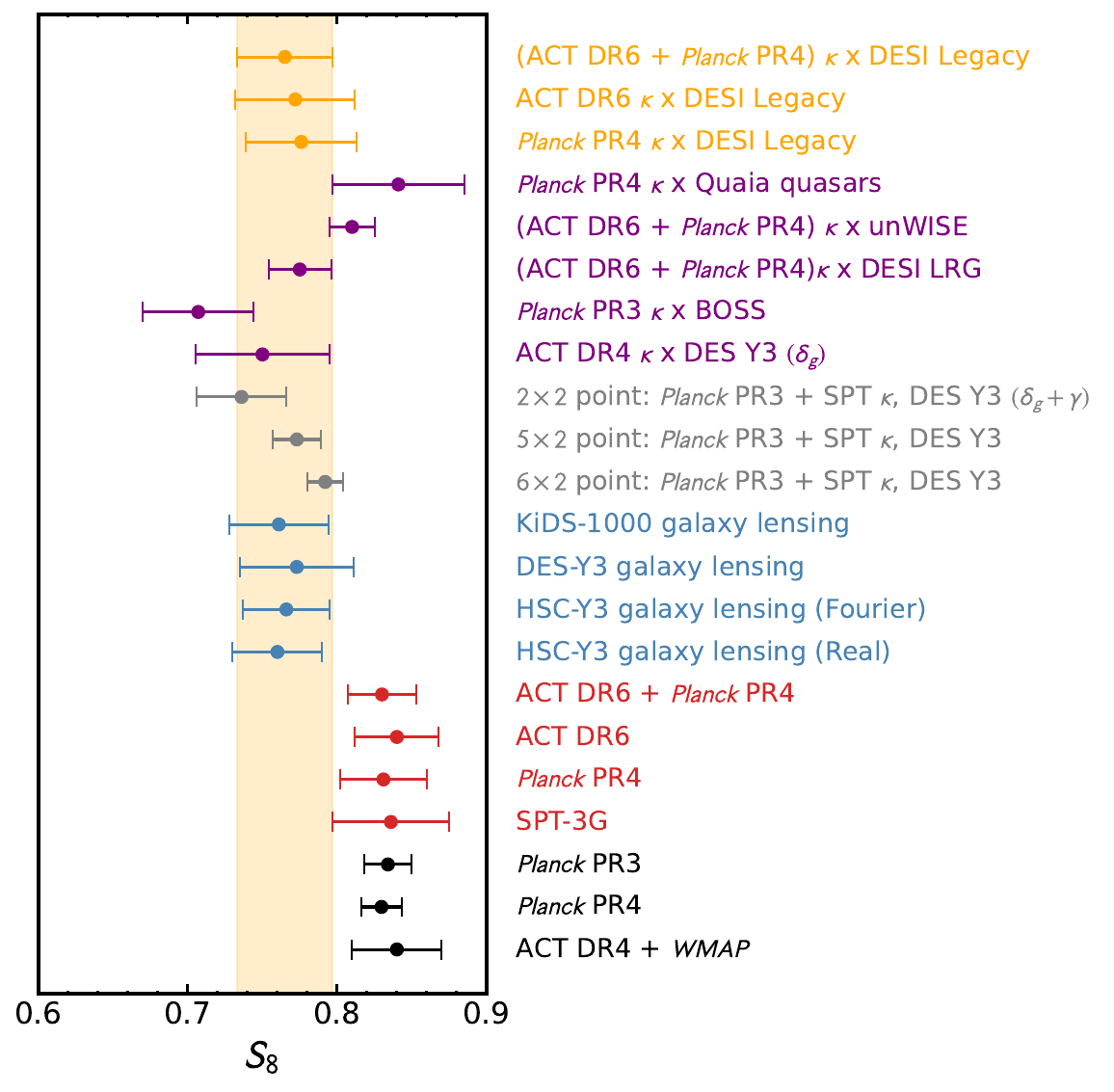} \\ 
    \caption{Compilation of $S_8$ measurements from this work in gold. Other cross-correlation measurements (purple), measurements combining CMB lensing and galaxy lensing (grey), galaxy weak lensing in blue, CMB lensing in red and extrapolation from CMB anisotropies (black). Gold shaded region consists of the $1\sigma$ uncertainty of our baseline ACT DR6+ \textit{Planck} PR4 lensing $\times$ DESI Legacy measured at $4\%$ level.}
    \label{fig: compilation_S8}
\end{figure*}


\begin{figure}
    \centering
    \includegraphics[width=\linewidth]{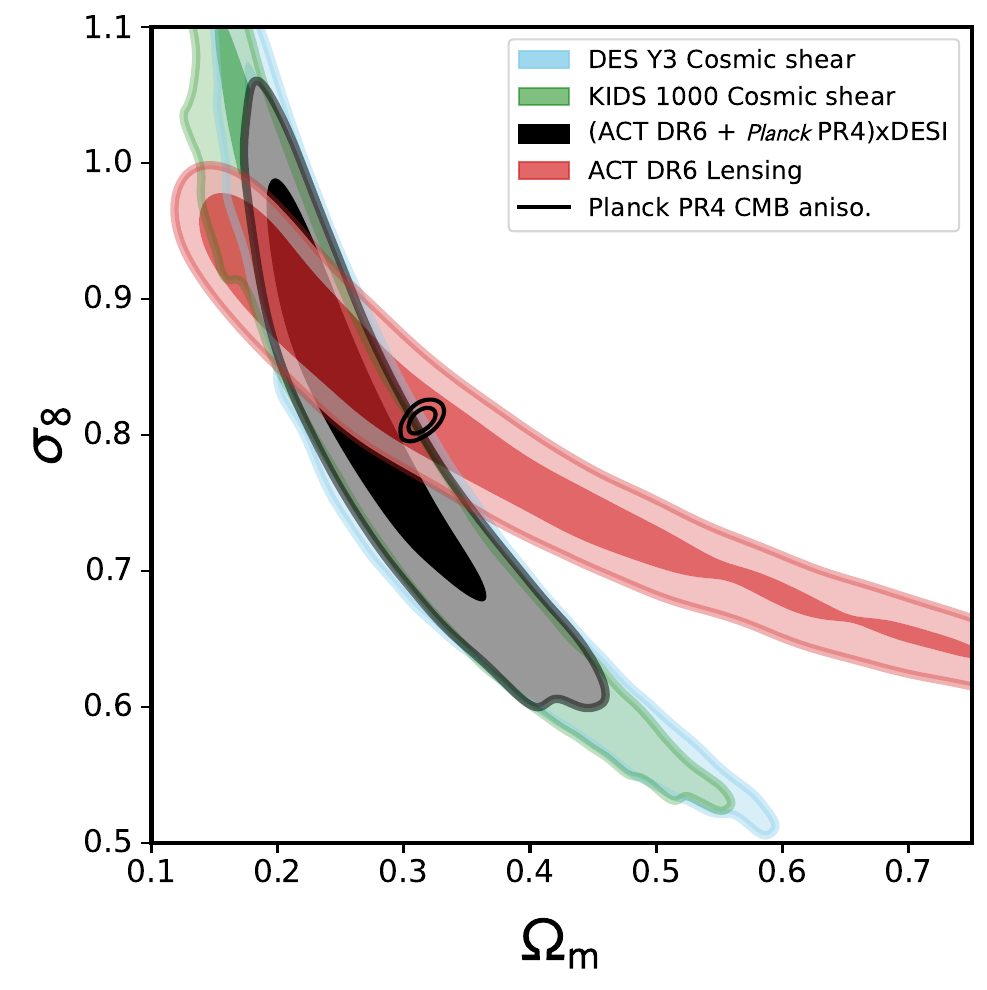} \\ 
    \caption{The $\Omega_m-\sigma_8$ contours for this work (black) compared to other lensing-only results: DES Y3 Cosmic shear (blue), KiDS-1000 Cosmic shear (green), and ACT DR6 lensing (red). The  black, unfilled contour shows the constraints from the Planck PR4 CMB anisotropy measurements for comparison. }
    \label{fig: compilation_lensing}
\end{figure}

\begin{figure}
    \centering
    \includegraphics[width=\linewidth]{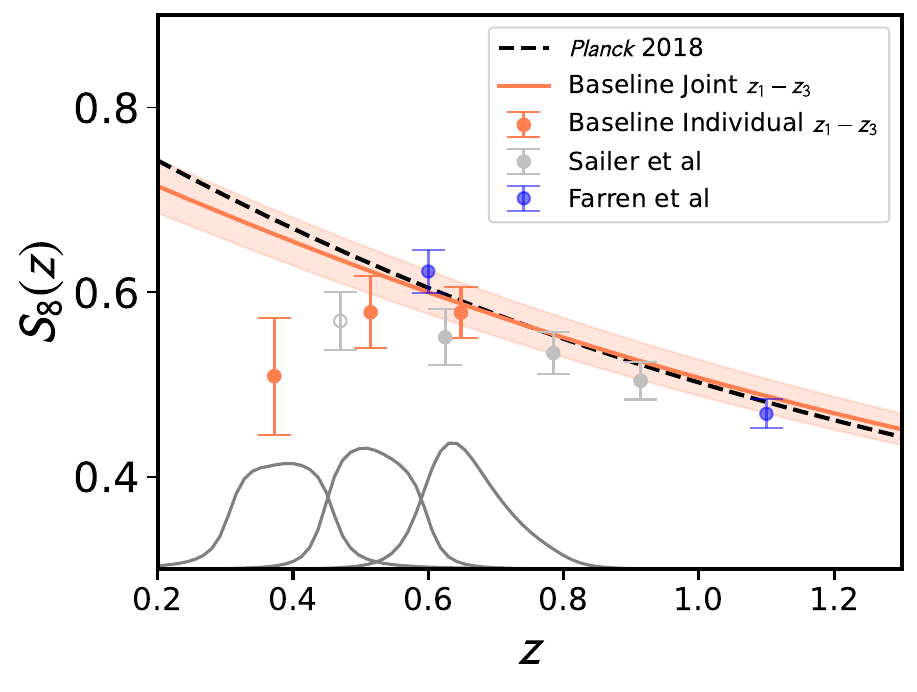} \\ 
    \caption{Results of the measured tomographic structure growth at each effective redshift (orange data points) and the joint redshift measurement (solid orange line).
    The blacked dashed line corresponds to the predicted growth evolution assuming a \textit{Planck} 2018 cosmology. In gray, we also plot $\alpha{dN/dz}$ of each galaxy sample, with $\alpha$ being an arbitrary constant for display purposes. We also show in silver for comparison the analysis from \cite{Sailer2024,Kim2024}, the cross-correlation of DESI LRG and ACT DR6 and PR4 lensing, the filled grey circles are their constraints using linear theory and extends to higher redshifts than the galaxy samples used in our analysis but are nevertheless consistent with our analysis. The first open circle is the result they obtain using their baseline hybrid effective field theory model. In blue we show the results obtained from the cross-correlation of the blue and green unWISE sample, correlated with ACT DR6 lensing \citep{farren2023atacama}.}
    \label{Fig. redevol}
\end{figure}

\subsection{Comparison with other measurements of large scale structure}

Fig.~\ref{fig: compilation_lensing} compares our results with other weak lensing surveys and the \textit{Planck} CMB anisotropy results. 
We show in the $\Omega_m - \sigma_8$ plane the constraints from cosmic shear measurements from Dark Energy Survey Year 3 \citep[DES Y3; ][]{2022PhRvD.105b3514A,2022PhRvD.105b3515S}, Kilo-Degree Survey \citep[KiDS 1000; ][]{2021A&A...645A.104A} and ACT DR6 lensing \citep{qu2023atacama}. Our measurements are very consistent with the other weak lensing-only results with a similar degeneracy direction. 
On the same plot, we also show in unfilled black contours the \textit{Planck} constraints from CMB anisotropy that we discussed in Sec.~\ref{sec.lcdm}.

We further discuss here comparisons of our results to measurements of structure growth from other large-scale structure observables. Fig.~\ref{fig: compilation_S8} shows the compilation of these measurements in terms of $S_8$ constraints. 
We include results from other cross-correlations with CMB lensing from ACT, \textit{Planck} and SPT in purple. Galaxy weak lensing surveys (DES, KiDS and HSC) are in blue and CMB lensing autospectrum analyses are in red.

\noindent We note that the posterior values of  HSC, DES, and the CMB lensing auto spectra measurements from ACT and \textit{Planck} PR4 are analysed with consistent prior choices to this work, while the other values are taken from the published results\footnote{\cite{ACT:2023kun} showed that the impact of using priors matching those in the ACT DR6 lensing measurements is small for the galaxy survey datasets.}.

We compare our results with measurements derived from cross-correlations of CMB lensing and galaxy positions, shown as purple points in Fig.~\ref{fig: compilation_S8}. Our results are very consistent with the work in \cite{Kim2024} and  \cite{Sailer2024}, that employed the same lensing map, but a different DESI galaxy sample containing only LRGs (thus one expect there to be significant overlap in the samples of galaxies used, in an analysis that extends to smaller scales using an EFT-based model.) They report  $S_8=0.775\pm0.021$, differing from our fiducial analysis by $0.26\sigma$. Our more aggressive scale cuts focusing on linear scales result in error bars that are $34\%$ larger than that obtained from the DESI LRG sample. The linear analysis in \cite{Sailer2024,Kim2024} is comparable to the analysis presented here, albeit differences in galaxy sample and redshift distribution discussed in Sec.~\ref{Sec. galaxy}.    In Fig. \ref{Fig. redevol}, we show a comparison of the constraints of $S_8$ as a function of redshift between our measurements and those from \cite{farren2023atacama} in blue and \cite{Sailer2024,Kim2024} in grey. It is worth noting that our analysis extends to lower redshifts than those reported in the unWISE and LRG analyses and at the overlapping redshifts the analyses are in good agreement with each other.  The linear analysis of \cite{Sailer2024,Kim2024} (filled grey circles) is in good agreement in terms of redshift and scales probes. This is not surprising since  we use the same lensing maps and the scales analyzed are similar for the linear analysis; we expect this level of agreement between the two analyses in the absence of independent systematic effects.

Another similar EFT-based analysis using the unWISE sample found $S_8=0.810\pm0.015$; this analysis at higher redshifts of 0.6 and 1.1 for the blue and green sample is consistent with our fiducial result within $1.3\sigma$. 
We find a similar $1.4\sigma$ consistency with the measurement of the cross-correlation of PR4 with Quaia quasars \citep{piccirilli2024growth} with $S_8=0.841\pm0.044$. Our result is in very good agreement ($0.27\sigma$), with the cross-correlation of the DES-Y3 MagLim galaxies with the ACT DR4 Lensing map \citep{2024JCAP...01..033M}. \cite{2024arXiv240704795C} measured the galaxy-galaxy lensing of the DESI Bright Galaxy Sample \citep[BGS; ][]{Hahn_2023} and LRG targets with the DES Y3 shear catalog. They found a lensing amplitude of $S_8=0.850^{+0.042}_{-0.050}$, consistent with the values from the primary CMB. Interestingly, they do not find the `lensing is low' tension to be redshift dependent, but rather could be due mismodeling of galaxy bias, or the deviation of the evolution of the intrinsic alignment signal from the usually assumed functional form.

Our results are also in good agreement with the measurements from the cross-correlation of CMB lensing and galaxy lensing. The first gray data-point shows the $2\times2$ point analysis of the correlation between CMB lensing from \textit{Planck} PR3 and SPT and DES Y3 galaxy clustering $\delta_g$ and galaxy shear $\gamma$ $\langle\delta_g\kappa\rangle+\langle\gamma\kappa\rangle$ \citep{2023PhRvD.107b3530C} which yields $S_8=0.736^{+0.032}_{-0.028}$ ($\sim0.7\sigma$). \cite{Abbott_2023} additionally includes correlations between galaxy positions and shear, as well as the autospectrum of the respective tracers, in the `$5\times2$' point analysis yielding $S_8=0.773\pm0.016$ $(\sim0.22\sigma)$. Adding the CMB lensing autospectrum from \textit{Planck} PR3 in the `$6\times2$' point analysis \citep{2023PhRvD.107b3530C} results in a  $S_8$ of $0.792\pm0.012$ $(0.82\sigma)$.

As already shown in Fig.~\ref{fig: compilation_lensing}, we also find good agreement between our results and those obtained from galaxy lensing. The blue points in Fig.~\ref{fig: compilation_S8} showing the constraints from a reanalysis of  KiDS-1000 DES-Y3 and HSC (Fourier and real space) data agree within $0.2\sigma$ with our baseline measurement in gold.
Our work is more similar to the combination of galaxy-galaxy lensing and clustering $2\times2$pt analysis, such as presented in \cite{2022PhRvD.106j3530P}, although with a different lensing kernel. Our results are again in good agreement with \cite{2022PhRvD.106j3530P} using the DES Y3 \texttt{MagLim} lens galaxies and shear, where $S_8=0.778^{+0.037}_{-0.031}$.

Finally, our results are consistent within 1.65,1.76,1.52 and $1.41\sigma$ to the DR6+PR4, ACT DR6, \textit{Planck} PR4 and SPT-3G results of the CMB lensing power spectrum respectively, shown as the red points in Fig.~\ref{fig: compilation_S8}. Comparison with the prediction from the primary CMB values are shown in black; refer to Sec.~\ref{sec.lcdm} for the detailed comparison with the cosmology preferred by the \textit{Planck} 2018 and NPIPE CMB anisotropies. Our measurement is also in $1.7\sigma$ agreement with the CMB anisotropies from ACT DR4 + \textit{WMAP}.

\section{Discussions and Conclusions} \label{sec.discussion}

We have presented cosmological results from the cross-correlation of the DESI Legacy Survey galaxies with the ACT DR6 and \textit{Planck} PR4 CMB lensing maps. In an analysis focused on linear scales and following a `blinding' procedure, we provide a $4.2\%$ constraint on $S_8$ with the combined ACT+\textit{Planck} correlation. 

Our analysis passes a suite of null and systematic tests, ensuring the measurement is robust to extragalactic foreground biases and other galaxy systematic tests. The blinding procedure prevented us from inferring cosmological parameters before the null tests were passed, reducing the risk of confirmation bias. At the price of some reduction in SNR, we restricted the analysis to linear scales $k<0.17h\si{Mpc}^{-1}$ where linearity of bias and of the matter power spectrum apply.


Our result for $S_8$ lies $1.9\sigma$ below the prediction from primary CMB results, constraining structure growth in the redshift range $0.3\leq{z}\leq0.8$. In Fig. \ref{Fig. redevol}, we exhibit the redshift dependence of $S_8$, by showing our constraint on this parameter within each redshift bin. While the joint constraint and analysis removing the lowest redshift bin are formally consistent with \textit{Planck} CMB predictions, hints of suppression of structure growth at lower redshifts probed by the lowest redshift bins are emerging, consistent with other independent analyses \citep[i.e., ][]{Sailer2024, Kim2024}.  \cite{2022MNRAS.516.5355A, 2023MNRAS.525.5554P} argued that the lower $S_8$ measured from weak lensing surveys relative to the \textit{Planck} primary CMB estimate could be due to the small-scale modeling not fully capturing the non-linearity and astrophysical effects such baryonic feedback, rather than a tension between the late and the early universe. Our analysis, however, focuses on mostly linear scales, but still finds consistent $S_8$ values with existing galaxy weak lensing measurements. Although the significance of the $S_8$ `tension' is lower in our case, due to larger uncertainties, our results might hint at the possibility of a redshift dependent effect affecting the large linear scales and not purely a modification affecting only the small scale power spectrum (See also discussions in the literature  about scale-independent suppression of growth \citep{2023PhRvL.131k1001N}). If this suppression is indeed due to new physics, it becomes particularly intriguing because the lowest redshift bin lies within the dark energy-dominated regime. Any deviations from expected structure growth on these redshifts could potentially point towards new insights on the behaviour of dark energy (i.e, \citep{desicollaboration2024desi2024vicosmological}.)

More statistical power and improved handling of the galaxy sample used for tomography will be needed before we can make concrete statements about the nature of this suppression on large scales -- new physics, statistical fluctuation, or perhaps systematics affecting particularly the lowest redshifts. For example, we have not considered changes to the Legacy Survey  galaxy sample compared to the previous analysis \citep{hang2021}. Despite the sample's robustness, there are several areas of potential improvement that we leave for future work, such as improved photometric redshift calibration using the newly available DESI spectra.
Some recent work in this direction was carried out by \cite{saraf2024effect}, who re-calibrated the photo-$z$ error distribution for the DESI Legacy Survey sample.
Their cross-correlation results with \textit{Planck} PR3 lensing map show overall a similar $1-3\sigma$ deviation below the theoretical expectation from the \textit{Planck} 2018 cosmology in bin 0-2, although the highest tomographic bin 3 has a
higher cross-correlation amplitude, closely consistent with the {\it Planck\/} prediction. 
Further efforts in understanding the exact calibration of the photo-$z$ data in this catalog will clearly be of interest.
Correlations of CMB lensing with low redshift galaxy samples, such as done by \cite{2024arXiv240704795C} with the photometric DESI BGS targets, are also an important alternative avenue in verifying our findings regarding $S_8$. In the future, the spectroscopic DESI BGS data will no doubt provide more insight into structure growth at low redshifts and its consistency with standard cosmology.

\section*{Acknowledgements}

FJQ and BDS acknowledge support from the European Research Council (ERC) under the European Union’s Horizon 2020 research and innovation programme (Grant agreement No. 851274). 
FJQ is grateful for Dr Anil Seal and Trinity College Henry Barlow Scholarship for their support. 
Computations were performed on the Niagara supercomputer at the SciNet HPC Consortium. SciNet is funded by Innovation, Science and Economic Development Canada; the Digital Research Alliance of Canada; the Ontario Research Fund: Research Excellence; and the University of Toronto.

CS acknowledges support from the Agencia Nacional de Investigaci\'on y Desarrollo (ANID) through Basal project FB210003. CEV received the support of a fellowship from “la Caixa” Foundation (ID 100010434). The fellowship code is LCF/BQ/EU22/11930099. JK acknowledges support from NSF grants AST-2307727 and AST-2153201. EC acknowledges support from the European Research Council (ERC) under the European Union’s Horizon 2020 research and innovation programme (Grant agreement No. 849169).

Support for ACT was through the U.S.~National Science Foundation through awards AST-0408698, AST-0965625, and AST-1440226 for the ACT project, as well as awards PHY-0355328, PHY-0855887 and PHY-1214379. Funding was also provided by Princeton University, the University of Pennsylvania, and a Canada Foundation for Innovation (CFI) award to UBC. The development of multichroic detectors and lenses was supported by NASA grants NNX13AE56G and NNX14AB58G. Detector research at NIST was supported by the NIST Innovations in Measurement Science program. 
ACT operated in the Parque Astron\'omico Atacama in northern Chile under the auspices of the Agencia Nacional de Investigaci\'on y Desarrollo (ANID). We thank the Republic of Chile for hosting ACT in the northern Atacama, and the local indigenous Licanantay communities whom we follow in observing and learning from the night sky.

This work has undergone DESI collaboration wide review. The authors thank the review chairs, Martin White and Joe DeRose, for their thorough and insightful comments. 

This material is based upon work supported by the U.S. Department of Energy (DOE), Office of Science, Office of High-Energy Physics, under Contract No. DE–AC02–05CH11231, and by the National Energy Research Scientific Computing Center, a DOE Office of Science User Facility under the same contract. Additional support for DESI was provided by the U.S. National Science Foundation (NSF), Division of Astronomical Sciences under Contract No. AST-0950945 to the NSF’s National Optical-Infrared Astronomy Research Laboratory; the Science and Technology Facilities Council of the United Kingdom; the Gordon and Betty Moore Foundation; the Heising-Simons Foundation; the French Alternative Energies and Atomic Energy Commission (CEA); the National Council of Humanities, Science and Technology of Mexico (CONAHCYT); the Ministry of Science, Innovation and Universities of Spain (MICIU/AEI/10.13039/501100011033), and by the DESI Member Institutions: \url{https://www.desi.lbl.gov/collaborating-institutions}. Any opinions, findings, and conclusions or recommendations expressed in this material are those of the author(s) and do not necessarily reflect the views of the U. S. National Science Foundation, the U. S. Department of Energy, or any of the listed funding agencies.

The authors are honored to be permitted to conduct scientific research on Iolkam Du’ag (Kitt Peak), a mountain with particular significance to the Tohono O’odham Nation.

%

\vspace{5mm}




\appendix








\section{Stellar density correction}
\label{sec: stellar}

In this section we show the galaxy density map correlation with the stellar density, before and after the correction. This correlation, $C_{\ell}^{gS}$, is shown in Fig.~\ref{fig.stellar} for bins 1 - 3. Given the ${\rm DESI}\times{\rm ACT}$ footprint, the impact of stellar density relatively small, about $1\%$ in the $\ell$ ranges adopted in the baseline analysis. The correction (see \cite{hang2021} for details) brings down the stellar density correlation by a further order of magnitude.

\begin{figure*}
    \centering
    \includegraphics[width=\linewidth]{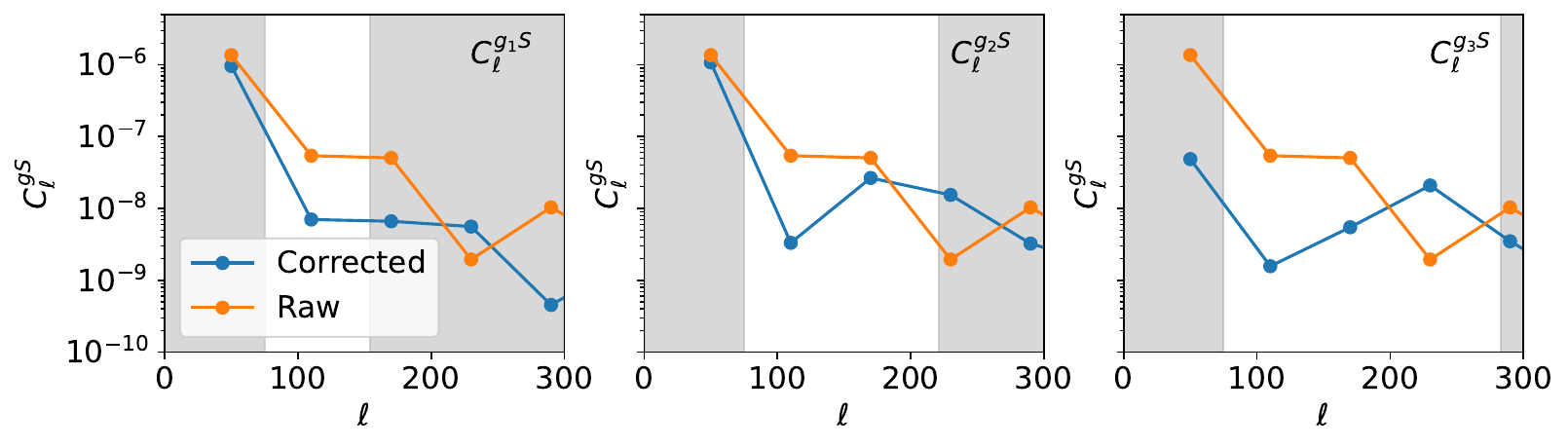}
    \caption{The correlation between galaxy density maps and the normalized ALLWISE total stellar density, $C_{\ell}^{gS}$, for the three tomographic bins with (blue) and without (orange) correction. We use the same measurement pipeline as the baseline $C_{\ell}^{gg}$ with the ${\rm DESI}\times{\rm ACT}$ footprint.} \label{fig.stellar}
\end{figure*}

\section{DESI galaxies $\times$ \textit{Planck} PR3 lensing reanalysis} \label{sec. planck}
In this section we provide a reanalysis of the cross-correlation of the \textit{Planck} PR3 lensing and the DESI Imaging galaxies as well as the detailed changes leading to the evolution from 
the original result in \citep{hang2021} to our baseline constrains. We use the priors listed in Table \ref{table:priors}. Among the pipeline improvements we include the correction $A^\mathrm{MC}_\ell$ for the lensing reconstruction misnormalization due to the presence of the mask to obtain an unbiased CMB lensing cross-correlation as $C^{\hat{\kappa}^{\mathrm{MC}}g}_\ell=A^\mathrm{MC}_{\ell}C^{\hat{\kappa}g}_\ell$. This normalization correction $A^\mathrm{MC}_{\ell}$ is computed with 480 Gaussian simulations and corresponding lensing reconstructions provided by \cite{Carron:2022}.

Using the scale cuts discussed in Sec.~\ref{ref.measurement} with $\ell_\mathrm{\min}(z)=50$, $\ell^{\mathrm{cross}}_\mathrm{\max}(z)=(170,230,290)$ and $\ell^{\mathrm{auto}}_\mathrm{\max}(z)=(170,230,290)$
we find $S_8=0.759\pm0.039$ using the full non linear power spectrum.  Excluding the first redshift bin which only contains one bandpower and is prior dominated on the shot noise and linear bias shifts the constraints up to $S_8=0.776\pm0.043$. We find a minimum $\chi^2=5.6$ and $\chi^2=7.79$ for the joint sample and for the fit excluding the first galaxy bin respectively. These corresponds to a PTE of 0.94 and 0.65 for the 12 and 10 bandpowers used. The inclusion of BAO to break the degeneracy of $\sigma_8$ and $\Omega_m$ results in $\sigma_8=0.719\pm0.031$ for the joint galaxy samples  and $\sigma_8=0.73\pm0.03$ for the analysis excluding redshift bin 1.

The original \cite{hang2021} paper measures the combination $S^{\times}_8\sim\sigma_8({\Omega_m/0.3})^{0.79}=0.760\pm0.023$. This analysis used a pseudo-C$\ell$ analysis with scales in the range from $10\leq\ell\leq500$ and a theoretical covariance matrix with no off-diagonal covariance matrix. In terms of non linear modeling, \citep{hang2021} prescribes a different bias parameter to the linear and non-linear regimes:

\begin{equation} C^{gg}_\ell{\sim}b^2_\mathrm{linear}P^{mm,\mathrm{linear}}_\ell+b^2_\mathrm{non-linear}(P^{mm}_\ell-P^{mm,\mathrm{linear}}_\ell),
\end{equation}

\begin{equation} C^{\kappa{g}}_\ell{\sim}b_\mathrm{linear}P^{mm,\mathrm{linear}}_\ell+b_\mathrm{non-linear}(P^{mm}_\ell-P^{mm,\mathrm{linear}}_\ell)
\end{equation}

where $P^{mm}$ are computed using the CAMB power spectrum.

Since then, many improvements to the original analysis pipeline have been made, as described in the main text -- and it is instructive to demonstrate how the adopted changes alter the derived values of  $S^{\times}_8$.  A summary of these changes can be found in Fig.~\ref{fig.s8evol}. Starting with the leftmost data point in black, we have the $S^\times_8=0.760\pm0.023$ constrained by \citep{hang2021}.

\begin{figure*}
    \centering
    \includegraphics[width=\linewidth]{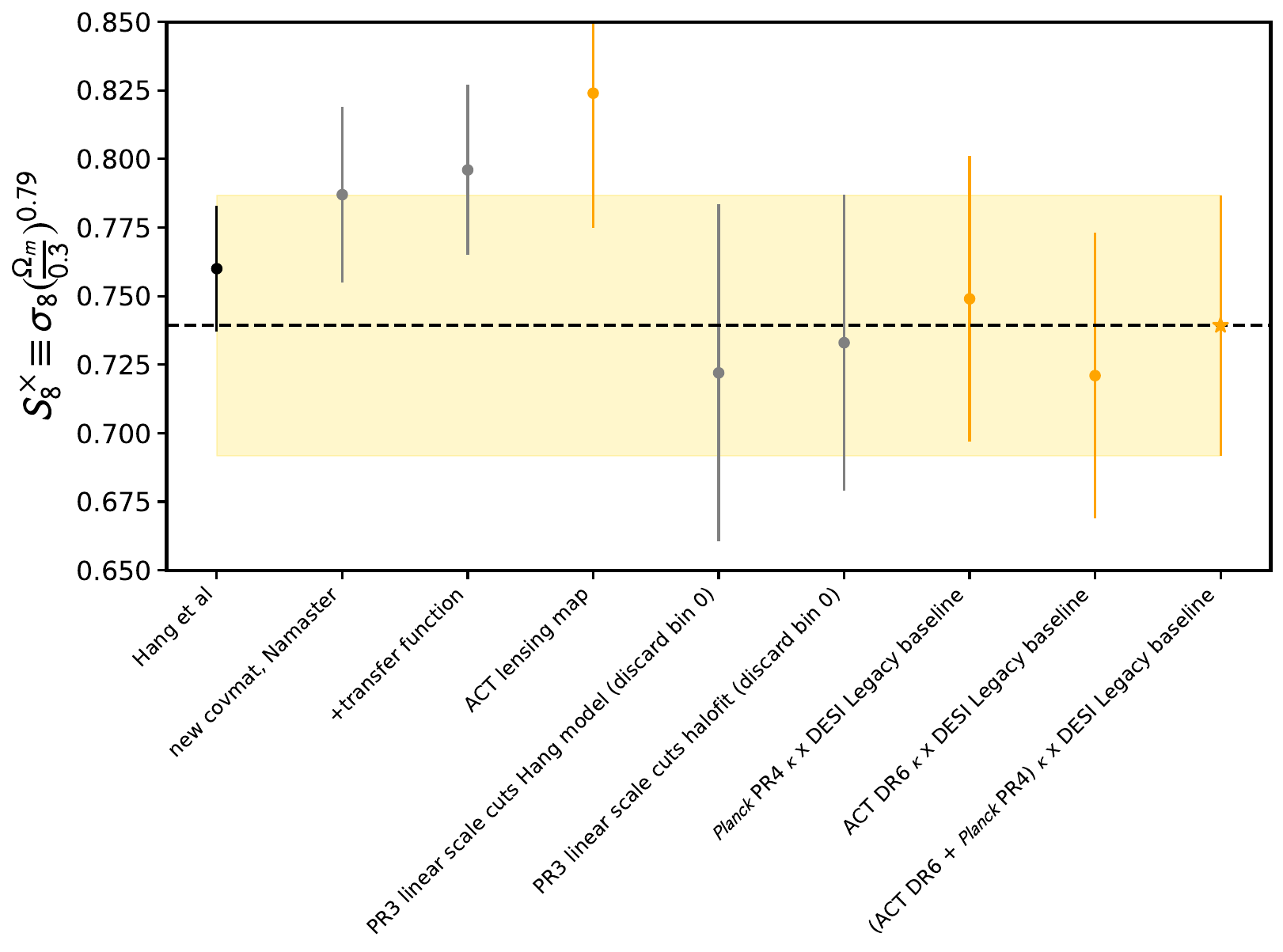}
    \caption{
    We show systematically the impact on the changes implemented in this work on $S^\times_8$, smoothly connecting the values obtained in \citep{hang2021} to the constraints obtained in this work.} \label{fig.s8evol}
\end{figure*}

\begin{itemize}
    \item Keeping similar scale cuts to the original analysis with $50\leq\ell\leq470$, but with broader binning $\Delta\ell=60$ instead of the original $\Delta\ell=10$ to reduce bin to bin correlations. $\Delta{S^{\times}_8}=+0.027$
    \item Inclusion of a multiplicative transfer function on the \textit{Planck} PR3 lensing map. $\Delta{S^{\times}_8}=+0.009$
    \item Restrict to linear scales, including discarding altogether the first redshift bin $z_0$. $\Delta{S^{\times}_8}=-0.06$. This corresponds to the fifth gray datapoint in Fig. \ref{fig.s8evol}. The shift in constraint suggests that the model employed in \citep{hang2021} might not be accurately describing the non-linear scales.
    \item Switching the model from the one prescribed in \citep{hang2021} to the baseline  Halofit model used in this work results in a shift of $\Delta{S^{\times}_8}=+0.01$ ($S^{\times}_8=0.733\pm0.054$). This small shift compared to the results using the model of \citep{hang2021} that accounts for non linear bias shows that within the scale cuts used we are insensitive to the details of the modeling of the non linear bias.
\end{itemize}

\subsection{Testing the effects of non-linear bias} \label{app.nonlinear}
We further verify that the scale cuts used in this work corresponding to spatial scales of $k_\mathrm{max}=0.15,0.16,0.17h/\si{Mpc}$ are insensitive to the effects on non linear bias.

First, as mentioned in the previous section, the small shift of $\Delta{S^{\times}_8}=+0.01$ when switching from the model of \citep{hang2021} to a constant bias model independent of scale, supports that the effect of non linear bias is small. We conduct an additional test where we analyse the ACT DR6 x DESI Legacy data using more conservative scale cuts that discards the first redshift bin and limits the $k_\mathrm{max}$ of $z_2$ and $z_3$ to $k_\mathrm{max}=0.12,0.13h/\si{Mpc}$ respectively. The constraint in $S_8$ shifts by $0.03\sigma$ as shown by Fig.~\ref{fig:planckXcorner-cut}. Hence, we conclude that the scales chosen are robust to non linear bias modeling.

\begin{figure}
    \centering
\includegraphics{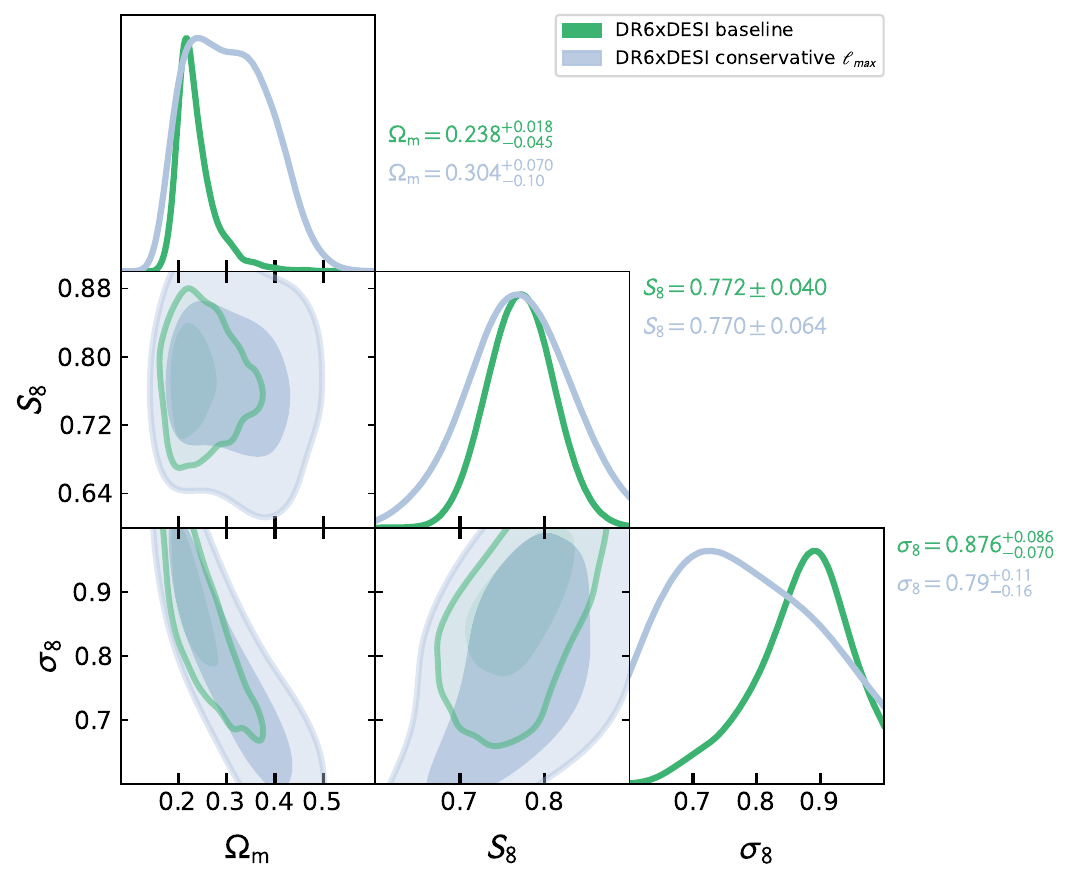}
    \caption{Cosmological constraints of the ACT DR6 $\times$ DESI Legacy Imaging data, using more conservative scale cuts for $\ell_{\rm max}$.}
    \label{fig:planckXcorner-cut}
\end{figure}


\begin{figure}
    \centering
\includegraphics{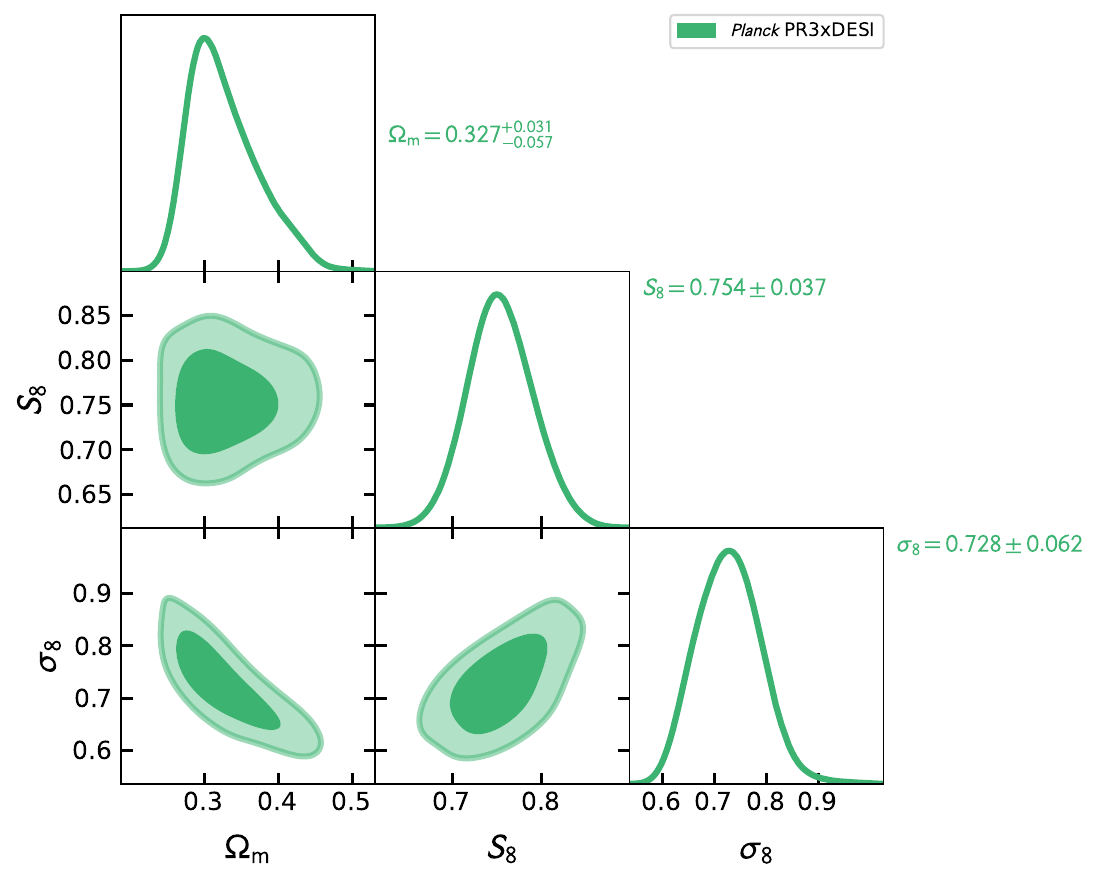}
    \caption{Cosmological constraints of the DESI Imaging galaxies and \textit{Planck} PR3 lensing. }
    \label{fig:planckXcorner}
\end{figure}

\begin{figure*}
    \centering
    \includegraphics[width=0.5\linewidth]{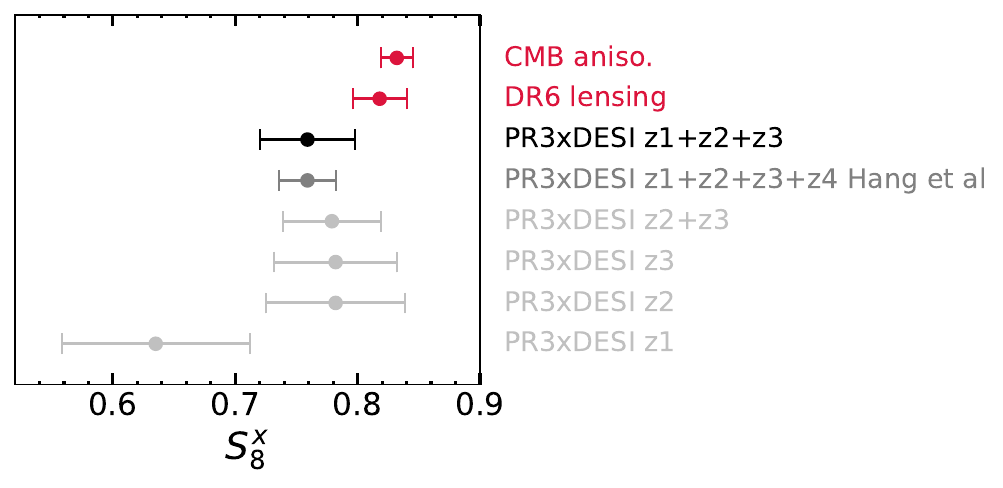}%
    \includegraphics[width=0.5\linewidth]{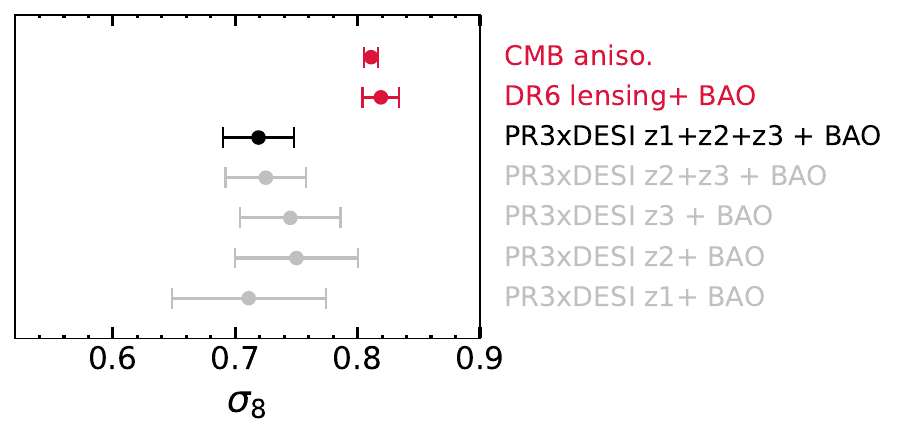}
    \caption{Whisker plots for the DESI Imaging galaxies and \textit{Planck} PR3 lensing. \textbf{Left} panel shows $S_8$ constraints using $C^{gg}_\ell$ and $C^{\kappa{g}}_\ell$ only. \textbf{Right} panel shows the $\sigma_8$ constraints when including BAO.}
    \label{fig:planckXcorner_fiducial}
\end{figure*}




    
    

\section{Null bandpowers for the highlighted null tests in the main text} \label{appendix:null}
We show specifically the cross-correlation between the signal nulled $\kappa$ map obtained by differencing the $150\si{GHz}$ and $90\si{GHz}$ maps and the galaxy fields and the cross correlation of the curl with the galaxy fields that have marginal failures of 0.03 in the second bin and 0.04 in the third bin respectively. It can be seen that those low PTE's are consistent with fluctuations.

\begin{figure}
    \centering
\includegraphics[width=\linewidth]{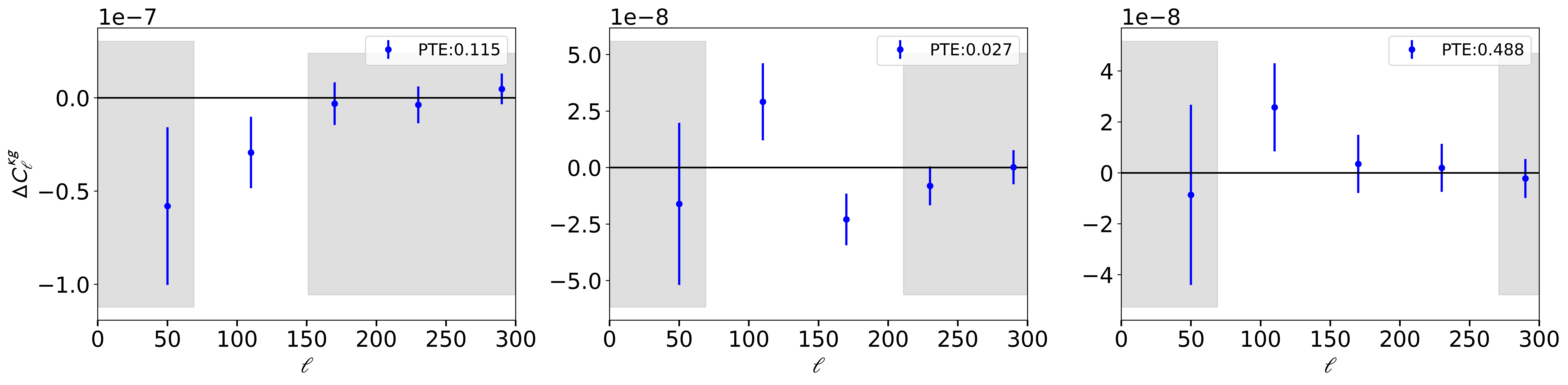}
    \caption{Cross-correlation of the individual galaxy redshift bins with the temperature-only lensing null map obtained by performing lensing reconstruction on the difference between the $150\si{GHz}$ and the $90\si{GHz}$  CMB maps}
    \label{fig:nullfog}
\end{figure}

\begin{figure}
    \centering
\includegraphics[width=\linewidth]{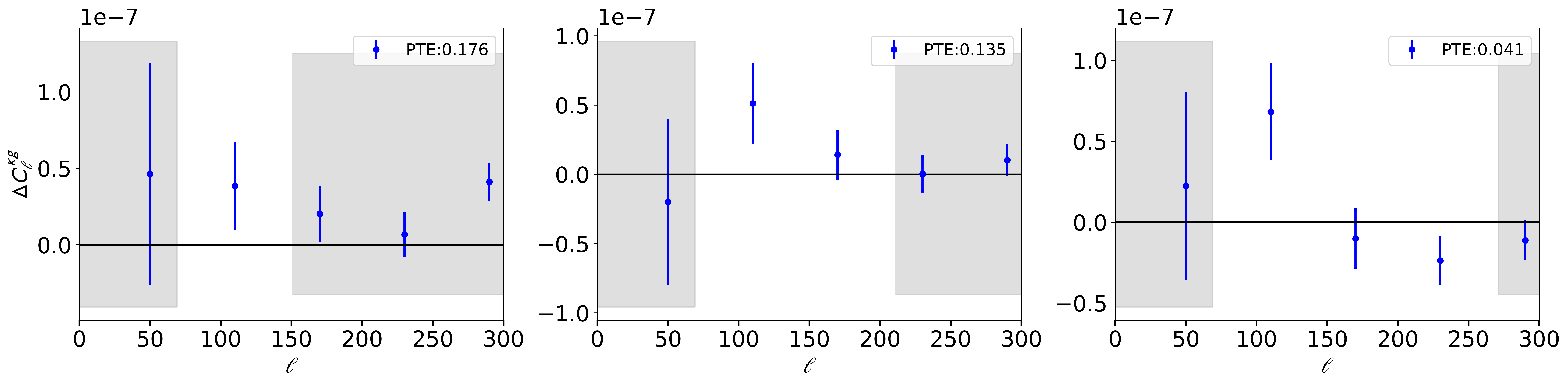}
    \caption{Cross-correlation of the individual galaxy redshift bins with the reconstructed lensing curl maps. }
    \label{fig:nullcurl}
\end{figure}

\section{Shifts in redshift distribution}
\label{apdx: dndz shifts}

\begin{figure}
    \centering
\includegraphics[width=\linewidth]{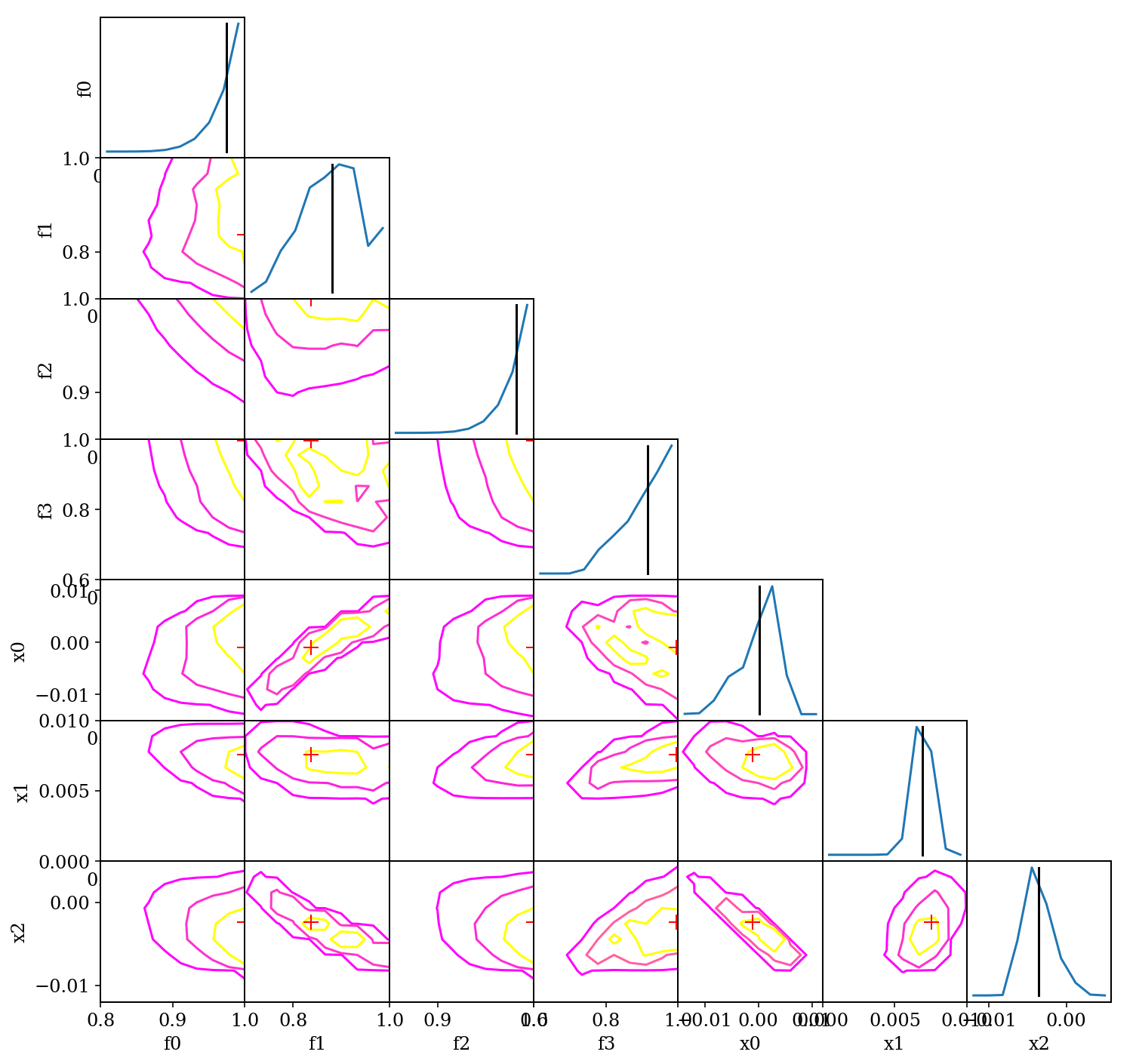}
    \caption{The posteriors of the redshift distribution calibration nuisance parameters in \cite{hang2021} using the cross-correlation of galaxies in different tomographic bins with the \textit{Planck} 2018 cosmology.}
    \label{fig:nz-param-posteriors}
\end{figure}

\begin{figure}
    \centering
\includegraphics[width=\linewidth]{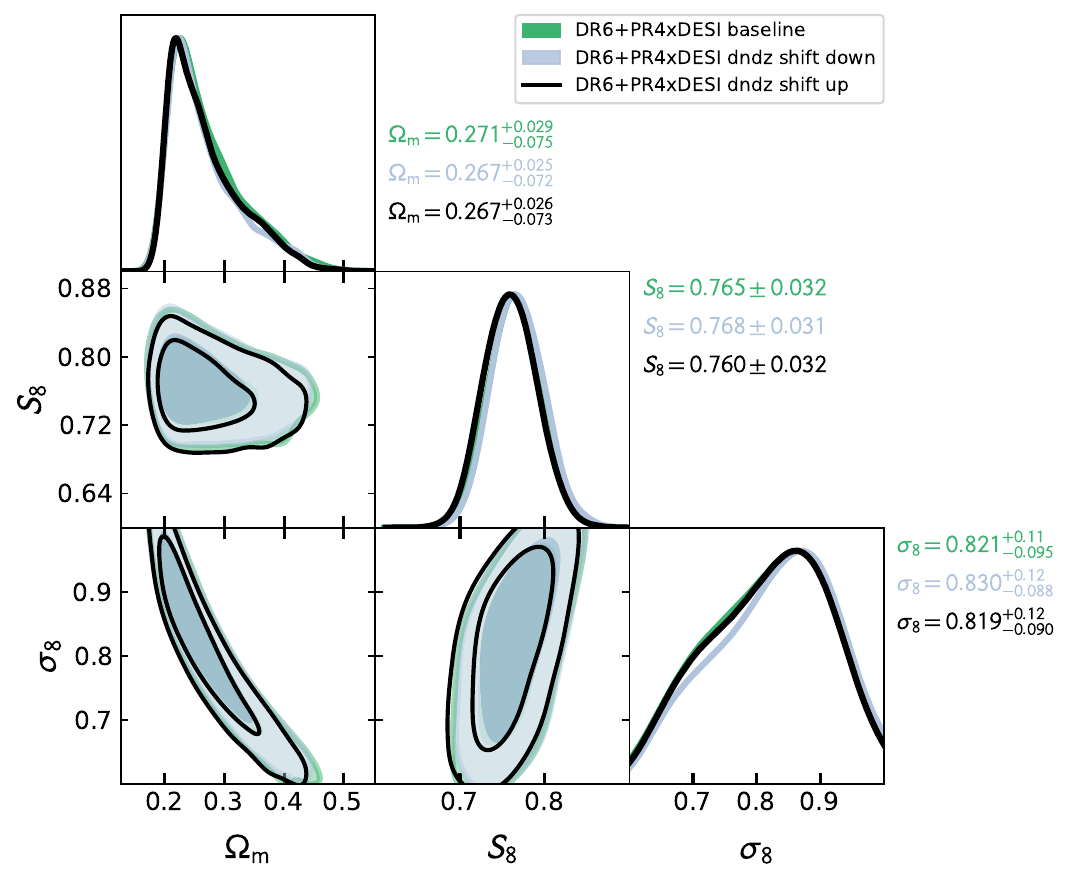}
    \caption{The change in cosmological parameter constraints by shifting the mean redshifts of the three tomographic bins. The green contours and curves show the baseline scenario, whereas blue and black show cases where the mean redshift of all bins being shifted downwards and upwards, by $\Delta z=0.0025,0.005,0.01$ for $z1-z3$ respectively.}
    \label{fig:dndz_shifts}
\end{figure}

In this Appendix, we show the impact of the photometric redshift uncertainty of the DESI Legacy Survey on the cosmological results. Fig.~\ref{fig:nz-param-posteriors} shows the posterior of the $n(z)$ calibration nuisance parameters in 
\cite{hang2021}. The parameters $f_0 - f_3$ scales the tail the redshift distribution, such that $a_i=f_i a_i^{\rm spec}$ in Eq.~\ref{eq: lorentz}, where $a_i^{\rm spec}$ is the best-fit value for the calibration sample, and $f_i\leq1$ such that the tails are larger for the photometric sample. The parameters $x_i$ are the shift parameters of the tomographic bin centres, and satisfy $\sum_{i=0}^{4} x_i=0$ such that mean redshift of the four tomographic bins are unchanged.

From the posteriors of these parameters, we identify the $3\sigma$ range of the shift parameters for bins 1 - 3: $\Delta z = 0.0025, 0.005, 0.01$. We test scenarios where the mean redshifts of the three bins are simultaneously shifted upward, i.e. $z\rightarrow z+\Delta z$, and downwards, i.e. $z\rightarrow z-\Delta z$. Fig.~\ref{fig:dndz_shifts} shows the change in cosmological parameter constraints for these two cases, compared to the baseline. We find that the shifts in are $<0.2\sigma$.

\bibliography{sample631}{}
\bibliographystyle{aasjournal}



\end{document}